\documentclass[pdflatex,sn-mathphys-num]{sn-jnl}
\usepackage[T1]{fontenc}
\usepackage{lmodern}
\usepackage{amsmath,amssymb,amsthm,mathtools}
\usepackage{enumitem}
\usepackage{booktabs}
\usepackage{tikz}
\usepackage{refcount}
\allowdisplaybreaks

\newtheorem{theorem}{Theorem}[section]
\newtheorem{lemma}[theorem]{Lemma}
\newtheorem{proposition}[theorem]{Proposition}
\newtheorem{corollary}[theorem]{Corollary}
\newtheorem{mainthm}{Theorem}

\newtheorem{conjecture}[mainthm]{Conjecture}
\theoremstyle{definition}
\newtheorem{definition}[theorem]{Definition}
\newtheorem{example}[theorem]{Example}
\theoremstyle{remark}
\newtheorem{remark}[theorem]{Remark}

\newcommand{\R}{\mathbb R}
\newcommand{\C}{\mathbb C}
\newcommand{\F}{\mathbb F}
\newcommand{\Z}{\mathbb Z}
\newcommand{\E}{\mathbb E}
\newcommand{\Pb}{\mathbb P}
\newcommand{\cL}{\mathcal L}
\newcommand{\cA}{\mathcal A}
\newcommand{\cD}{\mathcal D}
\newcommand{\cS}{\mathcal S}
\newcommand{\cZ}{\mathcal Z}
\newcommand{\Gr}{\mathrm{Gr}}
\newcommand{\op}{\mathrm{op}}
\newcommand{\rect}{\square}
\newcommand{\nop}[1]{\|#1\|_{\mathrm{op}}}
\newcommand{\nF}[1]{\|#1\|_{F}}
\DeclareMathOperator{\codim}{codim}
\DeclareMathOperator{\lct}{lct}
\DeclareMathOperator{\conv}{conv}
\DeclareMathOperator{\Res}{Res}

\hypersetup{pdftitle={Small-ball asymptotics of Gaussian matrix products: Open Fermi shells and the Grassmannian},bookmarksdepth=2}
\newcommand{\numericsurl}{\url{https://github.com/jiahuatian/gaussian-product-small-balls}}
\begin{document}

\title[Small balls of Gaussian matrix products]{Small-ball asymptotics of Gaussian matrix products: Open Fermi shells and the Grassmannian}
\author{\fnm{Jiahua} \sur{Tian}}\email{jtian1905@gmail.com}
\affil{\orgdiv{School of Physics}, \orgname{East China Normal University},
\orgaddress{\city{Shanghai}, \country{China}}}

\abstract{\unboldmath Let $P=A_N\cdots A_1$ be a product of independent standard real or complex Gaussian matrices of arbitrary fixed sizes. For complex entries we determine the exact leading terms, of the form $\kappa\,\varepsilon^C(\log1/\varepsilon)^q$, of the operator-norm and Frobenius small-ball probabilities of $P$, together with the corresponding asymptotics of the Laplace transform of $\nF P^2$. The exponent $C$ is the known diversity order, while the logarithmic power follows a filling rule: when fermions are placed on ladders of states determined by the sizes and the last occupied level has degeneracy $\mu$ and contains $h$ particles, $q=h(\mu-h)$. The determinant of this last shell expands in Pl\"ucker coordinates graded by the Schubert cells of the Grassmannian $\Gr(h,\mu)$, with leading coefficients proportional to normalized degrees of Schubert varieties. For real entries we prove that the Frobenius small-ball probability has an exact positive leading term of order $\varepsilon^{C/2}(\log1/\varepsilon)^q$ and that the operator-norm probability has the same order, which gives a proof of the known real log canonical threshold and its multiplicity that needs no explicit resolution of singularities. We evaluate the real leading coefficient in closed form for every shell with at most one logarithm, a class that contains all products of two factors, and we conjecture its general form, which agrees with the classical constant of vector chains. Applications include the outage probability of cascaded multi-antenna channels at high signal-to-noise ratio and the Bayesian evidence of deep linear networks.}

\maketitle

\tableofcontents

\section{Introduction}\label{sec:intro}

Let $\F\in\{\R,\C\}$ and let $A_j\in\F^{d_j\times d_{j-1}}$, $j=1,\dots,N$, be independent matrices with independent standard Gaussian entries: $N(0,1)$ if $\F=\R$, and density $\pi^{-1}e^{-|z|^2}$ on $\C$ if $\F=\C$. Put
\[
 P=A_N\cdots A_1\in\F^{d_N\times d_0},\qquad n=\min_{0\le j\le N}d_j .
\]
Almost surely $P$ has rank $n$ and we let $x_1,\dots,x_n>0$ be its nonzero squared singular values. For a fixed size vector $d=(d_0,\dots,d_N)$ we study
\begin{equation}\label{eq:observables}
 F_\op(\varepsilon)=\Pb\bigl(\nop P^2\le\varepsilon\bigr),\qquad
 F_F(\varepsilon)=\Pb\bigl(\nF P^2\le\varepsilon\bigr),\qquad
 Z(t)=\E\,e^{-t\nF P^2},
\end{equation}
as $\varepsilon\downarrow0$ and $t\to\infty$. Thus $F_\op(\varepsilon)$ is the probability that \emph{all} nonzero squared singular values are at most $\varepsilon$, and $F_F$ is the probability that their sum is. These are small-ball probabilities of the product around its most singular point $P=0$. They are finite-size quantities, complementary to the hard-edge gap probability, which concerns the smallest singular value, and to the large-$n$ and large-$N$ limits that dominate the literature on products \cite{AIK13,KZ14,LWZ16,ABK14}.

On the other hand, the Laplace transform $Z(t)$ is a common object behind several of the structures studied here. Up to the rescaling $t\mapsto t/2$, it is the classical partition function, at inverse temperature $t$, of the energy $H(A)=\tfrac12\nF{A_N\cdots A_1}^2$ with respect to the Gaussian reference measure, and for the zero-target deep linear networks recalled below this partition function is the Bayesian evidence. The limit $t\to\infty$ is a zero-temperature limit that probes the geometry of the zero set $\{P=0\}$ of the energy. Heuristically, the exponent is half the number of real directions transverse to this set, as in equipartition. In the same picture the logarithms come from continuous families of equally costly degenerations of the product that interpolate between competing components of $\{P=0\}$, and the scales along these families are organized by Schubert combinatorics, and we develop this picture in Section~\ref{sec:picture}. Our results show that these singular asymptotics admit a fermionic filling description and a Schubert-graded Grassmannian structure. We note that the fermionic picture is combinatorial: the energy of a filling is the codimension of a rank stratum of $\{P=0\}$ (Proposition~\ref{prop:geometry}), and for complex entries the same filling arises from the determinantal structure of the product ensemble.

The same three quantities appear in three rather different fields. In wireless communication, products of Gaussian matrices model double- and multiple-scattering, keyhole and cascaded MIMO channels \cite{GBGP02,Mul02,YB11,YB07}. The outage probability of optimal beamforming is $F_\op$ evaluated at a multiple of the inverse signal-to-noise ratio, and that of orthogonal space-time block codes is $F_F$ (Section~\ref{sec:wireless}). Yang and Belfiore determined the diversity order, that is, the exponent of the small-ball probability, for all sizes \cite{YB11,YB07}. For two factors the distribution of the largest eigenvalue is known in determinantal form \cite{JMWG08,ZJWM10}, but logarithmic corrections and coding gains have been extracted from it only in special cases, essentially when one node has a single antenna \cite{JMWG08,GL07,Zho10,ZRJW12}, and for the closely related dual-hop relay channel Zhong et al.\ describe the general case as challenging \cite{ZRJW12}. For multi-layer scattering channels, Taricco studied the diversity--multiplexing tradeoff \cite{Tar20}, and Wei, Zheng, Corander and Taricco approximated the outage of orthogonal space-time block codes in closed form by matching exact moments \cite{WZCT15}. In singular learning theory, $Z(t/2)$ is the Bayesian evidence at sample size $t$ of a zero-target deep linear network with Gaussian prior. In Watanabe's theory \cite{Wat09} its decay $t^{-\lambda}(\log t)^{m-1}$ is governed by the real log canonical threshold $\lambda$, or learning coefficient, and by its multiplicity $m$. For deep linear networks both were computed by Aoyagi through resolution of singularities \cite{Aoy24,Aoy25}, and P\'epin Lehalleur and Rim\'anyi, using quiver representations, identified $\lambda$ with half the codimension of the zero fibre of the multiplication map \cite[Thm.~8.6]{PLR24}. The constant in front of $t^{-\lambda}(\log t)^{m-1}$, which enters the free energy at order one, is known for networks with one-dimensional output, whose evidence Hanin and Zlokapa computed exactly in terms of Meijer-$G$ functions \cite{HZ23}. Finally, in the geometry of matrix multiplication, the zero fibre $\Sigma_d=\{P=0\}$ and the ideal $I_d$ generated by the entries of $P$ have been studied in \cite{PLR24,KR26,Ter26}. Their codimension $C$ and the number $\theta$ of top-dimensional components are invariant under permutations of $d$, and $\E\,\nF P^{-2z}$ is the Gaussian average of the archimedean zeta function of $I_d$.

In Section~\ref{sec:dln} we record the resulting learning coefficients, multiplicities and leading constants of deep linear networks, together with the exact integration over the last layer.

Our results are organized by a filling rule, which we explain in Section~\ref{sec:statements}. Deleting one entry equal to $n$ from $d$ and subtracting $n$ from the others gives offsets $\nu_1,\dots,\nu_N$. The Mellin factors $\Gamma(s+\nu_j)$ of the product ensemble define $N$ ladders of one-particle states at energies $\nu_j,\nu_j+1,\dots$, which we fill with $n$ fermions. The highest occupied level, the Fermi level, has degeneracy $\mu$ and occupation $h$, and we put $q=h(\mu-h)$ (Figure~\ref{fig:ladders}). For complex entries we prove in Theorem~\ref{thm:A} that
\[
 F_\op(\varepsilon)\sim\kappa_\op\varepsilon^C(\log1/\varepsilon)^q,\qquad F_F(\varepsilon)\sim\kappa_F\varepsilon^C(\log1/\varepsilon)^q,\qquad Z(t)\sim\cA\,t^{-C}(\log t)^q,
\]
with explicit, generally distinct Gamma-product constants $\kappa_\op$, $\kappa_F$ and $\cA$, which Corollary~\ref{cor:lct} reformulates as $\lct(I_d)=C=\codim\Sigma_d$ with multiplicity $q+1$. The logarithms come from a shell determinant. In Theorem~\ref{thm:grass} we show that this determinant is a Pl\"ucker expansion whose terms are indexed by the Schubert cells of $\Gr(h,\mu)$, with logarithmic degrees equal to the cell dimensions and leading coefficients proportional to normalized Schubert degrees. Consequently the number of Pl\"ucker terms is $\binom\mu h=\chi(\Gr(h,\mu))$, which by \cite{PLR24,KR26,Ter26} is also the number $\theta$ of top components of $\Sigma_d$, the top degree is $q=\dim\Gr(h,\mu)$, and the shell factor of $\kappa_\op$ and $\cA$ is $\deg\Gr(h,\mu)/q!$.

For real entries we prove in Theorem~\ref{thm:real} that $F_F(\varepsilon)\sim\kappa_\R\varepsilon^{C/2}(\log1/\varepsilon)^q$ with $\kappa_\R>0$, and that $F_\op$ has the same two-sided order, so that $(C/2,q+1)$ is the real log canonical threshold and multiplicity. Our proof is independent of Aoyagi's theorem, on which the identification of the threshold in \cite{PLR24} rests, and of any explicit resolution of the multiplication map. It combines a positive-probability event built from the Bartlett decomposition with Theorem~\ref{thm:A}, through a real--complex comparison of principalizations. The constant $\kappa_\R$ is harder to reach, because the singular values of real products have no tractable joint law. We find it by first rewriting the complex constant as a product over pairs formed by an occupied state and the first empty state of another ladder (Proposition~\ref{prop:particle}). Each factor in that product is a Gamma function $\Gamma(f_i-\varepsilon_p)$ of the gap between an occupied energy $\varepsilon_p$ and the first empty energy $f_i$ on another ladder. In Theorem~\ref{thm:closed} we show that for a closed shell, $h=\mu$, and for a shell with a single logarithm, $q=1$, the real constant contains the same pair product with every Gamma argument halved,
\[
 \kappa_\R=\frac{2^{-NC/2}}{\Gamma(C/2+1)}\,\frac{\prod_jS_1(e_j)}{S_1(n)}\,\frac{\prod^*_{(i,p)}\Gamma\bigl(\tfrac{f_i-\varepsilon_p}2\bigr)}{\prod_j\prod_{e_j<u\le n}\Gamma\bigl(\tfrac{\nu_j+u}2\bigr)},
\]
where $e_j$ is the number of particles on ladder $j$, $S_1(m)=\prod_{t\le m}\Gamma(t/2)$, and $\prod^*$ runs over the ladders $i$ and the particles $p$ on other ladders with $f_i\ne\varepsilon_p$. The precise statement is in Section~\ref{sec:statementsreal}. For a closed shell the proof evaluates a Morse--Bott integral along the unique top component of $\Sigma_d$, in a slice adapted to the layered structure of $P$ in which the transversal Hessian is block diagonal. It then shows that no other divisor of a principalization contributes: over $\C$ this follows from Theorem~\ref{thm:A}, and over $\R$ it follows from the real--complex comparison. For an open shell the same Morse--Bott integrand has exactly $q$ logarithmically divergent factors. Replacing each of them by the coefficient of its pole and multiplying by the shell factor reproduces Theorem~\ref{thm:A} exactly, and on the real side it reproduces our closed form for two factors and the classical one for vector chains, the products in which some $d_j=1$. For $q=1$ we prove this residue rule by an exact local normal form of the ideal of entries along the intersection of the two top components, and we conjecture that it gives $\kappa_\R$ in general (Conjecture~\ref{conj:open}).

\begin{example}[Four $2\times2$ factors]\label{ex:intro}
For $d=(2,2,2,2,2)$ and complex entries,
\[
 \Pb(\nop P^2\le\varepsilon)\sim\tfrac1{48}\,\varepsilon^3(\log1/\varepsilon)^4,\qquad
 \Pb(\nF P^2\le\varepsilon)\sim\tfrac1{72}\,\varepsilon^3(\log1/\varepsilon)^4 .
\]
Here $\Sigma_d$ has $\theta=6$ components of codimension $3$, but there are only four powers of the logarithm. In shell language $\mu=4$ and $h=2$, and the Grassmannian $\Gr(2,4)$ has Euler characteristic $6$, dimension $4$ and degree $2$, so that the shell factor is $2/4!=1/12$ (Section~\ref{sec:fourfactors}). For real entries the residue rule predicts $\Pb(\nF P^2\le\varepsilon)\sim\frac1{576}\,\varepsilon^{3/2}(\log1/\varepsilon)^4$. For three real $3\times3$ factors the shell is closed ($h=\mu=3$), and Theorem~\ref{thm:closed} gives $\Pb(\nF P^2\le\varepsilon)\sim\frac{\pi^2}{192}\,\varepsilon^{3}$.
\end{example}

Our proofs use the joint law of the squared singular values of complex Gaussian products as a polynomial ensemble with Meijer-$G$ weights \cite{AIK13,KZ14}, its independence of the ordering of the sizes \cite{IK14}, the Bartlett decomposition, and principalization of ideals in characteristic zero \cite{Hir64,Wlo05,Kol07}. Several statements that we need are known, and we recall them with attribution. These are the exponent $C$ as the diversity order and its invariance under permutations of the sizes \cite{YB11,YB07}, the identity $C=\codim\Sigma_d$, the quadratic integer program behind the filling rule \cite[Thm.~6.1]{PLR24} and the count $\theta=\binom\mu h$ \cite{PLR24,KR26,Ter26}, and the real pair $(C/2,q+1)$, which is due to Aoyagi \cite{Aoy24}, after the two-factor case of Aoyagi and Watanabe \cite{AW05}, and was related to $\codim\Sigma_d$ in \cite[Thm.~8.6]{PLR24}; Lemma~\ref{lem:aoyagi} translates between Aoyagi's indices and ours. They also include the two-factor keyhole case $d=(2,1,2)$ \cite{JMWG08} and the cases $N=1$ (the Selberg integral \cite{FW08}) and $n=1$ (products of Gamma variables \cite{ST70}). For complex entries, the logarithmic power $q$ and the constants for all $N$ and all sizes are new, as is the Pl\"ucker/Schubert structure of the shell. Also new are the probabilistic proof on the real side, the real constant for every shell with at most one logarithm together with its conjectured extension, and the outage asymptotics. P\'epin Lehalleur and Rim\'anyi express the number of top components as a binomial coefficient in the same integers that determine the multiplicity \cite[Prop.~7.6, Rem.~8.7]{PLR24}, and Theorem~\ref{thm:grass} shows that the Grassmannian enters through the Schubert grading of the maximal minors of a Wronskian matrix (Section~\ref{sec:proofC}).

This paper is organized as follows. In Section~\ref{sec:statements} we define the shell data and state our results, including the particle form of the constants. In Section~\ref{sec:examples} we work out examples and exact consistency checks, among them the Schubert structure of Theorem~\ref{thm:grass} for four $2\times2$ factors and the first values of $\kappa_\R$ for three factors and $n\ge2$, and we explain the form of the results through a low-temperature physical picture. In Sections~\ref{sec:proofA}, \ref{sec:proofC} and \ref{sec:proofB} we prove Theorems~\ref{thm:A}, \ref{thm:grass} and \ref{thm:real}. In Section~\ref{sec:proofB} we also derive the two-factor constant directly from the real Wishart density, which provides an independent check of what follows. In Section~\ref{sec:proofD} we prove Theorem~\ref{thm:closed} and Proposition~\ref{prop:particle}, and we explain the residue rule behind Conjecture~\ref{conj:open}. In Section~\ref{sec:applications} we discuss the consequences for MIMO outage and for deep linear networks and close with a discussion of open problems. In Appendix~\ref{app:invariance} we sketch an elementary proof of the width-permutation invariance used throughout.

\section{Shell data and statements}\label{sec:statements}

\subsection{Shell data and constants}

\begin{definition}[Shell data]\label{def:shell}
Let $d=(d_0,\dots,d_N)\in\Z_{\ge1}^{N+1}$ with $N\ge1$, and let $n=\min_jd_j$. Delete one entry equal to $n$ from the multiset $\{d_0,\dots,d_N\}$ and subtract $n$ from each of the $N$ remaining entries; the result is the multiset of \emph{offsets} $\{\nu_1,\dots,\nu_N\}\subset\Z_{\ge0}$. For $L\in\Z_{\ge0}$ put
\[
 \mu_L=\#\{j:\nu_j\le L\},\qquad N_<(L)=\sum_{j=1}^N(L-\nu_j)_+=\sum_{M<L}\mu_M .
\]
The \emph{Fermi level} $L_F$ is the unique integer with $N_<(L_F)<n\le N_<(L_F+1)$, and we put
\[
 \mu=\mu_{L_F},\qquad h=n-N_<(L_F),\qquad q=h(\mu-h),
\]
so that $1\le h\le\mu$. The shell is \emph{closed} if $h=\mu$, so that $q=0$, and \emph{open} otherwise. The \emph{occupation numbers} are $n_L=\mu_L$ for $L<L_F$, $n_{L_F}=h$ and $n_L=0$ for $L>L_F$. A level is \emph{occupied} if $n_L>0$, which happens exactly for $\min_j\nu_j\le L\le L_F$, and $\sum_Ln_L=n$. Finally we put
\[
 C=\frac{n(n+1)}2+\sum_L L\,n_L .
\]
\end{definition}

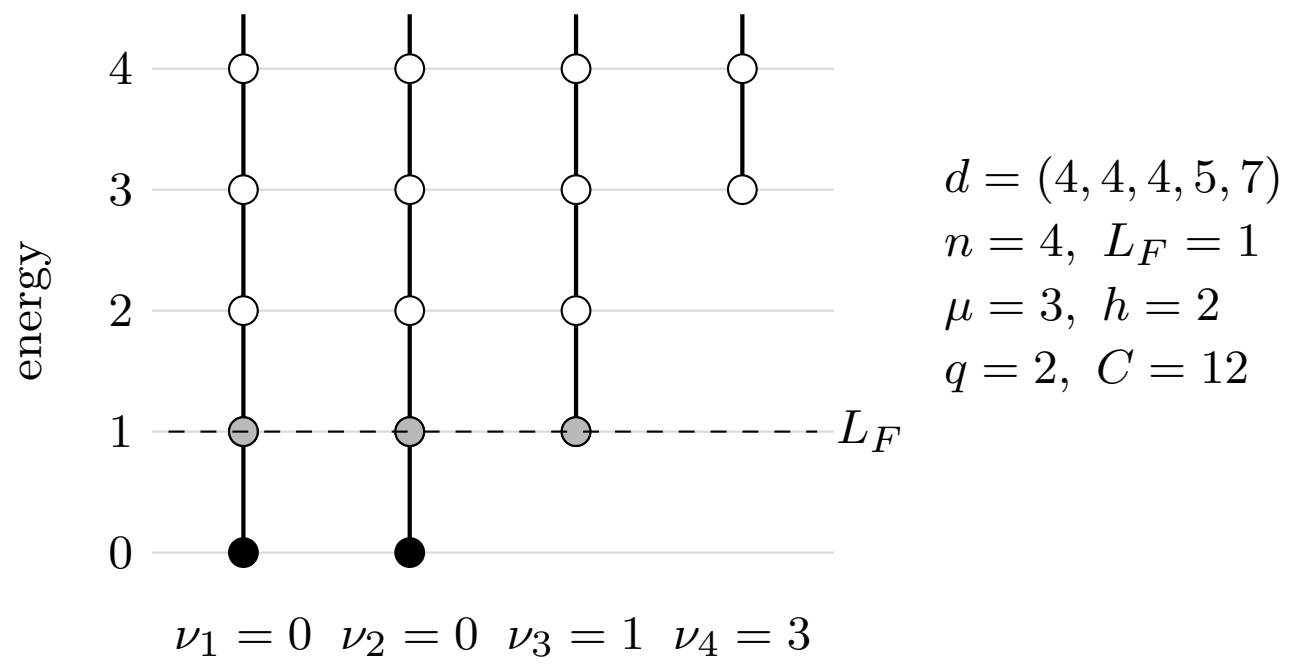
\begin{figure}[t]
\centering
\begin{tikzpicture}[x=1.1cm,y=0.8cm]
 \foreach \L in {0,...,4} {\draw[gray!30] (0.45,\L) -- (4.55,\L); \node[left,font=\scriptsize] at (0.45,\L) {$\L$};}
 \node[rotate=90,font=\scriptsize] at (-0.25,2) {energy};
 \foreach \j/\v in {1/0,2/0,3/1,4/3} {
  \draw[thick] (\j,\v) -- (\j,4.45);
  \foreach \L in {\v,...,4} {\draw[fill=white] (\j,\L) circle (2.7pt);}
  \node[font=\scriptsize] at (\j,-0.7) {$\nu_{\j}=\v$};
 }
 \fill (1,0) circle (2.7pt); \fill (2,0) circle (2.7pt);
 \foreach \j in {1,2,3} {\draw[fill=gray!55] (\j,1) circle (2.7pt);}
 \draw[dashed] (0.55,1) -- (4.45,1) node[right,font=\scriptsize] {$L_F$};
 \node[right,align=left,font=\small] at (5.1,2.3) {$d=(4,4,4,5,7)$\\[1pt] $n=4,\ L_F=1$\\[1pt] $\mu=3,\ h=2$\\[1pt] $q=2,\ C=12$};
\end{tikzpicture}
\caption{The filling rule for $d=(4,4,4,5,7)$, with offsets $\nu=(0,0,1,3)$. Each ladder carries one state (circle) per energy, starting at its offset. The four fermions fill the two states at energy $0$ (black) and any two of the three states at the Fermi level $L_F=1$ (grey), so that the shell is open with $\mu=3$, $h=2$ and $q=h(\mu-h)=2$. The $\binom32=3$ choices are the minimizers of Lemma~\ref{lem:filling} and the top components of $\Sigma_d$. For the minimizer that fills the grey states on ladders $1$ and $2$, the first empty energies are $f=(2,2,1,3)$.}
\label{fig:ladders}
\end{figure}

The terminology reflects the following picture, which we draw in Figure~\ref{fig:ladders}. Ladder $j$ carries one-particle states at energies $\nu_j,\nu_j+1,\nu_j+2,\dots$. Level $L$ has degeneracy $\mu_L$, and $N_<(L)$ states lie strictly below it. Filling $n$ fermions into the lowest states, with one particle per state, fills every level below $L_F$ and puts $h$ particles into the $\mu$ states at $L_F$, so that the ground state is unique precisely when the shell is closed.

\begin{lemma}[Ground states]\label{lem:filling}
For $e\in\Z_{\ge0}^N$ with $\sum_je_j=n$ put
\begin{equation}\label{eq:Egas}
 E(e)=\frac{n^2+\sum_je_j^2}{2}+\sum_j\nu_je_j
 =\frac{n(n+1)}2+\sum_{j=1}^N\sum_{k=0}^{e_j-1}(\nu_j+k).
\end{equation}
Then $\min_eE(e)=C$, and the minimum is attained exactly at the $\binom\mu h$ vectors
\[
 e_j=(L_F-\nu_j)_++\eta_j,\qquad \eta\in\{0,1\}^{\{j:\nu_j\le L_F\}},\ \textstyle\sum\eta_j=h .
\]
\end{lemma}

\begin{proof}
The second expression in \eqref{eq:Egas} follows from $\sum_{k<e}(\nu+k)=e\nu+e(e-1)/2$. The double sum is the total energy of the $n$ states that are occupied when ladder $j$ holds its $e_j$ lowest states. This total is minimal exactly when the occupied states are $n$ lowest states of the union of the ladders. Such a choice must contain all $N_<(L_F)$ states below $L_F$ and $h$ of the $\mu$ states at level $L_F$.
\end{proof}

Proposition~\ref{prop:geometry} below shows that $E(e)$ is the codimension of a rank stratum of $\Sigma_d$. In this form, $\min E$ is the quadratic integer program of \cite[Thm.~6.1]{PLR24}, and the minimizers of Lemma~\ref{lem:filling} are those described in \cite[Prop.~7.6]{PLR24}. The constants in our results are built from the shell data as follows.

\begin{definition}[Constants]\label{def:constants}
All products below over $L$ run over occupied levels, and empty products equal $1$. We put
\[
 R_L=\prod_{j:\,\nu_j>L}\Gamma(\nu_j-L),\qquad
 V=\prod_{L<M}(M-L)^{n_Ln_M},\qquad
 B=\prod_{s=0}^{h-1}\frac{s!}{(\mu-1-s)!},
\]
\[
 G=\prod_{i=0}^{n-1}\prod_{j=1}^N\Gamma(i+\nu_j+1),\qquad S_n=\prod_{i=0}^{n-1}i! ,
\]
and define
\begin{equation}\label{eq:constants}
\begin{aligned}
 \kappa_\op&=\frac{BV}{G}\prod_L\Bigl[\frac{R_L\,\Gamma(L+1)}{\Gamma(L+n+1)}\Bigr]^{n_L},\\
 \cA&=\frac{BV}{G\,S_n}\prod_L\bigl[R_L\,\Gamma(L+1)\bigr]^{n_L},\qquad
 \kappa_F=\frac{\cA}{\Gamma(C+1)} .
\end{aligned}
\end{equation}
\end{definition}

\begin{example}\label{ex:constants}
For the size vector $d=(4,4,4,5,7)$ of Figure~\ref{fig:ladders} the offsets are $\nu=(0,0,1,3)$. Thus $\mu_0=2$ and $\mu_1=3$, and $N_<(1)=2<4\le N_<(2)=5$ gives $L_F=1$, $\mu=3$, $h=2$ and $q=2$. The occupied levels are $L=0$ and $L=1$, with $n_0=n_1=2$, so that $C=10+2=12$. The factors of $R_L$ come from the ladders that have no state at level $L$, each contributing $\Gamma$ of its distance above $L$. At $L=0$ these are ladders $3$ and $4$, at distances $1$ and $3$, and at $L=1$ only ladder $4$, at distance $2$. Hence
\[
 R_0=\Gamma(1)\,\Gamma(3)=2,\qquad R_1=\Gamma(2)=1 .
\]
The two occupied levels are one unit apart, so that $V=1^{n_0n_1}=1$, and $B=\frac{0!}{2!}\cdot\frac{1!}{1!}=\frac12$, which is $\deg\Gr(2,3)/2!$ because $\Gr(2,3)\cong\mathbb P^2$ has degree one. Further $S_4=0!\,1!\,2!\,3!=12$, and $G$ has one factor for each ladder,
\[
 G=(0!\,1!\,2!\,3!)^2\cdot(1!\,2!\,3!\,4!)\cdot(3!\,4!\,5!\,6!) .
\]
By \eqref{eq:constants},
\[
 \cA=\frac{BV}{G\,S_4}\,R_0^2R_1^2=\frac1{6G},\qquad
 \kappa_F=\frac{\cA}{12!},\qquad
 \kappa_\op=\frac{BV}{G}\Bigl(\frac{R_0}{4!}\Bigr)^2\Bigl(\frac{R_1}{5!}\Bigr)^2=693\,\kappa_F ,
\]
in agreement with~\eqref{eq:ratio} below. For complex entries Theorem~\ref{thm:A} below therefore gives $\Pb(\nop P^2\le\varepsilon)\sim\kappa_\op\,\varepsilon^{12}(\log1/\varepsilon)^2$.
\end{example}

\subsection{Complex entries and the Grassmannian}\label{sec:statementC}

\begin{mainthm}\label{thm:A}
Let $\F=\C$. For every size vector $d$, as $\varepsilon\downarrow0$ and $t\to\infty$,
\begin{align}
 \Pb\bigl(\nop P^2\le\varepsilon\bigr)&=\kappa_\op\,\varepsilon^C(\log1/\varepsilon)^q\bigl(1+O(1/\log(1/\varepsilon))\bigr),\label{eq:A1}\\
 \E\,e^{-t\nF P^2}&=\cA\,t^{-C}(\log t)^q\bigl(1+O(1/\log t)\bigr),\label{eq:A2}\\
 \Pb\bigl(\nF P^2\le\varepsilon\bigr)&=\kappa_F\,\varepsilon^C(\log1/\varepsilon)^q\bigl(1+o(1)\bigr).\label{eq:A3}
\end{align}
The constants are positive, and
\begin{equation}\label{eq:ratio}
 \frac{\kappa_\op}{\kappa_F}=\frac{\Gamma(C+1)\,S_n}{\prod_L\Gamma(L+n+1)^{n_L}} .
\end{equation}
\end{mainthm}

Since $\nop P\le\nF P$, the ratio \eqref{eq:ratio} is at least $1$. It equals $1$ for $n=1$, where the two events coincide, and exceeds $1$ for every size vector with $n\ge2$ in Table~\ref{tab:examples}.

\begin{corollary}[Complex threshold and multiplicity]\label{cor:lct}
Let $\F=\C$.
\begin{enumerate}[label=(\roman*)]
\item The function $z\mapsto\E\,\nF P^{-2z}$ is holomorphic on $\Re z<C$ and extends meromorphically to $\C$. At $z=C$ it has a pole of order exactly $q+1$, and
\[
 \E\,\nF P^{-2z}=\frac{\cA\,q!}{\Gamma(C)\,(C-z)^{q+1}}+O\bigl((C-z)^{-q}\bigr).
\]
\item Let $I_d$ be the ideal generated by the entries of $P$ in the polynomial ring $\C[A_1,\dots,A_N]$ of the $D=\sum_{j=1}^Nd_jd_{j-1}$ matrix entries. Then
\[
 \lct(I_d)=\lct_0(I_d)=C=\codim_\C\Sigma_d .
\]
Here $\lct_0$ is the threshold at the origin. For every ball $U$ centred at $0$, the integral $\int_U\bigl(\sum_{a,b}|P_{ab}|^2\bigr)^{-z}dA$ is finite for $z<C$ and has a pole of order exactly $q+1$ at $z=C$.
\end{enumerate}
\end{corollary}

Note that $\lct\le\codim$ holds for every ideal, since near a general point of a top-dimensional component the ideal is contained in the reduced ideal of that component. Corollary~\ref{cor:lct} says that the product ideal attains this bound.

\begin{corollary}\label{cor:square}
Let $d=(n,n,\dots,n)$, i.e.\ a product of $N$ independent $n\times n$ complex Ginibre matrices. Write $n=Nk+h$ with $k\ge0$ and $1\le h\le N$. Then
\[
 C=\frac{n(n+1)}2+N\binom k2+hk,\qquad q=h(N-h),\qquad \theta=\binom Nh .
\]
In particular $F_\op$ and $F_F$ carry a logarithm unless $N$ divides $n$. If $N\ge n$, then $C=n(n+1)/2$, $q=n(N-n)$, and
\[
 \kappa_\op=\frac{1}{S_n^N\,(n!)^n}\prod_{s=0}^{n-1}\frac{s!}{(N-1-s)!}.
\]
For two factors ($N=2$), a single logarithm is present exactly when $n$ is odd.
\end{corollary}

\begin{proof}
Here all $\nu_j=0$, so $\mu_L=N$ and $N_<(L)=NL$. Hence $L_F=k$ and $\mu=N$, while $R_L=1$ and $G=S_n^N$. If $N\ge n$, then $k=0$, $h=n$, $V=1$, and a single level is occupied.
\end{proof}

We now describe the full polynomials behind Theorem~\ref{thm:A}. Let $\rect=((\mu-h)^h)$ be the rectangular partition with $h$ rows of length $\mu-h$, and let $f^{\lambda/\rho}$ be the number of standard Young tableaux of skew shape $\lambda/\rho$. For a partition $\rho\subset\rect$ let $X_\rho\subset\Gr(h,\mu)$ be the Schubert variety of codimension $|\rho|$, so that $\dim X_\rho=q-|\rho|$. Its degree in the Pl\"ucker embedding is $\deg_{\rm Pl}X_\rho=f^{\rect/\rho}$ \cite[\S9.4]{Ful97}. For a subvariety $X\subset\Gr(h,\mu)$ we call
\[
 \operatorname{vol}_{\rm Pl}(X)=\frac{\deg_{\rm Pl}X}{(\dim X)!}
\]
its \emph{normalized Pl\"ucker degree-volume}, so that $\operatorname{vol}_{\rm Pl}(X_\rho)=f^{\rect/\rho}/(q-|\rho|)!$. To an $h$-subset $T=\{s_1<\dots<s_h\}\subset\{0,\dots,\mu-1\}$ associate the partition
\[
 \rho(T)_c=s_{h+1-c}-(h-c),\qquad c=1,\dots,h .
\]
This is a bijection onto the partitions contained in $\rect$, and $T_0=\{0,\dots,h-1\}$ corresponds to $\rho=\varnothing$. In the fermionic picture we read $T$ as the set of orbitals occupied by the $h$ particles of the open shell, with its $\mu$ orbitals labelled $0,\dots,\mu-1$, and the remaining $\mu-h$ orbitals as holes. Since these orbitals have the same energy, we order them by their labels, so that one lies below another when its label is smaller. Row $c$ of $\rho(T)$ then counts the holes below the $c$-th highest particle. Hence the box in row $c$ and column $k$ of $\rect$ corresponds to the pair formed by the $c$-th highest particle and the $k$-th lowest hole, and it lies in $\rho(T)$ exactly when the hole lies below the particle. The $q$ boxes of $\rect$ are thus the $q$ particle--hole pairs of the shell, the boxes of $\rho(T)$ are the pairs with the hole below the particle, the $q-|\rho(T)|$ boxes of $\rect/\rho(T)$ are the pairs with the particle below the hole, and adding a box moves one particle up past one hole. In Section~\ref{sec:fourfactors} we work this out for four $2\times2$ factors.

\begin{mainthm}\label{thm:grass}
Let $\F=\C$, and let $\cL$ denote $\log(1/\varepsilon)$ or $\log t$.
\begin{enumerate}[label=(\alph*)]
\item There are polynomials $\mathcal P_\op,\mathcal P_{\rm Lap}\in\R[\cL]$ such that, for every $\eta\in(0,1)$,
\begin{align*}
 \Pb(\nop P^2\le\varepsilon)&=\varepsilon^C\mathcal P_\op(\log1/\varepsilon)+O(\varepsilon^{C+\eta}),\\
 \E\,e^{-t\nF P^2}&=t^{-C}\mathcal P_{\rm Lap}(\log t)+O(t^{-C-\eta}).
\end{align*}
Both have the Pl\"ucker form
\[
 \mathcal P_\bullet(\cL)=\sum_{T}\Delta^\bullet_T\,p_T(\cL),\qquad \bullet\in\{\op,\mathrm{Lap}\}.
\]
The sum runs over $h$-subsets $T\subset\{0,\dots,\mu-1\}$; the $\Delta_T^\bullet$ are constants; and $p_T$ are the maximal minors (Pl\"ucker coordinates) of a $\mu\times h$ matrix $K(\cL)$ with polynomial entries, determined by the Fermi shell and given in \eqref{eq:Kblock}. The matrix $K$ is the same for both observables.
\item Each $p_T$ has degree exactly $\dim X_{\rho(T)}=q-|\rho(T)|$ in $\cL$, with leading coefficient
\[
 (-1)^{\binom h2}\,R_{L_F}^h\,\operatorname{vol}_{\rm Pl}\bigl(X_{\rho(T)}\bigr).
\]
\item Consequently $\#\{T\}=\binom\mu h=\chi(\Gr(h,\mu))$, the generating function $\sum_Tx^{\deg p_T}$ is the Poincar\'e polynomial $\sum_\rho x^{|\rho|}$ of $\Gr(h,\mu)$, the top degree $q=\dim\Gr(h,\mu)$ is attained only by the big cell $T_0$, and the shell factor satisfies
\[
 B=\operatorname{vol}_{\rm Pl}(\Gr(h,\mu))=\frac{\deg_{\rm Pl}\Gr(h,\mu)}{q!}=\frac{f^{\rect}}{q!} .
\]
Under the standard unit-cube Lebesgue normalization it also equals the Euclidean volume of the order polytope of $[h]\times[\mu-h]$ \cite{Sta86}.
\end{enumerate}
\end{mainthm}

\begin{remark}[Fermionic reading]
The last shell consists of $h$ fermions in $\mu$ degenerate orbitals. Slater determinants correspond to decomposable vectors in $\Lambda^h\C^\mu$, whose projectivization is the Pl\"ucker image of $\Gr(h,\mu)$. The $\binom\mu h$ occupation-number states are its torus-fixed points, and at each of them the $q=h(\mu-h)$ particle--hole pairs index a basis of the tangent space of $\Gr(h,\mu)$. Theorem~\ref{thm:grass} makes this precise at the level of the small-ball asymptotics: \emph{each particle--hole pair inside the open shell contributes one power of the logarithm}. Filled levels contribute none. Term by term, Theorem~\ref{thm:grass}(b) says that the degree of $p_T$ in $\cL$ is the number of particle--hole pairs of the occupation pattern $T$ in which the particle lies below the hole, so that the full power $q$ comes from $T_0$ alone, where every particle lies below every hole. The shell factor is the normalized Pl\"ucker degree-volume of $\Gr(h,\mu)$. How $\Gr(h,\mu)$ appears in the geometry of $(\C^D,I_d)$ itself is an open question (Section~\ref{sec:open}).
\end{remark}

\subsection{Real entries}\label{sec:statementsreal}

\begin{mainthm}\label{thm:real}
Let $\F=\R$. For every size vector $d$:
\begin{enumerate}[label=(\roman*)]
\item $s\mapsto\E\,\nF P^{-2s}$ is holomorphic on $\Re s<C/2$, extends meromorphically to $\C$, and has a pole of order exactly $q+1$ at $s=C/2$. Equivalently, the real log canonical threshold of $\nF P^2$ and its multiplicity are $(C/2,\,q+1)$, at the origin and globally.
\item There is $\kappa_\R=\kappa_\R(d)\in(0,\infty)$ such that
\[
 \Pb\bigl(\nF P^2\le\varepsilon\bigr)\sim\kappa_\R\,\varepsilon^{C/2}(\log1/\varepsilon)^q,\qquad
 \E\,e^{-t\nF P^2}\sim\Gamma(C/2+1)\,\kappa_\R\,t^{-C/2}(\log t)^q .
\]
\item $\Pb(\nF P^2\le\varepsilon)\le\Pb(\nop P^2\le\varepsilon)\le\Pb(\nF P^2\le n\varepsilon)$. Hence $\Pb(\nop P^2\le\varepsilon)\asymp\varepsilon^{C/2}(\log1/\varepsilon)^q$.
\item If $n=1$, let $d'_1,\dots,d'_N$ be the sizes that remain after deleting one entry equal to $1$. Then $C=\min_jd'_j$, $\mu$ is the number of $j$ with $d'_j=C$, and
\[
 \kappa_\R=\frac{2^{-NC/2}}{(\mu-1)!\,\Gamma(C/2+1)\,\Gamma(C/2)^{\mu-1}}\prod_{j:\,d'_j>C}\frac{\Gamma\bigl((d'_j-C)/2\bigr)}{\Gamma(d'_j/2)} .
\]
\item If $N=2$, then $\kappa_\R$ is the product of multivariate Gamma functions given in Proposition~\ref{prop:twofactors}.
\end{enumerate}
\end{mainthm}

Statement (i) is known: it is Aoyagi's formula \cite{Aoy24}, translated by Lemma~\ref{lem:aoyagi}, and \cite[Thm.~8.6]{PLR24} identifies the threshold with half the codimension of $\Sigma_d$ by comparison with it. However, our proof is independent of both. The lower bound comes from an explicit event of positive probability (Proposition~\ref{prop:lower}), and the upper bound comes from Theorem~\ref{thm:A} by a real--complex comparison (Lemma~\ref{lem:comparison}). Statement (iv) is classical, since for $n=1$ the variable $\nF P^2$ is a product of independent chi-square variables \cite{ST70}. The closed forms (iv) and (v) are special cases of a single formula for $\kappa_\R$.

To describe $\kappa_\R$ for more factors, we first rewrite the constants of Theorem~\ref{thm:A} in terms of the occupied states themselves. Fix a minimizer $e$ of Lemma~\ref{lem:filling}, so that ladder $j$ holds its $e_j$ lowest states. The \emph{particles} of $e$ are these occupied states $p$, each with a ladder $\ell(p)=j$ and an energy $\varepsilon_p=\nu_j+l$, $0\le l<e_j$. We write $f_j=\nu_j+e_j$ for the first empty energy on ladder $j$. For $\beta\in\{1,2\}$ put the superfactorials $S_\beta(m)=\prod_{t=1}^m\Gamma(\beta t/2)$, so that $S_2(m)=S_m$, and
\begin{equation}\label{eq:Kbeta}
 K_\beta(e)=\frac{(\beta/2)^{N\beta C/2}}{\Gamma(\beta C/2+1)}\;\frac{\prod_{j=1}^NS_\beta(e_j)}{S_\beta(n)}\;
 \frac{\prod^{*}_{i\ne\ell(p)}\Gamma\bigl(\beta(f_i-\varepsilon_p)/2\bigr)}{\prod_{j=1}^N\prod_{u=e_j+1}^{n}\Gamma\bigl(\beta(\nu_j+u)/2\bigr)} .
\end{equation}
Here the product $\prod^*$ runs over all ladders $i$ and all particles $p$ on other ladders, omitting the pairs with $f_i=\varepsilon_p$. Such pairs occur only in open shells, where $f_i=\varepsilon_p=L_F$ means that $p$ sits at the Fermi level while ladder $i$ has an empty state there. For $\beta=2$ every argument is an integer and the prefactor is $1/C!$; for $\beta=1$ the arguments outside the prefactor are halved, and the prefactor is $2^{-NC/2}/\Gamma(C/2+1)$.

\begin{proposition}[Particle form of the complex constant]\label{prop:particle}
For every size vector $d$ and every minimizer $e$, the constants of Theorem~\ref{thm:A} satisfy $\kappa_F=\cA/C!=B\,K_2(e)$. In particular $K_2(e)$ does not depend on the choice of minimizer.
\end{proposition}

The proof is a regrouping of Definition~\ref{def:constants}, which we give in Section~\ref{sec:evaluation}: the Vandermonde factor $V$ and the factors $R_L$ are both recovered from the gaps $f_i-\varepsilon_p$. For $q\le1$ the real constant is the case $\beta=1$ of \eqref{eq:Kbeta}:

\begin{mainthm}\label{thm:closed}
Let $\F=\R$ and suppose that $q\le1$, that is, the shell is closed or $(\mu,h)=(2,1)$. Then for every minimizer $e$
\[
 \kappa_\R=K_1(e)=\frac{2^{-NC/2}}{\Gamma(C/2+1)}\;\frac{\prod_jS_1(e_j)}{S_1(n)}\;\frac{\prod^*_{i\ne\ell(p)}\Gamma\bigl((f_i-\varepsilon_p)/2\bigr)}{\prod_j\prod_{u=e_j+1}^{n}\Gamma\bigl((\nu_j+u)/2\bigr)} .
\]
\end{mainthm}

We prove Theorem~\ref{thm:closed} in Section~\ref{sec:proofD}, for closed shells in Section~\ref{sec:divisors} and for $q=1$ in Section~\ref{sec:onelog}. For an open shell the Morse--Bott integral that gives the closed-shell constant diverges through exactly $q$ chi-square moments of the borderline form $\E(\chi^2_\nu)^{-\nu/2}$ (Section~\ref{sec:picture}), and in \eqref{eq:Kbeta} these are the omitted pairs, so that $\prod^*$ replaces each divergent moment by the coefficient of its pole. The resulting $K_1(e)$ is again independent of $e$ (Proposition~\ref{prop:etaindep}), and this suggests the following.

\begin{conjecture}\label{conj:open}
For every size vector $d$ and every minimizer $e$, $\kappa_\R=B\,K_1(e)$, with $B$ as in Definition~\ref{def:constants}.
\end{conjecture}

Conjecture~\ref{conj:open} reduces to Theorem~\ref{thm:closed} when $q\le1$, since then $B=1$. It also holds for vector chains and for two factors, because $B\,K_1(e)$ coincides with Theorem~\ref{thm:real}(iv) when $n=1$ and with Proposition~\ref{prop:twofactors} when $N=2$, by the computation in Section~\ref{sec:conjevidence}. Its complex counterpart is Proposition~\ref{prop:particle}.

\subsection{Geometry of the zero fibre}

\begin{proposition}[\cite{PLR24,KR26,Ter26}]\label{prop:geometry}
For every $d$, $\codim_\C\Sigma_d=C$, and the number $\theta$ of irreducible components of $\Sigma_d$ of maximal dimension is $\binom\mu h$. If $d_0=n$, so that $\nu_j=d_j-n$, then for each $e$ as in Lemma~\ref{lem:filling} the rank stratum
\[
 S_e=\Bigl\{\operatorname{rank}(A_j\cdots A_1)=n-\textstyle\sum_{i\le j}e_i\ \text{for all } j\Bigr\}
\]
is irreducible, locally closed and of codimension $E(e)$. The top-dimensional components of $\Sigma_d$ are the closures of the $S_e$ with $e$ a minimizer.
\end{proposition}

\begin{proof}[Proof for $d_0=n$]
Put $r_j=n-\sum_{i\le j}e_i$. Given $A_{j-1},\dots,A_1$, the image of $A_{j-1}\cdots A_1$ has dimension $r_{j-1}$, and $A_j$ restricted to it is a $d_j\times r_{j-1}$ matrix of rank $r_j$. The rest of $A_j$ is free. Matrices of size $d_j\times r_{j-1}$ and rank $r_j$ form an irreducible set of codimension $(r_{j-1}-r_j)(d_j-r_j)$. Hence
\[
 \codim S_e=\sum_je_j(d_j-r_j)=\sum_j\nu_je_j+\sum_je_j\sum_{i\le j}e_i=E(e),
\]
using $d_j=n+\nu_j$. The variety $\Sigma_d$ is the finite union of the $S_e$. Each stratum is open and dense in its closure, so distinct strata of the same dimension have distinct closures, and Lemma~\ref{lem:filling} gives the count. For other orderings we refer to \cite{PLR24,Ter26}.
\end{proof}

The fibres of the multiplication map and their rank strata, with dimensions and tangent and normal spaces, are also studied in \cite{SB24}, and its differential in \cite{TKB20}.

\section{Examples and a low-temperature picture}\label{sec:examples}

\subsection{Small cases}
For one factor, $N=1$, we have $\nu=|d_1-d_0|$. The levels $\nu,\dots,\nu+n-1$ are singly occupied, so $h=\mu=1$, $q=0$ and $C=n(n+\nu)=d_0d_1$. Moreover $B=R_L=1$, $V=S_n$ and $G=\prod_{i<n}\Gamma(i+\nu+1)$, so
\[
 \kappa_\op=\prod_{i=0}^{n-1}\frac{i!}{\Gamma(i+\nu+n+1)},\qquad \cA=1 .
\]
The first value agrees with the Selberg integral. The density of the eigenvalues of $A^*A$ is $Z_n^{-1}\Delta(x)^2\prod_ix_i^\nu e^{-x_i}$ with $Z_n=\prod_{i<n}\Gamma(i+2)\Gamma(i+\nu+1)$. Hence $\Pb(\nop A^2\le\varepsilon)\sim\varepsilon^{n(n+\nu)}Z_n^{-1}\mathsf S_n(\nu+1,1,1)$, where $\mathsf S_n$ is the Selberg integral, and $\mathsf S_n(\nu+1,1,1)=\prod_{i<n}\Gamma(\nu+1+i)\,i!\,(i+1)!/\Gamma(\nu+n+i+1)$ \cite{FW08}. The second value is exact: $\E\,e^{-t\nF A^2}=(1+t)^{-d_0d_1}$.

If some $d_k=1$, then $P$ is a vector chain of rank one and $\nF P^2=\prod_{j\ne k}Y_j$ with independent factors, where $Y_j\sim\Gamma(d_j,1)$ in the complex case and $Y_j\sim\chi^2_{d_j}$ in the real case. This follows from the Bartlett decomposition (Lemma~\ref{lem:bartlett}), or directly. The shell data are $L_F=C-1$ with $C=\min_{j\ne k}d_j$, $h=1$, and $\mu$ equal to the multiplicity of $C$ among the $d_j$ with $j\ne k$; so $q=\mu-1$. Theorem~\ref{thm:A} gives
\[
 \kappa_\op=\kappa_F=\frac{1}{(\mu-1)!\,\Gamma(C)^{\mu-1}\Gamma(C+1)}\prod_{j\ne k:\,d_j>C}\frac{\Gamma(d_j-C)}{\Gamma(d_j)} .
\]
This agrees, through Corollary~\ref{cor:lct}(i), with the leading Laurent coefficient of $\E\prod Y_j^{-z}=\prod_j\Gamma(d_j-z)/\Gamma(d_j)$ at $z=C$, a pole of order $\mu$. The real constant in Theorem~\ref{thm:real}(iv) is obtained in the same way from $\E(\chi^2_d)^{-s}=2^{-s}\Gamma(d/2-s)/\Gamma(d/2)$.

The simplest open shell is the keyhole channel $d=(2,1,2)$. Here $\nF P^2=XY$ with $X,Y\sim\Gamma(2,1)$ independent. The density of $XY$ is $2u\,\mathrm K_0(2\sqrt u)$, with $\mathrm K_0,\mathrm K_1,\mathrm K_2$ the modified Bessel functions of the second kind, and the distribution function is
\[
 F(x)=1-2x\,\mathrm K_2(2\sqrt x)-2x^{3/2}\mathrm K_1(2\sqrt x)=\frac{x^2}2\Bigl[\log\frac1x+\frac12-2\gamma_E\Bigr]+O(x^3|\log x|),
\]
where $\gamma_E$ is Euler's constant. The shell data are $\nu=(1,1)$, $L_F=1$, $\mu=2$, $h=1$, $q=1$, $C=2$, giving $\kappa_\op=\kappa_F=\tfrac12$ and $\cA=1$. This case is contained in the analysis of Jin--McKay--Wong--Gao \cite{JMWG08} of beamforming over Rayleigh-product channels with unit normalization.

For two factors the smallest closed shells with $\mu=2$ are $d=(2,2,2)$, with $C=3$, $q=0$, $\kappa_\op=\tfrac14$, $\cA=1$ and $\kappa_F=\tfrac16$, and $d=(2,3,3)$, with $\nu=(1,1)$, $C=5$, $q=0$, $G=4$, $\kappa_\op=\tfrac1{144}$, $\cA=\tfrac14$ and $\kappa_F=\tfrac1{480}$. For $d=(3,3,3)$ we have $L_F=1$ and $h=1$ with $\mu=2$, so this shell is open, with $C=7$, $q=1$, $\kappa_\op=\tfrac1{3456}$, $\cA=\tfrac18$ and $\kappa_F=\tfrac1{40320}$. The values $C=3$ for $(2,2,2)$ and $C=7$ for $(3,3,3)$ agree with \cite{YB11}. In these cases the expression $n_Tn_Sn_R/\max(n_T,n_S,n_R)$ for the double-scattering diversity in \cite{SW08}, where $(n_T,n_S,n_R)=(d_0,d_1,d_2)$ are the numbers of transmit antennas, scatterers and receive antennas, exceeds the exact exponent (it gives $4$ and $9$), and Yang and Belfiore showed that it is attained only when $2\max(n_T,n_S,n_R)+1\ge n_T+n_S+n_R$ \cite{YB11}. The three-factor product $d=(3,3,3,3)$ has $C=6$ (Table~\ref{tab:examples}), one less than $d=(3,3,3)$.

\begin{table}[t]
\centering
\begin{tabular}{@{}lccccccccc@{}}
\toprule
$d$ & $n$ & $L_F$ & $(\mu,h)$ & $C$ & $q$ & $\theta$ & $\kappa_\op$ & $\kappa_F$ & $\kappa_\R$\\
\midrule
$(1,1,1)$ & 1 & 0 & $(2,1)$ & 1 & 1 & 2 & $1$ & $1$ & $1/\pi$\\
$(1,1,2)$ & 1 & 0 & $(1,1)$ & 1 & 0 & 1 & $1$ & $1$ & $1$\\
$(2,1,2)$ & 1 & 1 & $(2,1)$ & 2 & 1 & 2 & $1/2$ & $1/2$ & $1/4$\\
$(2,2,2)$ & 2 & 0 & $(2,2)$ & 3 & 0 & 1 & $1/4$ & $1/6$ & $\pi/6$\\
$(2,2,3)$ & 2 & 1 & $(2,1)$ & 4 & 1 & 2 & $1/24$ & $1/48$ & $1/16$\\
$(2,2,4)$ & 2 & 1 & $(1,1)$ & 4 & 0 & 1 & $1/144$ & $1/288$ & $1/16$\\
$(2,3,3)$ & 2 & 1 & $(2,2)$ & 5 & 0 & 1 & $1/144$ & $1/480$ & $1/15$\\
$(3,3,3)$ & 3 & 1 & $(2,1)$ & 7 & 1 & 2 & $1/3456$ & $1/40320$ & $1/(105\pi)$\\
$(2,2,2,3)$ & 2 & 0 & $(2,2)$ & 3 & 0 & 1 & $1/8$ & $1/12$ & $\pi^{3/2}/(6\sqrt2)$\\
$(2,2,3,4)$ & 2 & 1 & $(2,1)$ & 4 & 1 & 2 & $1/288$ & $1/576$ & $1/32$\\
$(3,3,3,3)$ & 3 & 0 & $(3,3)$ & 6 & 0 & 1 & $1/1728$ & $1/11520$ & $\pi^2/192$\\
$(2,2,2,2)$ & 2 & 0 & $(3,2)$ & 3 & 2 & 3 & $1/8$ & $1/12$ & $\sqrt{2\pi}/48^{\,*}$\\
$(2,2,2,2,2)$ & 2 & 0 & $(4,2)$ & 3 & 4 & 6 & $1/48$ & $1/72$ & $1/576^{\,*}$\\
\bottomrule
\end{tabular}
\caption{Shell data and constants for small size vectors. The complex constants are from Theorem~\ref{thm:A}. For real entries the exponent is $C/2$ with the same $q$, and $\kappa_\R$ is from Theorem~\ref{thm:real}(iv), Proposition~\ref{prop:twofactors} and Theorem~\ref{thm:closed}. The values marked $^*$ are those predicted by Conjecture~\ref{conj:open}.}
\label{tab:examples}
\end{table}

\subsection{\texorpdfstring{Four $2\times2$ factors}{Four 2x2 factors}}\label{sec:fourfactors}
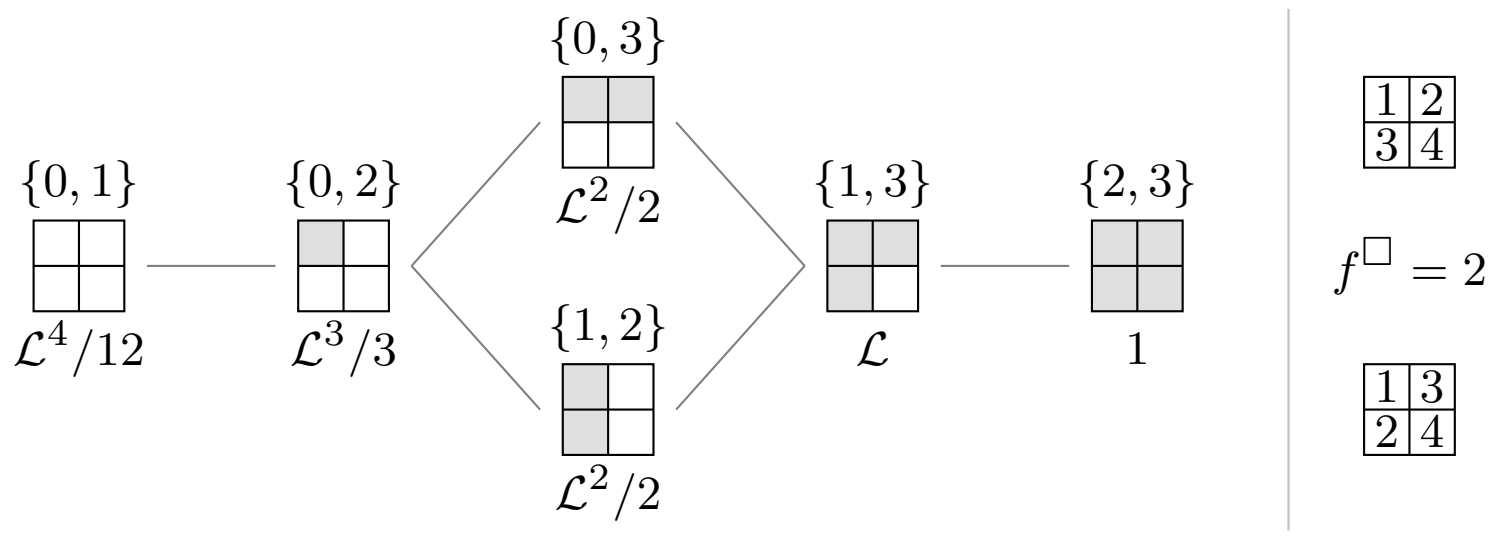
\begin{figure}[t]
\centering
\newcommand{\shellnode}[6]{%
 \begin{scope}[shift={(#1,#2)}]
  \ifnum#3>0 \foreach \k in {1,...,#3} {\fill[gray!25] ({(\k-2)*0.3},0.3) rectangle ({(\k-1)*0.3},0.6);}\fi
  \ifnum#4>0 \foreach \k in {1,...,#4} {\fill[gray!25] ({(\k-2)*0.3},0) rectangle ({(\k-1)*0.3},0.3);}\fi
  \draw (-0.3,0) rectangle (0.3,0.6); \draw (0,0) -- (0,0.6); \draw (-0.3,0.3) -- (0.3,0.3);
  \node[font=\scriptsize] at (0,0.84) {$\{#5\}$};
  \node[font=\scriptsize] at (0,-0.25) {#6};
 \end{scope}}
\newcommand{\shelltab}[6]{%
 \begin{scope}[shift={(#1,#2)},font=\scriptsize]
  \draw (-0.3,0) rectangle (0.3,0.6); \draw (0,0) -- (0,0.6); \draw (-0.3,0.3) -- (0.3,0.3);
  \node at (-0.15,0.45) {#3}; \node at (0.15,0.45) {#4};
  \node at (-0.15,0.15) {#5}; \node at (0.15,0.15) {#6};
 \end{scope}}
\begin{tikzpicture}
 \foreach \a/\b in {{(0.45,0.3)}/{(1.3,0.3)},{(2.2,0.3)}/{(3.05,1.25)},{(2.2,0.3)}/{(3.05,-0.65)},
   {(3.95,1.25)}/{(4.8,0.3)},{(3.95,-0.65)}/{(4.8,0.3)},{(5.7,0.3)}/{(6.55,0.3)}}
  {\draw[gray] \a -- \b;}
 \shellnode{0}{0}{0}{0}{0,1}{$\cL^4/12$}
 \shellnode{1.75}{0}{1}{0}{0,2}{$\cL^3/3$}
 \shellnode{3.5}{0.95}{2}{0}{0,3}{$\cL^2/2$}
 \shellnode{3.5}{-0.95}{1}{1}{1,2}{$\cL^2/2$}
 \shellnode{5.25}{0}{2}{1}{1,3}{$\cL$}
 \shellnode{7}{0}{2}{2}{2,3}{$1$}
 \draw[gray!50] (8.0,-1.45) -- (8.0,2.0);
 \shelltab{8.8}{0.95}{1}{2}{3}{4}
 \shelltab{8.8}{-0.95}{1}{3}{2}{4}
 \node[font=\scriptsize] at (8.8,0.3) {$f^{\rect}=2$};
\end{tikzpicture}
\caption{The partitions in $\rect=(2,2)$ for four $2\times2$ factors, $d=(2,2,2,2,2)$ (Section~\ref{sec:fourfactors}). Left: each node shows a subset $T\subset\{0,1,2,3\}$, its partition $\rho(T)$ (shaded) and the leading term of $-p_T$, and each edge adds one box. Right: the two standard Young tableaux of shape $\rect$.}
\label{fig:young}
\end{figure}

We now work out Example~\ref{ex:intro}, where $d=(2,2,2,2,2)$, and explain Figure~\ref{fig:young}. Here $\nu=(0,0,0,0)$, $L_F=0$, $\mu=4$, $h=2$, hence $q=4$ and $C=3$. Moreover $n_0=2$, $B=\tfrac{0!}{3!}\tfrac{1!}{2!}=\tfrac1{12}$ and $V=R_0=G=1$. Therefore
\[
 \kappa_\op=\tfrac1{12}\,\Gamma(3)^{-2}=\tfrac1{48},\qquad\cA=\tfrac1{12},\qquad\kappa_F=\tfrac1{72} .
\]
The real exponent is $3/2$, with the same four logarithms, and Conjecture~\ref{conj:open} predicts the real constant $\kappa_\R=\frac1{12}K_1(e)=\frac1{576}$.

Theorem~\ref{thm:grass} describes the full polynomials $\mathcal P_\op$ and $\mathcal P_{\rm Lap}$, which here have degree $4$ in $\cL=\log(1/\varepsilon)$ or $\log t$. They are assembled along the chain
\[
 \text{pole of $\Gamma^4$ at $0$}\to\xi_0,\xi_1\to K(\cL)\to\text{six minors}\to\text{Pl\"ucker coordinates on }\Gr(2,4),
\]
which we now follow. Here the function $\Phi$ of Section~\ref{sec:proofA} is $\Gamma^4$, with one Gamma function for each of the four ladders, which are all tied, so that $\Phi$ has a pole of order $\mu=4$ at $0$. The polynomials $\xi_0$ and $\xi_1$ of Section~\ref{sec:proofC} are read off from the principal parts at $0$ of the functions $F_{0,0}=\Phi$ and $F_{0,1}$ of Section~\ref{sec:proofA}, which have poles of orders $4$ and $3$ there. Hence $\xi_0$ and $\xi_1$ have degrees $3$ and $2$, with leading terms $\cL^3/6$ and $\cL^2/2$ since $R_0=1$. The matrix $K(\cL)$ has $\mu=4$ rows, indexed by $s=0,1,2,3$, and $h=2$ columns. Its row $0$ is $(\xi_0,\xi_1)$ and its row $s$ is the $s$-th derivative of row $0$, so that the leading terms of its entries are
\[
 \begin{pmatrix}\cL^3/6&\cL^2/2\\ \cL^2/2&\cL\\ \cL&1\\ 1&0\end{pmatrix},
\]
and the last row is exact. A two-element subset $T\subset\{0,1,2,3\}$ selects two rows, and $p_T$ is the $2\times2$ minor on these rows; we write $p_{ij}$ for $p_{\{i,j\}}$. For $T_0=\{0,1\}$ the minor $p_{01}=\xi_0\xi_1'-\xi_0'\xi_1$ is the Wronskian of $\xi_0$ and $\xi_1$, with leading term $\cL^4/6-\cL^4/4=-\cL^4/12$. In the same way $p_{02}$, $p_{03}$, $p_{12}$, $p_{13}$ and $p_{23}$ have leading terms $-\cL^3/3$, $-\cL^2/2$, $-\cL^2/2$, $-\cL$ and $-1$. Since $p_{23}=-1$, the columns of $K(\cL)$ span a plane in $\C^4$ for every $\cL$, and the six minors are the Pl\"ucker coordinates of this point of $\Gr(2,4)$. They therefore satisfy the Pl\"ucker relation
\[
 p_{01}p_{23}-p_{02}p_{13}+p_{03}p_{12}=0,
\]
which defines $\Gr(2,4)$ as a quadric in $\mathbb P^5$; in degree $4$ it reads $\frac1{12}-\frac13+\frac14=0$.

Theorem~\ref{thm:grass}(b) explains these degrees through the partitions $\rho(T)$ of Section~\ref{sec:statementC}. For $T=\{s_1<s_2\}$ we have $\rho(T)=(s_2-1,s_1)$, which we draw as a Young diagram in $\rect=(2,2)$, with $s_2-1$ boxes in the first row and $s_1$ boxes in the second. The subsets $\{0,1\}$, $\{0,2\}$, $\{0,3\}$, $\{1,2\}$, $\{1,3\}$ and $\{2,3\}$ give $\varnothing$, $(1)$, $(2)$, $(1,1)$, $(2,1)$ and $(2,2)$, the shaded diagrams of Figure~\ref{fig:young}. Each term of $p_T$ carries $s_1+s_2=|\rho(T)|+1$ derivatives, so that each box of $\rho(T)$ is one derivative more than in the Wronskian, and each derivative lowers the degree by one. Hence $\deg p_T=4-|\rho(T)|$ is the number of unshaded boxes. In the fermionic reading of Section~\ref{sec:statementC}, $T$ lists the occupied orbitals among $0,1,2,3$. For $T=\{1,3\}$ the holes are $0$ and $2$, the particle at $3$ has two holes below it and the particle at $1$ has one, so that $\rho(T)=(2,1)$, and the only pair in which the particle lies below the hole is formed by the particle at $1$ and the hole at $2$, so that $\deg p_{13}=1$. For $T_0$ every particle lies below every hole, which gives the top degree $q=4$, and for $\{2,3\}$ no particle does, which gives degree $0$. An edge of the figure raises one element of $T$ by one, which moves one particle up past one hole and adds one box to $\rho(T)$. The degrees $4,3,2,2,1,0$ give $\sum_Tx^{\deg p_T}=1+x+2x^2+x^3+x^4$, the Poincar\'e polynomial of $\Gr(2,4)$, as in Theorem~\ref{thm:grass}(c).

By Theorem~\ref{thm:grass}(b) the leading coefficient of $p_T$ is $(-1)^{\binom22}R_0^2\operatorname{vol}_{\rm Pl}(X_{\rho(T)})=-f^{\rect/\rho(T)}/(4-|\rho(T)|)!$. The number $f^{\rect/\rho}$ of standard Young tableaux of the skew shape $\rect/\rho$ counts the ways of adding the missing boxes one at a time, that is, the paths from $\rho$ to $\rect$ in Figure~\ref{fig:young} that move from left to right. There are two such paths from $\varnothing$ and from $(1)$, through $(2)$ or through $(1,1)$, and one from each of the other partitions. This gives the coefficients $\frac2{4!}=\frac1{12}$, $\frac2{3!}=\frac13$, $\frac12$, $\frac12$, $1$ and $1$ under the diagrams, in agreement with the minors above. Geometrically, $X_\varnothing=\Gr(2,4)$ is the quadric defined by the Pl\"ucker relation, $X_{(1)}$ is a three-dimensional quadric cone, $X_{(2)}$ and $X_{(1,1)}$ are planes, $X_{(2,1)}$ is a line and $X_{(2,2)}$ is a point, in agreement with $\deg_{\rm Pl}X_\rho=f^{\rect/\rho}$. The $f^\rect=2$ standard Young tableaux of shape $\rect$ on the right of the figure record the order in which the boxes are added along the two paths from $\varnothing$ to $\rect$. They are the two linear extensions of the poset $[2]\times[2]$ of boxes, and they cut its order polytope, the set of points of $[0,1]^4$ whose coordinates, indexed by the boxes, increase along rows and columns, into two simplices of volume $1/4!$. Its volume $B=\frac1{12}$ is the shell factor in the constants above. Only $T_0$ reaches the top degree, so that the leading terms of $\mathcal P_\op$ and $\mathcal P_{\rm Lap}$ come from $p_{01}$ alone, and the other five minors contribute only to lower powers of $\cL$.

\subsection{Three real factors}\label{sec:threefactors}
Theorem~\ref{thm:closed} gives the first closed forms with more than two factors and $n\ge2$. For $d=(2,2,2,3)$ the offsets are $\nu=(0,0,1)$, and the shell is closed with $L_F=0$ and $\mu=h=2$. The unique minimizer is $e=(1,1,0)$, with one particle at energy $0$ on each of the first two ladders and first empty energies $f=(1,1,1)$. Each of the four pairs in $\prod^*$ contributes $\Gamma(\frac12)$, the superfactorials give $S_1(1)^2/S_1(2)=\sqrt\pi$, and the only nontrivial denominator comes from the third ladder, $\Gamma(1)\Gamma(\frac32)=\frac{\sqrt\pi}2$. Hence
\[
 \kappa_\R(2,2,2,3)=\frac{2^{-9/2}}{\Gamma(5/2)}\cdot\sqrt\pi\cdot\pi^2\cdot\frac2{\sqrt\pi}=\frac{\pi^{3/2}}{6\sqrt2},
\]
so that the product of two real $2\times2$ matrices and a real $3\times2$ matrix satisfies $\Pb(\nF P^2\le\varepsilon)\sim\frac{\pi^{3/2}}{6\sqrt2}\,\varepsilon^{3/2}$. Compared with $\kappa_\R(2,2,2)=\pi/6$, the third factor multiplies the constant by $\sqrt{\pi/2}=\E(\chi^2_3)^{-1}\,\E(\chi^2_2)^{-1/2}$. This is typical: a layer that kills no direction enters only through chi-square moments (Section~\ref{sec:evaluation}). For three real $3\times3$ factors, $d=(3,3,3,3)$, the shell is again closed, with $\nu=(0,0,0)$, $e=(1,1,1)$ and six pairs of gap $1$, and Theorem~\ref{thm:closed} gives $\kappa_\R=\pi^2/192$. The simplest case with one logarithm, three factors and $n\ge2$ is $d=(2,2,3,4)$, for which Theorem~\ref{thm:closed} gives $\Pb(\nF P^2\le\varepsilon)\sim\frac1{32}\,\varepsilon^{2}\log(1/\varepsilon)$, half the two-factor value $\kappa_\R(2,2,3)=\frac1{16}$; we confirm this independently in Section~\ref{sec:onelog}.

\begin{figure}[t]
\centering
\begin{tikzpicture}[scale=2.4]
 \fill[gray!25] (1,0) -- (1,1) -- (0,1) -- cycle;
 \draw[->] (0,0) -- (1.25,0) node[right] {$s$};
 \draw[->] (0,0) -- (0,1.2) node[above] {$t$};
 \draw (1,0) -- (1,1) -- (0,1);
 \draw[dashed] (1,0) -- (0,1);
 \foreach \p in {(1,0),(1,1),(0,1)} {\fill \p circle (0.7pt);}
 \node[left,font=\scriptsize] at (-0.02,1) {$e=(0,1,1)$};
 \node[right,font=\scriptsize] at (1.02,1) {$e=(1,0,1)$};
 \node[below,font=\scriptsize] at (1,-0.03) {$e=(1,1,0)$};
 \node[below left,font=\scriptsize] at (0,0) {$0$};
 \node[font=\small] at (0.7,0.7) {$\tfrac12$};
\end{tikzpicture}
\caption{The window of scales for three real $2\times2$ factors, $d=(2,2,2,2)$. The small Bartlett entries are $z_1=\delta^s$, $z_2=\delta^{1-s}$, $x_3=\delta^t$ and $x_2=\delta^{1-t}$ with $\delta=\sqrt\varepsilon$, and the small-ball event forces $s+t\ge1$ (shaded). The vertices are the three minimizers, at which the pairs $\{z_2,x_3\}$, $\{z_1,x_3\}$ and $\{z_1,x_2\}$ carry the whole scale $\delta$. The triangle has area $\frac12=B=\deg_{\rm Pl}\Gr(2,3)/2!$, and its two dimensions produce the two logarithms.}
\label{fig:triangle}
\end{figure}
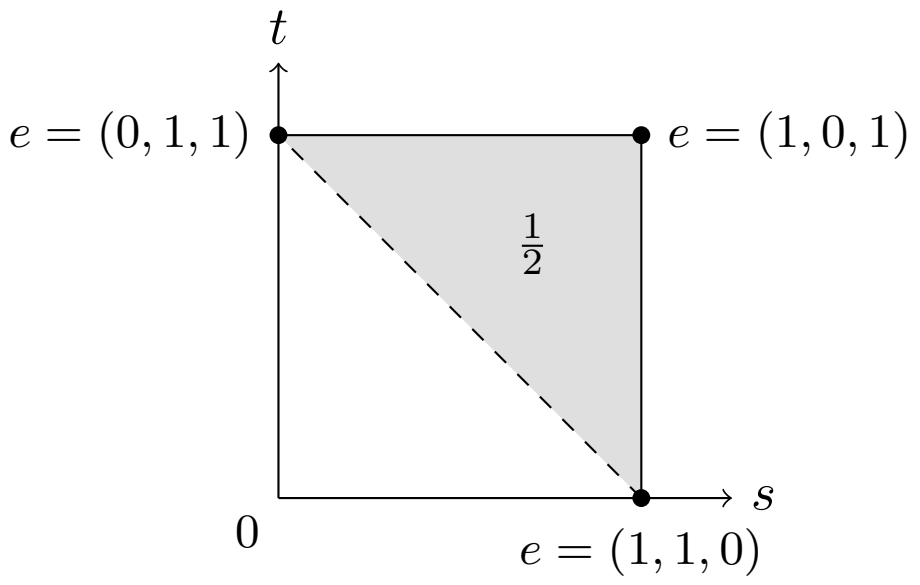

The shell of three real $2\times2$ factors, $d=(2,2,2,2)$, is open, with $\mu=3$ and $h=2$, and here Conjecture~\ref{conj:open} predicts $\kappa_\R=\sqrt{2\pi}/48$. It is instructive to see where the two logarithms and the factor $B=\frac12$ come from. In the Bartlett coordinates of Lemma~\ref{lem:bartlett} the three factors become upper triangular $2\times2$ matrices with diagonal entries $x_j\sim\chi_2$ and $z_j\sim\chi_1$ and off-diagonal entries $y_j\sim N(0,1)$, and
\[
 \nF P^2=(x_3x_2x_1)^2+(z_3z_2z_1)^2+(x_3x_2y_1+x_3y_2z_1+y_3z_2z_1)^2 .
\]
The three minimizers make $\nF P$ small through the pairs $\{z_2,x_3\}$, $\{z_1,x_3\}$ and $\{z_1,x_2\}$ of small entries. Writing $z_1=\delta^s$, $z_2=\delta^{1-s}$, $x_3=\delta^t$ and $x_2=\delta^{1-t}$ with $\delta=\sqrt\varepsilon$, the event $\nF P^2\le\varepsilon$ requires $s+t\ge1$, which is the triangle of Figure~\ref{fig:triangle}. Inside the triangle the middle path $x_3y_2z_1$ is $o(\delta)$, and $\nF P^2/\delta^2$ reduces to the quadratic form of the two-factor problem $d=(2,2,2)$ in the merged variables $x_2x_3/\delta$ and $z_1z_2/\delta$. Integrating over the triangle produces $\frac12(\log1/\delta)^2=\frac18(\log1/\varepsilon)^2$, and the densities of the four small entries at the origin contribute a factor $\sqrt{2/\pi}$ relative to the two-factor problem. Hence $\kappa_\R(2,2,2,2)=\frac18\sqrt{2/\pi}\,\kappa_\R(2,2,2)=\sqrt{2\pi}/48$, as predicted. The computation is heuristic at the corners of the triangle. Carried out with complex entries it gives $\frac12\kappa_F(2,2,2)=\frac1{12}$, which is the value of Theorem~\ref{thm:A}. Numerical evaluations of the Laplace transform for three and four real $2\times2$ factors agree with the predicted constants $\sqrt{2\pi}/48$ and $1/576$; the code and the results are available at \numericsurl.

\subsection{A low-temperature picture}\label{sec:picture}

We now explain why the results take this form. The arguments of this subsection are heuristic, and at the end we indicate where the proofs make them precise. Here and below $\beta=1$ for real and $\beta=2$ for complex entries.

Let $\gamma$ be the standard Gaussian measure on the space of tuples $A=(A_1,\dots,A_N)$, which is the law of the factors, and consider the classical energy
\[
 H(A)=\tfrac12\nF{A_N\cdots A_1}^2 .
\]
Its partition function at inverse temperature $t$ is
\[
 \cZ(t)=\int e^{-tH(A)}\,d\gamma(A)=\E\,e^{-\frac t2\nF P^2}=Z(t/2),
\]
and for zero-target deep linear networks $H$ is the population loss and $\cZ(t)$ is the Bayesian evidence (Section~\ref{sec:dln}). The rescaling $t\mapsto t/2$ multiplies the constants by $2^{\beta C/2}$ and leaves the exponent and the power of the logarithm unchanged, so we work with the Laplace transform $Z(t)=\cZ(2t)$ and call $t$ the inverse temperature.

The partition function and the small-ball probability measure the same neighbourhood of the zero fibre $\Sigma_d=\{P=0\}$. Integrating by parts,
\[
 Z(t)=\int_0^\infty e^{-tw}\,dF_F(w)=\int_0^\infty e^{-u}F_F(u/t)\,du ,
\]
so that at inverse temperature $t$ the partition function probes $F_F$ at the scale $\varepsilon\sim1/t$, with a smooth weight in place of a sharp cutoff. Thus $F_F(\varepsilon)\sim\kappa\,\varepsilon^{\beta C/2}(\log1/\varepsilon)^q$ gives $Z(t)\sim\Gamma(\beta C/2+1)\,\kappa\,t^{-\beta C/2}(\log t)^q$, and conversely, since $F_F$ is nondecreasing, Karamata's Tauberian theorem recovers $F_F$ from $Z$. Here $\kappa$ is $\kappa_\R$ or $\kappa_F$. This is how Theorems~\ref{thm:A} and~\ref{thm:real} pass between the two quantities, and we may argue with either.

As $t\to\infty$ the factor $e^{-t\nF P^2}$ suppresses everything except a neighbourhood of $\Sigma_d$, and the first question is which ways of making $P$ vanish dominate. Assume $d_0=n$, let $r_j$ be the rank of $A_j\cdots A_1$, with $r_0=n$, and let $e_j=r_{j-1}-r_j$ be the number of directions killed at layer $j$, so that $\sum_je_j=n$ on $\Sigma_d$. Killing them requires the restriction of $A_j$ to the image of $A_{j-1}\cdots A_1$, a $d_j\times r_{j-1}$ matrix, to have rank $r_j$, which imposes $e_j(d_j-r_j)$ conditions. As in the proof of Proposition~\ref{prop:geometry}, the codimension of the rank stratum $S_e$ is therefore
\[
 \codim S_e=\sum_{j=1}^Ne_j(d_j-r_j)=\frac{n(n+1)}2+\sum_{j=1}^N\sum_{k=0}^{e_j-1}(\nu_j+k)=E(e).
\]
The right-hand side is the energy \eqref{eq:Egas} of $n$ fermions when ladder $j$ holds its $e_j$ lowest states: the $k$th direction killed at layer $j$ costs the energy $\nu_j+k$ of the $k$th state of ladder $j$. The fermion gas is thus an auxiliary device that encodes the rank strata of $\Sigma_d$ and their codimensions.

A stratum of codimension $c$ contributes $t^{-\beta c/2}$ to $Z(t)$, one factor $t^{-1/2}$ for each real direction transverse to it, as in equipartition. The contribution $t^{-\beta E(e)/2}=\exp(-\frac\beta2E(e)\log t)$ of $S_e$ is a Boltzmann weight of the gas at inverse temperature $\frac\beta2\log t$, so that the low-temperature limit of $Z(t)$ is also a low-temperature limit of the gas. The ground states of the gas, given by the filling rule of Lemma~\ref{lem:filling}, label the top components of $\Sigma_d$, and their energy $C$ fixes the power of $t$. When the shell is open there are $\theta=\binom\mu h$ degenerate ground states, and the power of $\log t$ comes from the competition between them.

When the shell is closed there is a single top component. Near its generic points the energy is a nondegenerate quadratic form in the $C$ transverse directions, and Laplace's method gives the constant as the Gaussian integral, along the component, of the inverse square root of the transversal Hessian. In the slice of Section~\ref{sec:slice} this integral factorizes into chi-square moments (Section~\ref{sec:evaluation}), which regroup into the pairs of the particle form $K_\beta(e)$ of \eqref{eq:Kbeta}, each pair formed by a particle and the first empty state of another ladder. For real entries this is Theorem~\ref{thm:closed}.

\begin{figure}[t]
\centering
\begin{tikzpicture}[baseline=(current bounding box.center)]
 \node[font=\scriptsize] at (0,1.95) {$1$};
 \node[font=\scriptsize] at (0.6,1.95) {$2$};
 \draw[fill=gray!55] (0,1.6) circle (4pt);
 \draw[fill=white] (0.6,1.6) circle (4pt);
 \node[font=\scriptsize] at (0.3,1.18) {$e=(1,0)$, $x=0$};
 \draw[fill=white] (0,0.6) circle (4pt);
 \draw[fill=gray!55] (0.6,0.6) circle (4pt);
 \node[font=\scriptsize] at (0.3,0.18) {$e=(0,1)$, $y=0$};
\end{tikzpicture}\hspace{0.6cm}
\begin{tikzpicture}[scale=1.1,baseline=(current bounding box.center)]
 \fill[gray!25] (-1,-1) rectangle (1,1);
 \foreach \sx/\sy in {1/1,-1/1,-1/-1,1/-1}
  \fill[white,domain=0.15:1,smooth,variable=\u] plot ({\sx*\u},{\sy*0.15/\u}) -- (\sx,\sy) -- cycle;
 \draw (-1,-1) rectangle (1,1);
 \draw[very thick] (-1,0) -- (1,0) node[right,font=\scriptsize] {$y=0$};
 \draw[very thick] (0,-1) -- (0,1) node[above,font=\scriptsize] {$x=0$};
\end{tikzpicture}\hspace{0.6cm}
\begin{tikzpicture}[scale=1.5,baseline=(current bounding box.center)]
 \fill[gray!25] (1,0) -- (1.3,0) -- (1.3,1.3) -- (0,1.3) -- (0,1) -- cycle;
 \draw[->] (0,0) -- (1.45,0) node[right,font=\scriptsize] {$s$};
 \draw[->] (0,0) -- (0,1.45) node[above,font=\scriptsize] {$s'$};
 \draw[very thick] (1,0) -- (0,1);
 \fill (1,0) circle (0.8pt) (0,1) circle (0.8pt);
 \node[below,font=\scriptsize] at (1,-0.02) {$x=0$};
 \node[left,font=\scriptsize] at (-0.02,1) {$y=0$};
 \node[below left,font=\scriptsize] at (0,0) {$0$};
\end{tikzpicture}
\caption{The product $P=xy$ of two real standard Gaussian scalars, $d=(1,1,1)$. Left: the Fermi level $L_F=0$ has one state on each ladder, and the fermion occupies one of them (grey). The filling $e=(1,0)$ kills the direction at layer $1$, so that $x=0$, and $e=(0,1)$ kills it at layer $2$, so that $y=0$. Middle: the small-ball event $|xy|\le\delta$, with $\delta=\sqrt\varepsilon$ (shaded), concentrates along these two lines. Right: in the exponents $|x|=\delta^s$ and $|y|=\delta^{s'}$ the event is $s+s'\ge1$ (shaded), and the cheapest configurations form the segment $s+s'=1$, $0\le s\le1$, whose endpoints are the two fillings. Its length $1=B$ in the coordinate $s$ and its single dimension produce the factor and the logarithm in $\Pb(x^2y^2\le\varepsilon)\sim\frac1\pi\,\varepsilon^{1/2}\log(1/\varepsilon)$.}
\label{fig:toy}
\end{figure}
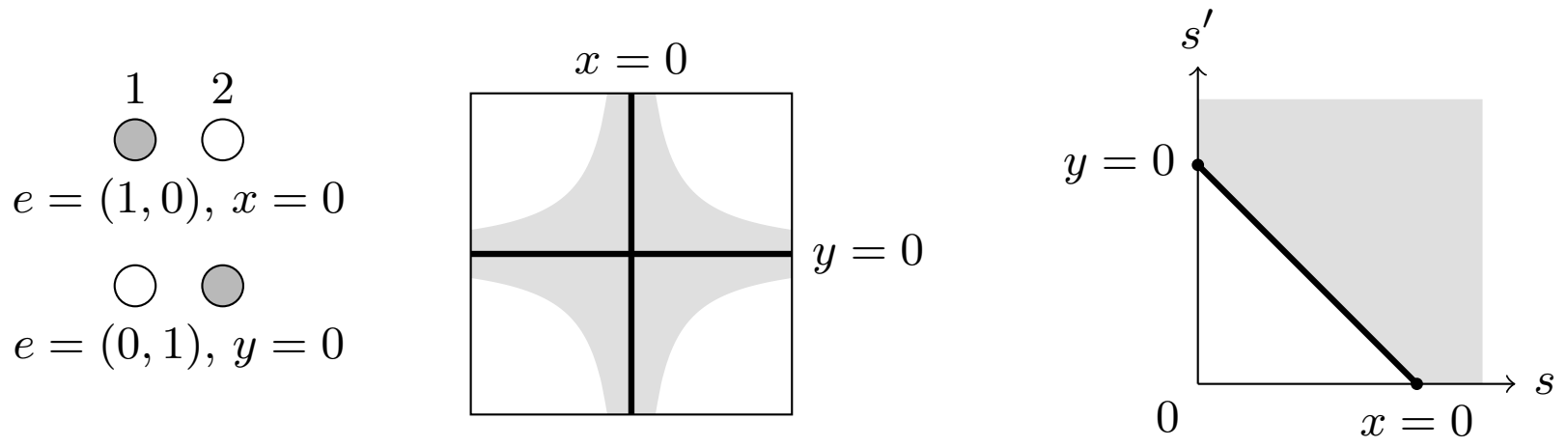

When the shell is open, the $\theta$ top components meet, and along their intersection the transversal Hessian of each of them degenerates. If the components were isolated, their contributions would simply add. A particle--hole exchange at the Fermi level moves one killed direction from one layer to another and passes from one top component to another, and near their intersection the small-ball event can be realized in a continuum of ways that interpolate between the two, by spreading the smallness over entries of several layers at different logarithmic scales. In the Gaussian integral along one component this appears as exactly $q$ divergent chi-square moments, one for each pair formed by a particle at the Fermi level and an empty state at the same level, that is, for each particle--hole pair inside the shell. The simplest case is $d=(1,1,1)$, where $P=xy$ with independent standard real Gaussian $x$ and $y$, and $C=1$, $\mu=2$, $h=1$ and $q=1$. The two degenerate fillings $e=(1,0)$ and $e=(0,1)$ correspond to the two lines $x=0$ and $y=0$ of $\Sigma_d$, on which $A_1=x$ or $A_2=y$ vanishes. Write $|x|=\delta^s$ and $|y|=\delta^{s'}$ with $\delta=\sqrt\varepsilon$. The event $x^2y^2\le\varepsilon$ is $s+s'\ge1$, and since $|x|$ and $|y|$ have density $\sqrt{2/\pi}$ at the origin, configurations with exponents $(s,s')$ carry probability of order $\delta^{s+s'}$. The cheapest ones fill the segment $s+s'=1$, $0\le s\le1$, of Figure~\ref{fig:toy}. Its endpoints are the two fillings, its interior interpolates between them, with $x$ and $y$ sharing the smallness, and its length in the coordinate $s$ is $1=B$. Integrating over the segment produces one logarithm,
\[
 \Pb(x^2y^2\le\varepsilon)\sim\tfrac2\pi\,\delta\log\tfrac1\delta=\tfrac1\pi\,\varepsilon^{1/2}\log\tfrac1\varepsilon ,
\]
which is Theorem~\ref{thm:real}(iv) for $d=(1,1,1)$. The partition function shows the same logarithm directly, and here it can be computed exactly. Integrating out $y$ and substituting $x=t^{-1/2}\sinh(v/2)$,
\begin{align*}
 \cZ(t)&=\E\,e^{-tx^2y^2/2}=\E\,(1+tx^2)^{-1/2}=\frac{e^{1/(4t)}}{\sqrt{2\pi t}}\int_0^\infty e^{-\cosh(v)/(4t)}\,dv\\
 &=\frac{\log(8t)-\gamma_E+o(1)}{\sqrt{2\pi t}}\sim\frac1{\sqrt{2\pi}}\,t^{-1/2}\log t,
\end{align*}
since the last integral is the modified Bessel function of the second kind of order zero at $1/(4t)$. The logarithm comes from the range $t^{-1/2}\ll|x|\ll1$, where $(1+tx^2)^{-1/2}\approx(\sqrt t\,|x|)^{-1}$, so that $\int dx/|x|$ produces $\log t$. In this range $|x|=t^{-s/2}$ and the relevant $|y|$ is of order $t^{-(1-s)/2}$, so that $s$ runs along the same segment. For three real $2\times2$ factors the same computation over the triangle of Figure~\ref{fig:triangle} gives the two logarithms and the factor $B=\frac12$ found in Section~\ref{sec:threefactors}.

In general the cheapest assignments of scales form a window of dimension $q$, and integrating over it multiplies the contribution of a single component by $(\log1/\varepsilon)^q$ and by the volume of the window. For three real $2\times2$ factors the window is the order polytope of $[2]\times[1]$, and the residue rule of Section~\ref{sec:conjevidence} assigns to every shell the volume $B$ of the order polytope of $[h]\times[\mu-h]$ (Theorem~\ref{thm:grass}(c)). Together with the replacement of each divergent moment by the coefficient of its pole, this gives $\kappa_F=B\,K_2(e)$ and $\kappa_\R=B\,K_1(e)$. For complex entries the formula is exact, by Proposition~\ref{prop:particle}. For real entries it is Theorem~\ref{thm:closed} when $q\le1$ and Conjecture~\ref{conj:open} in general.

In thermodynamic language, Theorems~\ref{thm:A} and~\ref{thm:real} give
\[
 -\log Z(t)=\tfrac{\beta C}2\log t-q\log\log t+O(1).
\]
Write $-\log Z=tU-S$, with the mean energy $U=-\partial_t\log Z(t)$ and the entropy $S$ relative to the Gaussian reference measure. Since $\Pb(\nF P^2\le\varepsilon)$ is regularly varying of index $\beta C/2$ at $0$, Karamata's theorem gives $tU\to\beta C/2$, the equipartition value, so that both logarithms belong to the entropy. The first is the logarithm of the volume of the Gaussian fluctuations transverse to $\Sigma_d$, whose width shrinks like $t^{-1/2}$ in each of the $\beta C$ real transverse directions. The second is the logarithm of the volume $(\log t)^q$ of the window of competing scales, an entropy that closed shells, with their single top component, do not have. For $N>n$ square $n\times n$ factors, $C=n(n+1)/2$ and $q=n(N-n)$ by Corollary~\ref{cor:square}, so that layers beyond the width leave the power of $t$ unchanged and each adds $n$ logarithmic directions.

The proofs follow this picture. For complex entries the ladders are the poles of the Mellin factors $\Gamma(s+\nu_j)$ of the product ensemble, and its determinantal structure allows one particle per state, so that the filling rule arises a second time, from the exact integral. The determinantal structure then turns the picture into an exact computation (Sections~\ref{sec:proofA} and~\ref{sec:proofC}). For real entries the lower bound of Theorem~\ref{thm:real} integrates over a $q$-dimensional simplex of scales of the kind shown in Figures~\ref{fig:toy} and~\ref{fig:triangle}, the upper bound transfers the complex result through a comparison of principalizations (Section~\ref{sec:proofB}), and for $q\le1$ we evaluate in Section~\ref{sec:proofD} the Gaussian integral along the top components and show that nothing else contributes.

\section{\texorpdfstring{Proof of Theorem~\ref{thm:A} and Corollary~\ref{cor:lct}}{Proof of Theorem A and Corollary \getrefnumber{cor:lct}}}\label{sec:proofA}

\subsection{The product ensemble and a Newton basis}

\begin{lemma}\label{lem:ensemble}
Let $\F=\C$, and let $x_1,\dots,x_n$ be the nonzero squared singular values of $P$, that is, the nonzero eigenvalues of $P^*P$.
\begin{enumerate}[label=(\alph*)]
\item The joint law of $x_1,\dots,x_n$ depends on $d$ only through the multiset $\{\nu_j\}$ \cite{IK14} (Appendix~\ref{app:invariance}).
\item Let $\Phi(s)=\prod_{j=1}^N\Gamma(s+\nu_j)$, and let $w_k$ be the Meijer $G$-function with Mellin transform
\[
 \int_0^\infty x^{s-1}w_k(x)\,dx=(s+\nu_1)_k\,\Phi(s),
\]
where $(u)_k$ is the rising factorial. Then the density of $(x_1,\dots,x_n)$ on $(0,\infty)^n$ is \cite{AIK13,KZ14}
\[
 \rho(x)=\frac{1}{n!\,\cD}\,\Delta(x)\det\bigl[w_k(x_i)\bigr]_{i,k},\qquad \Delta(x)=\prod_{i<i'}(x_{i'}-x_i),
\]
with
\[
 \cD=\det\bigl[(i+1+\nu_1)_k\,\Phi(i+1)\bigr]_{i,k<n}=S_n\,G .
\]
\end{enumerate}
\end{lemma}

The value of $\cD$ follows by extracting $\Phi(i+1)$ from row $i$ and evaluating the Vandermonde determinant of the monic polynomials $(u)_k$ at $u=i+1+\nu_1$. The span of $\{(s+\nu_1)_k\Phi(s)\}_{k<n}$, namely $\{p\,\Phi:\deg p<n\}$, does not depend on which offset carries the index $k$. Andr\'eief's identity gives
\begin{align}
 \Pb(\nop P^2\le\varepsilon)&=\cD^{-1}\det\Bigl[\int_0^\varepsilon x^iw_k(x)\,dx\Bigr]_{i,k<n},\label{eq:andreief1}\\
 \E\,e^{-t\nF P^2}&=\cD^{-1}\det\Bigl[\int_0^\infty e^{-tx}x^iw_k(x)\,dx\Bigr]_{i,k<n}.\label{eq:andreief2}
\end{align}
A determinantal formula for the moment generating function of $\nF P^2$, as a formal power series, also appears in \cite[Prop.~1]{WZCT15}.

For an occupied level $L$ and $0\le j<n_L$ put
\[
 p_{L,j}(s)=(s+L)^j\prod_{M<L}(s+M)^{\mu_M},\qquad F_{L,j}=p_{L,j}\Phi,\qquad a_{L,j}=\mu_L-1-j .
\]

\begin{lemma}\label{lem:newton}
\begin{enumerate}[label=(\alph*)]
\item The degrees $\deg p_{L,j}=N_<(L)+j$ run through $0,1,\dots,n-1$, each exactly once.
\item $F_{L,j}(s)=(s+L)^j\prod_{i=1}^N\Gamma\bigl(s+\max(\nu_i,L)\bigr)$.
\item $F_{L,j}$ is holomorphic on $\Re s>-L$. At $s=-L$ it has a pole of order $a_{L,j}+1\ge1$, and
\[
 F_{L,j}(s)=R_L\,(s+L)^{-a_{L,j}-1}\bigl(1+O(s+L)\bigr).
\]
\end{enumerate}
\end{lemma}

\begin{proof}
(a) For $L<L_F$ the degrees run from $N_<(L)$ to $N_<(L)+\mu_L-1=N_<(L+1)-1$, and for $L=L_F$ from $N_<(L_F)$ to $n-1$. The lowest occupied level is $\min\nu_j$, where $N_<=0$.

(b) $\prod_{M<L}(s+M)^{\mu_M}=\prod_{i:\nu_i<L}\prod_{M=\nu_i}^{L-1}(s+M)$, and $\Gamma(s+\nu)\prod_{M=\nu}^{L-1}(s+M)=\Gamma(s+L)$.

(c) The poles of (b) lie at $s\le-L$. At $s=-L$ exactly $\mu_L$ factors equal $\Gamma(s+L)$, each with residue $1$. The remaining factors give $R_L$, and $(s+L)^j$ removes $j$ orders, with $j<n_L\le\mu_L$.
\end{proof}

Order the pairs $(L,j)$ by $\deg p_{L,j}$. Since both $\{(s+\nu_1)_k\}$ and $\{p_{L,j}\}$ are monic and degree-ordered bases of the polynomials of degree $<n$, the change of basis is unitriangular. Hence, with $f_{L,j}$ the inverse Mellin transform of $F_{L,j}$ (a linear combination of the $w_k$), one may replace $w_k$ by $f_{L,j}$ in \eqref{eq:andreief1}--\eqref{eq:andreief2} without changing either determinant.

\subsection{Mellin--Barnes evaluation of the entries}

Write the principal part of $F_{L,j}(z)$ at $z=-L$ as
\[
 \sum_{b=0}^{a}c_b\,\frac{b!}{(z+L)^{b+1}},\qquad a=a_{L,j},\qquad c_a=\frac{R_L}{a!},
\]
and put
\begin{equation}\label{eq:Kentry}
 K^{(L,j)}_s(\cL)=\sum_{b=s}^{a}c_b\,\frac{b!}{(b-s)!}\,\cL^{b-s}\qquad(0\le s\le a),
\end{equation}
a polynomial of degree $a-s$ with leading coefficient $R_L/(a-s)!$.

\begin{lemma}\label{lem:entries}
Fix an occupied $L$, an index $j<n_L$, a row $i\in\{0,\dots,n-1\}$ and $\eta\in(0,1)$. Put $\alpha=i+L+1$. Then
\begin{align*}
 \int_0^\varepsilon x^if_{L,j}(x)\,dx&=\varepsilon^{\alpha}\sum_{s=0}^{a_{L,j}}K^{(L,j)}_s(\log\tfrac1\varepsilon)\,Y_{i,(L,s)}+O(\varepsilon^{\alpha+\eta}),\\
 \int_0^\infty e^{-tx}x^if_{L,j}(x)\,dx&=t^{-\alpha}\sum_{s=0}^{a_{L,j}}K^{(L,j)}_s(\log t)\,W_{i,(L,s)}+O(t^{-\alpha-\eta}),
\end{align*}
where $Y_{i,(L,s)}=\alpha^{-s-1}$ and $W_{i,(L,s)}=(-1)^s\,\Gamma^{(s)}(\alpha)/s!$.
\end{lemma}

\begin{proof}
By Stirling's formula, $F_{L,j}$ decays exponentially on vertical lines, uniformly in vertical strips. Fix $\sigma\in(-\min_j\nu_j,1)\cap(-L,1)$; then $f_{L,j}(x)=\frac1{2\pi i}\int_{(\sigma)}x^{-z}F_{L,j}(z)\,dz$. By Fubini,
\begin{align*}
 \int_0^\varepsilon x^if_{L,j}\,dx&=\frac1{2\pi i}\int_{(\sigma)}F_{L,j}(z)\,\frac{\varepsilon^{i+1-z}}{i+1-z}\,dz,\\
 \int_0^\infty e^{-tx}x^if_{L,j}\,dx&=\frac1{2\pi i}\int_{(\sigma)}F_{L,j}(z)\,\Gamma(i+1-z)\,t^{z-i-1}\,dz .
\end{align*}
Move the contour to $\Re z=-L-\eta$. The only pole crossed is $z=-L$, because the poles of the second factor lie at $z\ge i+1>\sigma$. On the new line the integrands are $O(\varepsilon^{\alpha+\eta})$ and $O(t^{-\alpha-\eta})$ times an integrable function. With $u=z+L$, the residues are
\[
 \Res_{u=0}\sum_bc_b\frac{b!}{u^{b+1}}\cdot\frac{\varepsilon^\alpha e^{u\cL}}{\alpha-u}
 \quad\text{and}\quad
 \Res_{u=0}\sum_bc_b\frac{b!}{u^{b+1}}\cdot t^{-\alpha}e^{u\cL}\Gamma(\alpha-u).
\]
Extracting the coefficient of $u^b$ from $e^{u\cL}(\alpha-u)^{-1}$, respectively from $e^{u\cL}\Gamma(\alpha-u)$, gives the stated sums.
\end{proof}

\subsection{The leading coefficient}

\begin{lemma}\label{lem:confluent}
Index columns by pairs $(L,s)$ with $L$ occupied and $0\le s<n_L$, ordered lexicographically, and rows by $i=0,\dots,n-1$. Put $\sigma=\sum_L\binom{n_L}2$. Then
\begin{align*}
 \det\bigl[Y_{i,(L,s)}\bigr]&=(-1)^\sigma S_nV\prod_L\Bigl[\frac{\Gamma(L+1)}{\Gamma(L+n+1)}\Bigr]^{n_L},\\
 \det\bigl[W_{i,(L,s)}\bigr]&=(-1)^\sigma V\prod_L\Gamma(L+1)^{n_L}.
\end{align*}
\end{lemma}

\begin{proof}
Let $u$ be an analytic $\C^n$-valued function and $X$ a fixed $n\times(n-m)$ matrix. As distinct points $y_1,\dots,y_m$ tend to $y$,
\[
 \det[u(y_1),\dots,u(y_m),X]\Big/\prod_{k<k'}(y_{k'}-y_k)\ \longrightarrow\ \det\bigl[u(y),u'(y),\dots,u^{(m-1)}(y)/(m-1)!,X\bigr].
\]
This follows from Newton's divided differences and the Hermite--Genocchi formula.

For the first determinant, apply this to the Cauchy matrix $[1/(i+1+y_k)]_{i,k}$. Cauchy's formula gives $\det=\prod_{i<i'}(i'-i)\prod_{k<k'}(y_{k'}-y_k)/\prod_{i,k}(i+1+y_k)$. Let the $y_k$ coalesce in groups of size $n_L$ at $y=L$. Dividing by the Vandermonde factors within groups leaves the factor $\prod_{i<i'}(i'-i)=S_n$, the cross-group factor $V$ and the denominator $\prod_i\prod_L(i+L+1)^{n_L}=\prod_L[\Gamma(L+n+1)/\Gamma(L+1)]^{n_L}$. Since $\partial_y^s(x+y)^{-1}/s!=(-1)^s(x+y)^{-s-1}$, the columns differ from $Y$ by the signs $(-1)^s$, whose product is $(-1)^\sigma$.

For the second determinant, take $u(y)=(\Gamma(i+y+1))_i=\Gamma(y+1)\bigl((y+1)_i\bigr)_i$. Then
\[
 \det[u(y_k)]=\prod_k\Gamma(y_k+1)\prod_{k<k'}(y_{k'}-y_k),
\]
because $(y+1)_i$ is monic of degree $i$ in $y$. The same limit gives $V\prod_L\Gamma(L+1)^{n_L}$, and $W$ carries the same signs $(-1)^s$.
\end{proof}

For each occupied level $L$ let $K^{(L)}\in\R[\cL]^{\mu_L\times n_L}$ be the matrix with entries $K^{(L)}_{s,j}=K^{(L,j)}_s$ for $s\le a_{L,j}$ and $0$ otherwise. For the last shell this is the matrix of Theorem~\ref{thm:grass}:
\begin{equation}\label{eq:Kblock}
 K(\cL)=K^{(L_F)}(\cL)\in\R[\cL]^{\mu\times h}.
\end{equation}

\begin{proposition}\label{prop:leading}
The polynomials $\mathcal P_\op$ and $\mathcal P_{\rm Lap}$ of Theorem~\ref{thm:grass}(a) exist, have degree exactly $q$, and have leading coefficients $\kappa_\op$ and $\cA$ respectively. Explicitly,
\[
 \mathcal P_\op(\cL)=\cD^{-1}\prod_{L<L_F}\det K^{(L)}\cdot\sum_T\det\bigl[Y^{\rm ext}_{\cdot,S_T}\bigr]\,p_T(\cL),
\]
where $Y^{\rm ext}$ has the columns $Y_{\cdot,(L,s)}$ for $0\le s<\mu_L$, the set $S_T$ consists of all columns of the filled levels together with the columns $(L_F,s)$, $s\in T$, and $p_T=\det K_{T,\cdot}$. Thus $\Delta^\op_T=\cD^{-1}\prod_{L<L_F}\det K^{(L)}\cdot\det\bigl[Y^{\rm ext}_{\cdot,S_T}\bigr]$ in Theorem~\ref{thm:grass}(a). The same holds for $\mathcal P_{\rm Lap}$ with $W$ in place of $Y$.
\end{proposition}

\begin{proof}
By Lemmas~\ref{lem:newton} and \ref{lem:entries}, entry $(i,(L,j))$ of the matrix in \eqref{eq:andreief1} equals
\[
 \varepsilon^{i+1}\varepsilon^{L}\bigl[(Y^{\rm ext}\widehat K)_{i,(L,j)}+O(\varepsilon^\eta)\bigr],
\]
where $\widehat K=\bigoplus_LK^{(L)}$ is block diagonal. Since $\sum_i(i+1)+\sum_Ln_LL=C$, the determinant is $\varepsilon^C\bigl[\det(Y^{\rm ext}\widehat K)+O(\varepsilon^{\eta'})\bigr]$ for every $\eta'<\eta$.

We first expand by Cauchy--Binet. By block diagonality, $\det(Y^{\rm ext}\widehat K)$ is a sum over choices of $n_L$ rows from each block $K^{(L)}$. For $L<L_F$ the block is square ($n_L=\mu_L$), so all rows are taken. In its Leibniz expansion only the permutation $j\mapsto s=a_{L,j}=\mu_L-1-j$ survives, because $K^{(L)}_{s,j}=0$ for $s>a_{L,j}$; the surviving entries equal $K^{(L,j)}_{a}=c_aa!=R_L$. Hence $\det K^{(L)}=(-1)^{\binom{\mu_L}2}R_L^{\mu_L}$, a constant. For $L=L_F$ the choice is an $h$-subset $T$, which gives the stated formula.

Next we determine the degrees. The entry $K_{s,j}$ has degree $a_{L_F,j}-s$. Hence
\[
 \deg p_T\le\sum_{j<h}(\mu-1-j)-\sum_{s\in T}s\le h(\mu-1)-h(h-1)=q,
\]
with equality in the second step only for $T=T_0$.

It remains to find the leading coefficient. The top coefficient of $p_{T_0}$ is
\[
 \det\bigl[R_{L_F}/(a_j-s)!\bigr]_{s,j<h}=R_{L_F}^h\prod_j\frac1{a_j!}\prod_{j<j'}(a_{j'}-a_j)=(-1)^{\binom h2}R_{L_F}^hB,
\]
with $1/m!=0$ for $m<0$, since $a!/(a-s)!=a(a-1)\cdots(a-s+1)$ is monic of degree $s$ in $a$ and $a_j=\mu-1-j$. By Lemma~\ref{lem:confluent}, $\det Y^{\rm ext}_{\cdot,S_{T_0}}=\det[Y]$.

Collecting the factors, we note that the signs $(-1)^{\sum_L\binom{n_L}{2}}$ from the blocks cancel those of Lemma~\ref{lem:confluent}. Dividing by $\cD=S_nG$ gives $\kappa_\op$. The Laplace transform is identical, with $\cL=\log t$ and $W$ in place of $Y$, and gives $\cA$.
\end{proof}

By Lemma~\ref{lem:ensemble}(a) we may assume $d_0=n$. Then \eqref{eq:A1} and \eqref{eq:A2} follow from Proposition~\ref{prop:leading}: the terms of lower degree in $\cL$ give the relative errors $O(1/\cL)$.

For \eqref{eq:A3}, the function $u\mapsto\Pb(\nF P^2\le u)$ is nondecreasing, and \eqref{eq:A2} is the asymptotics of its Laplace--Stieltjes transform. Since $(\log t)^q$ is slowly varying, Karamata's Tauberian theorem \cite[XIII.5, Thm.~3]{Fel71} gives $\Pb(\nF P^2\le\varepsilon)\sim\cA\varepsilon^C(\log1/\varepsilon)^q/\Gamma(C+1)$.

For the ratio \eqref{eq:ratio}, divide the two expressions in \eqref{eq:constants}. \qed

We turn to Corollary~\ref{cor:lct}. Write $g=\nF P^2$ and $D=\sum_jd_jd_{j-1}$. Lemma~\ref{lem:polar} below gives
\[
 \E\,g^{-z}=\frac{\Gamma(D-Nz)}{2\pi^{D}}\int_{S^{2D-1}}g(\omega)^{-z}\,d\omega .
\]
The sphere integral is meromorphic on $\C$ \cite{Ati70,BG69}. For real $z<C$,
\[
 \E\,g^{-z}=\int_0^\infty u^{-z}\,dF_F(u)=O(1)+z\int_0^1u^{-z-1}F_F(u)\,du .
\]
By \eqref{eq:A3} this is $\sim C\kappa_Fq!/(C-z)^{q+1}=\cA\,q!/(\Gamma(C)(C-z)^{q+1})$ as $z\uparrow C$. This identifies the order and the leading Laurent coefficient of the pole.

Part (ii) follows from Lemma~\ref{lem:polar}, which transfers the pole to integrals over balls, together with the analytic description of the log canonical threshold of an ideal \cite[Thm.~1.2]{Mus12}. Because $I_d$ is homogeneous, the origin realizes the minimum of the local thresholds. Finally $\codim\Sigma_d=C$ by Proposition~\ref{prop:geometry}. \qed

\section{\texorpdfstring{Proof of Theorem~\ref{thm:grass}}{Proof of Theorem B}}\label{sec:proofC}

Part (a) is Proposition~\ref{prop:leading}. The same block $K$ enters both observables because $K$ depends only on the Laurent coefficients $c_b$.

For (b), let $T=\{s_1<\dots<s_h\}$. The term of $p_T$ of top degree $\sum_ja_j-\sum_ks_k$ is
\[
 R_{L_F}^h\det\bigl[1/(a_j-s_k)!\bigr]_{j,k},
\]
with $1/m!=0$ for $m<0$. Aitken's formula \cite{Ait43}, \cite[Cor.~7.16.3]{Sta99} states
\[
 f^{\lambda/\rho}=|\lambda/\rho|!\,\det\bigl[1/(\lambda_i-\rho_c-i+c)!\bigr]_{i,c=1}^h .
\]
Take $\lambda=\rect$, $\rho=\rho(T)$ and $i=j+1$. Then $\lambda_i-\rho_c-i+c=a_j-s_{h+1-c}$. Reversing the column order costs $(-1)^{\binom h2}$, and $|\rect/\rho(T)|=\sum_ja_j-\sum_ks_k=q-|\rho(T)|$. In particular this determinant is nonzero, since $\rho(T)\subset\rect$, so $\deg p_T=q-|\rho(T)|$ exactly. Finally $f^{\rect/\rho}=\deg_{\rm Pl}X_\rho$, by iterating Pieri's rule $\sigma_\rho\sigma_1=\sum_{\rho'}\sigma_{\rho'}$ in $H^*(\Gr(h,\mu))$, where $\rho'\subset\rect$ runs over the partitions obtained from $\rho$ by adding one box \cite[\S9.4]{Ful97}.

Part (c) follows. The map $T\mapsto\rho(T)$ is a bijection onto the partitions in $\rect$, whose number is $\binom\mu h$. The Betti numbers of $\Gr(h,\mu)$ count these partitions by size, and the symmetry $\rho\mapsto\rect\setminus\rho$ turns $\sum_\rho x^{q-|\rho|}$ into $\sum_\rho x^{|\rho|}$. For $T_0$ we get $B=f^\rect/q!$, in agreement with the hook-length formula $f^{(m^h)}/(hm)!=\prod_{i<h}i!/(i+m)!$. Finally, the linear extensions of $[h]\times[\mu-h]$ are the standard Young tableaux of shape $\rect$, so that $f^\rect/q!$ is the volume of the order polytope \cite{Sta86}. \qed

The proof has a classical interpretation. For $j<h$ let $\xi_j(u)=\sum_bc_bu^b$, where the $c_b$ are the coefficients of the principal part of $F_{L_F,j}$ used in \eqref{eq:Kentry}, so that $\xi_j$ is a polynomial of degree $a_j$. By \eqref{eq:Kentry}, $K^{(L_F,j)}_s(\cL)=\xi_j^{(s)}(\cL)$, so that $K(\cL)$ is the matrix of derivatives of $\xi_0,\dots,\xi_{h-1}$ and $p_{T_0}$ is their Wronskian. By Taylor's formula $\xi_j(u+\cL)=\sum_s\xi_j^{(s)}(\cL)\,u^s/s!$, the $p_T(\cL)$ are the Pl\"ucker coordinates, in the basis $u^s/s!$ of $\C[u]_{<\mu}$, of the translate $\exp(\cL\,d/du)\Xi$ of the point $\Xi=\operatorname{span}\{\xi_j\}$ of $\Gr(h,\C[u]_{<\mu})$. On the Pl\"ucker basis the operator $d/du$ removes one box from $\rho(T)$ in all possible ways, so that, by the Cauchy--Binet formula and Aitken's formula,
\[
 p_T(\cL)=\sum_{\rho'\supseteq\rho(T)}\frac{f^{\rho'/\rho(T)}}{|\rho'/\rho(T)|!}\,\cL^{|\rho'/\rho(T)|}\,p_{T(\rho')}(0),
\]
where $T(\rho')$ is the subset with $\rho(T(\rho'))=\rho'$. Since $p_{T(\rect)}(0)=(-1)^{\binom h2}R_{L_F}^h$, the term $\rho'=\rect$ is the top-degree term found above. The same structure underlies the Wronski map $\Gr(h,\C[u]_{<\mu})\to\mathbb P^q$, a finite morphism of degree $f^\rect$ whose fibres and monodromy are governed by standard Young tableaux \cite{EH83,Pur10}.

\begin{remark}
The lower coefficients of $\mathcal P_\op$ also receive contributions from the subleading Laurent coefficients $c_b$, $b<a$, which enter the formula above through the Pl\"ucker coordinates $p_{T'}(0)$ with $\rho(T')\ne\rect$. Only the top coefficient of each $p_T$ is purely geometric.
\end{remark}

\section{\texorpdfstring{Proof of Theorem~\ref{thm:real}}{Proof of Theorem C}}\label{sec:proofB}

Throughout, $\F=\R$ and $g=\nF P^2$. By \cite[Sec.~III.B]{IK14} (see also Appendix~\ref{app:invariance}), the law of the singular values depends only on the multiset of sizes, so we may and do assume $n=d_0\le d_1\le\dots\le d_N$. Then $\nu_j=d_j-n$ is nondecreasing, and the states at the Fermi level lie on the consecutive ladders $j=1,\dots,\mu$.

\subsection{Bartlett coordinates and the lower bound}

\begin{lemma}[Bartlett]\label{lem:bartlett}
Let $Q_0=I_n$ and define recursively thin QR decompositions $A_jQ_{j-1}=Q_jR_j$, with $Q_j\in\R^{d_j\times n}$ having orthonormal columns and $R_j$ upper triangular with positive diagonal. Then $P=Q_NR_N\cdots R_1$ and $\nF P=\nF{R_N\cdots R_1}$. The $R_j$ are independent, and the entries of each $R_j$ are independent with
\[
 (R_j)_{aa}\sim\chi_{d_j-a+1},\qquad (R_j)_{ab}\sim N(0,1)\quad(a<b).
\]
\end{lemma}

\begin{proof}
Conditionally on $A_1,\dots,A_{j-1}$, the matrix $Q_{j-1}$ is a fixed orthonormal $n$-frame, so $A_jQ_{j-1}$ is a standard Gaussian $d_j\times n$ matrix. Apply the Bartlett decomposition \cite[Thm.~3.2.14]{Mui82}.
\end{proof}

This reduction of a Gaussian product to independent triangular factors is the standard device in the study of Lyapunov exponents \cite{New86,For13}.

Form a layered directed graph with vertices $(j,a)$ for $0\le j\le N$ and $1\le a\le n$, and an edge $\epsilon=(j;a,b)$ from $(j-1,b)$ to $(j,a)$ whenever $a\le b$, with weight $x_\epsilon=(R_j)_{ab}$. Each entry of $R_N\cdots R_1$ is the sum of the weight products over the complete paths from layer $0$ to layer $N$. Put
\[
 c_{(j;a,a)}=d_j-a+1,\qquad c_{(j;a,b)}=1\quad(a<b).
\]
On the event that every $x_\epsilon\in(0,e^{-1})$, the joint density of the weights is bounded above and below by constant multiples of $\prod_\epsilon x_\epsilon^{c_\epsilon-1}$.

For $e$ as in Lemma~\ref{lem:filling} put $r_j=n-\sum_{i\le j}e_i$ (so $r_0=n$ and $r_N=0$). Let $\mathrm{Cut}(e)$ be the set of edges $(j;a,b)$ with $b\le r_{j-1}$ and $a>r_j$, that is, the edges from the \emph{source side} $\{(j,a):a\le r_j\}$ to the \emph{sink side} $\{(j,a):a>r_j\}$, and let $v(e)$ be its indicator vector.

\begin{lemma}\label{lem:cuts}
Every complete path contains an edge of $\mathrm{Cut}(e)$, and $\sum_{\epsilon\in\mathrm{Cut}(e)}c_\epsilon=E(e)$.
\end{lemma}

\begin{proof}
A path starts at a vertex $(0,b)$ with $b\le r_0=n$ and ends at $(N,a)$ with $a>r_N=0$, so at some step it passes from $\{a\le r_{j-1}\}$ to $\{a>r_j\}$.

At layer $j$, the diagonal edges of the cut have $r_j<a\le r_{j-1}$ and cost $\sum_{a=r_j+1}^{r_{j-1}}(d_j+1-a)=e_j(d_j-r_j)-\binom{e_j}2$. The off-diagonal edges have $r_j<a<b\le r_{j-1}$; there are $\binom{e_j}2$ of them, each of cost $1$. Summing over $j$ and using the proof of Proposition~\ref{prop:geometry} gives $E(e)$.
\end{proof}

Yang and Belfiore read the diversity order as the cost of disconnecting a layered network \cite[after Thm.~3.4]{YB07}, and Lemma~\ref{lem:cuts} realizes this picture in Bartlett coordinates.

\begin{lemma}\label{lem:simplex}
There are minimizers $e^{(0)},\dots,e^{(q)}$ of $E$ whose cut vectors are affinely independent. Every point $\alpha$ of the $q$-simplex $\cS=\conv\{v(e^{(k)})\}$ satisfies
\[
 \alpha\ge0,\qquad \sum_{\epsilon\in\gamma}\alpha_\epsilon\ge1\ \text{for every complete path }\gamma,\qquad \sum_\epsilon c_\epsilon\alpha_\epsilon=C .
\]
\end{lemma}

\begin{proof}
Minimizers correspond to words $\eta\in\{0,1\}^\mu$ with $h$ ones, attached to the layers $1,\dots,\mu$. Start from $0^{\mu-h}1^h$ and apply $q$ moves $01\to10$ until $1^h0^{\mu-h}$ is reached. A move at positions $(p,p+1)$ lowers $r_p$ by one and leaves the other $r_j$ unchanged. The diagonal part of the difference of the two cut vectors is $\mathbf e_{(p;a,a)}-\mathbf e_{(p+1;a,a)}$, where $a$ is the old value of $r_p$. To see this, note that $(p-1,a)$ lies on the source side, and $(p+1,a)$ lies on the sink side because $e_{p+1}\ge1$ before the move.

Along the chain, the values of $r_p$ at each fixed $p$ strictly decrease, so the pairs $(p,a)$ are distinct. For fixed $a$, the vectors $\mathbf e_{(p;a,a)}-\mathbf e_{(p+1;a,a)}$ for distinct $p$ are independent edge vectors of a path graph, and distinct $a$ have disjoint supports. Hence the $q$ differences are linearly independent.

The inequalities hold at the vertices by Lemma~\ref{lem:cuts}, and therefore on the convex hull.
\end{proof}

\begin{remark}
A maximal chain of moves $01\to10$ is a standard Young tableau of shape $\rect$. Hence there are $f^\rect=\deg_{\rm Pl}\Gr(h,\mu)$ such simplices, matching Theorem~\ref{thm:grass}. The same argument works without sorting the sizes. When the $\mu$ layers that carry the words $\eta$ are not adjacent, a move between two consecutive ones $p<p'$ lowers $r_j$ by one for all $p\le j<p'$, and the diagonal part of the difference of the cut vectors becomes $\mathbf e_{(p;a,a)}-\mathbf e_{(p';a,a)}$, again an edge vector of a path graph.
\end{remark}

The simplex $\cS$ yields the lower bound.

\begin{proposition}\label{prop:lower}
There is $b>0$ such that $\Pb(\nF P^2\le\varepsilon)\ge b\,\varepsilon^{C/2}(\log1/\varepsilon)^q$ for $0<\varepsilon<1/2$.
\end{proposition}

\begin{proof}
Fix $\delta\in(0,1)$ and put $y_\epsilon=-\log x_\epsilon$ and $\mathsf L=\log(1/\delta)$. Then the density of $y$ is bounded below by $b_1e^{-c\cdot y}$ on $\{y_\epsilon>1\ \forall\epsilon\}$.

Choose a compact $q$-simplex $\cS_0$ in the relative interior of $\cS$. Complete a basis of its tangent space by $M-q$ coordinate vectors $\mathbf e_\epsilon$, $\epsilon\in H$, where $M$ is the number of edges, and set
\[
 \mathcal T_{\mathsf L}=\mathsf L\,\cS_0+s_0\mathbf 1+\Bigl\{\textstyle\sum_{\epsilon\in H}z_\epsilon\mathbf e_\epsilon:0<z_\epsilon<1\Bigr\},\qquad s_0>1 .
\]
This tube is injectively parametrized and has volume $c_1\mathsf L^q$. On it, $c\cdot y=\mathsf LC+O(1)$, and every complete path has $\sum_{\epsilon\in\gamma}y_\epsilon\ge\mathsf L$, so every path product is at most $\delta$. Hence $\nF{R_N\cdots R_1}\le N_{\rm path}\delta$, where $N_{\rm path}$ is the number of complete paths, and $\Pb(\nF P\le N_{\rm path}\delta)\ge b_2\delta^C\mathsf L^q$. Put $\varepsilon=N_{\rm path}^2\delta^2$.
\end{proof}

\subsection{A real--complex comparison and the conclusion of the proof}

\begin{lemma}\label{lem:comparison}
Let $f_1,\dots,f_r\in\R[x_1,\dots,x_D]$, and put $g_\R=\sum_kf_k^2$ on $\R^D$ and $g_\C=\sum_k|f_k|^2$ on $\C^D$. Let $\varphi\ge0$ on $\R^D$ and $\psi\ge0$ on $\C^D$ be compactly supported, smooth and not identically zero, with $\operatorname{supp}\varphi\subset\{\psi>0\}$. Let $\lambda_\R,m_\R$ and $c_\C,m_\C$ be the location and order of the first real pole of
\[
 \zeta_\R(s)=\int g_\R^{-s}\varphi\,dx,\qquad \zeta_\C(z)=\int g_\C^{-z}\psi\,dx\,d\bar x .
\]
Then $\lambda_\R\ge c_\C/2$, and if $\lambda_\R=c_\C/2$ then $m_\R\le m_\C$.
\end{lemma}

\begin{proof}
By principalization of ideals over the field $\R$ \cite{Hir64}, \cite[Thm.~1.0.1]{Wlo05}, \cite[Thm.~3.26]{Kol07}, there is a proper map $\pi:Y\to\R^D$, a composition of blow-ups along smooth centres defined over $\R$, together with an snc divisor $\sum E_i$, such that $\pi^*(f_1,\dots,f_r)=\mathcal O_Y(-\sum a_iE_i)$ and $K_{Y/\R^D}=\sum k_iE_i$. Base change to $\C$ gives a log principalization $\pi_\C$ of the complexified ideal.

The divisors through real points are real. Indeed, each $E_i$ is smooth over $\R$, so its complexification is smooth and its geometric components are pairwise disjoint. If a geometric component $E'$ of $E_i$ contains a real point $y$, then so does its complex conjugate $\overline{E'}$. Disjointness forces $\overline{E'}=E'$, so $E'$ is defined over $\R$.

As a consequence we have a common local form. Near every $y\in Y(\R)$ there are real-analytic coordinates $u$ in which the divisors through $y$ are $\{u_i=0\}$, $i\in I_y$, and
\[
 g_\R\circ\pi=\prod_{i\in I_y}u_i^{2a_i}\,v,\qquad |\det D\pi|=\prod_{i\in I_y}|u_i|^{k_i}\,J,
\]
with $v,J>0$. Here $v=\sum_kv_k^2$ with $f_k\circ\pi=u^av_k$, and some $v_k$ is a unit. Over $\C$ the same coordinates give $\prod|u_i|^{2a_i}\sum_k|v_k|^2$ and $\prod|u_i|^{2k_i}|J|^2$.

Consider now the poles. A partition of unity reduces $\zeta_\R$ to finitely many integrals of the form
\[
 \int\prod_i|u_i|^{k_i-2sa_i}\,(\text{positive})\,du
\]
near real points. Each has its first pole at $\min_{i\in I_y}(k_i+1)/(2a_i)$, of order equal to the number of indices attaining the minimum; positivity rules out cancellation. The complex integrals give the indices $(k_i+1)/a_i$ instead. The real points over $\operatorname{supp}\varphi$ lie where $\psi\circ\pi_\C>0$, and they carry the same $(a_i,k_i)$. This proves both claims.
\end{proof}

The multiplicity inequality can be strict. For $I=(x^2+y^2)$ one has $(\lambda_\R,m_\R)=(\tfrac12,1)$, whereas $I_\C=(x+iy)(x-iy)$ has $(c_\C,m_\C)=(1,2)$. The threshold inequality for a single function is due to Saito \cite[Cor.~2]{Sai07}, and multiplicity bookkeeping of this kind appears in \cite[Lemmas~3.5--3.6]{KW24}.

We also need to pass between Gaussian integrals and integrals over balls.

\begin{lemma}\label{lem:polar}
Let $D=\sum_jd_jd_{j-1}$ and $g=\nF P^2$, which is homogeneous of degree $2N$. Write $\zeta_{B_\rho,\F}$ for the Lebesgue integral of $g^{-z}$ over the ball of radius $\rho$, and $\zeta_{\gamma,\F}$ for the Gaussian expectation. Then
\[
 \zeta_{\gamma,\C}(z)=\pi^{-D}\Gamma(D-Nz+1)\,\rho^{-2(D-Nz)}\,\zeta_{B_\rho,\C}(z),
\]
\[
 \zeta_{\gamma,\R}(s)=(2\pi)^{-D/2}2^{D/2-Ns}\Gamma(D/2-Ns+1)\,\rho^{-(D-2Ns)}\,\zeta_{B_\rho,\R}(s).
\]
The first poles satisfy $C\le D/N$ and $\lambda_\R\le D/(2N)$, and the multipliers are holomorphic and nonzero there. Hence the Gaussian and ball zeta functions have the same first pole, with the same order.
\end{lemma}

\begin{proof}
Integrate in polar coordinates. The radial integrals are elementary. If the first pole exceeded $D/N$, the factor $\Gamma(D-Nz)$ in $\zeta_{\gamma,\C}=\tfrac12\pi^{-D}\Gamma(D-Nz)\int_{S^{2D-1}}g^{-z}$ would make $\zeta_{\gamma,\C}$ infinite at a point of holomorphy, since the sphere integral is positive. The real case is the same.
\end{proof}

We now prove Theorem~\ref{thm:real}, beginning with (i). By Proposition~\ref{prop:lower}, $\E\,g^{-s}=\int u^{-s}dF_F(u)$ has a pole at some $s\le C/2$. If the pole is at $C/2$, its order is at least $q+1$.

Conversely, fix $\rho>0$. Choose $\varphi\in C_c^\infty(\R^{D})$ with $\mathbf 1_{B_\rho}\le\varphi\le\mathbf 1_{B_{2\rho}}$, and $\psi\in C_c^\infty(\C^{D})$ with $0\le\psi\le1$ and $\psi=1$ near the complex ball of radius $2\rho$. By Lemma~\ref{lem:polar}, $\int g^{-s}\varphi$ has the same first pole and order as $\E_\R\,g^{-s}$, since it is squeezed between two ball integrals. The Gaussian density is bounded below on $\operatorname{supp}\psi$, so
\[
 \zeta_\C(z)\le\mathrm{const}\cdot\E_\C\,\nF P^{-2z}\qquad\text{for real }z<C .
\]
By Corollary~\ref{cor:lct}, therefore, $c_\C\ge C$, and the complex pole order at $C$ is at most $q+1$. Lemma~\ref{lem:comparison} gives $\lambda_\R\ge C/2$, and at equality $m_\R\le q+1$. Meromorphy follows from the polar formula as in Corollary~\ref{cor:lct}. This proves Theorem~\ref{thm:real}(i).

We now prove (ii) and (iii), using smooth cutoffs in place of balls. Put $\lambda=C/2$, and for $\psi\in C_c^\infty(\R^{D})$ write
\[
 \zeta_\psi(s)=\int g^{-s}\psi\,dA,\qquad F_\psi(u)=\int_{\{g\le u\}}\psi\,dA .
\]
By resolution of singularities, $F_\psi(u)$ has an asymptotic expansion as $u\downarrow0$ in the functions $u^r(\log1/u)^k$, with $k\ge0$ an integer and $r$ in a finite union of arithmetic progressions of positive rationals \cite[Ch.~4]{Wat09}. Since $|\psi|$ is bounded by a multiple of the indicator of a ball $B_\rho$, the proof of (i) shows that $\zeta_\psi$ is holomorphic on $\Re s<\lambda$ and has at most a pole of order $q+1$ at $\lambda$. The same bound gives $|F_\psi(u)|\le\|\psi\|_\infty u^{s'}\int_{B_\rho}g^{-s'}dA$ for every $s'<\lambda$, so no exponent $r<\lambda$ occurs. Inserting the expansion into $\zeta_\psi(s)=s\int_0^\infty u^{-s-1}F_\psi(u)\,du$, which holds for $0<s<\lambda$, and using $\int_0^1u^{\lambda-s-1}(\log1/u)^k\,du=k!\,(\lambda-s)^{-k-1}$, we see that the terms $u^\lambda(\log1/u)^k$ produce the pole at $\lambda$, so that the expansion begins with
\[
 F_\psi(u)=\frac{\mathcal R(\psi)}{\lambda\,q!}\,u^\lambda(\log1/u)^q+o\bigl(u^\lambda(\log1/u)^q\bigr),\qquad \mathcal R(\psi)=\lim_{s\uparrow\lambda}(\lambda-s)^{q+1}\zeta_\psi(s).
\]
The functional $\mathcal R$ is linear and positive, so it is a positive Radon measure on $\R^{D}$, and $\mathcal R(\psi)>0$ whenever $\psi\ge\mathbf 1_{B_\rho}$, because the ball integral has a pole of order exactly $q+1$ at $\lambda$.

To pass to the Gaussian density $\varphi_1(A)=(2\pi)^{-D/2}e^{-\|A\|^2/2}$, choose a radial $\psi_0\in C_c^\infty(\R^{D})$, nonincreasing in $\|A\|$, with $\psi_0=1$ on $B_1$ and support in $B_2$, and put $\psi_k=\psi_0(2^{-k}\,\cdot)-\psi_0(2^{1-k}\,\cdot)$ for $k\ge1$. Then $\psi_k\ge0$, $\sum_{k\ge0}\psi_k=1$, and $\psi_k=\psi_1(2^{1-k}\,\cdot)$ is supported in $\{2^{k-1}\le\|A\|\le2^{k+1}\}$. Homogeneity gives $F_{\psi_k}(u)=2^{(k-1)D}F_{\psi_1}(2^{-2N(k-1)}u)$, and the expansion of $F_{\psi_1}$ gives a constant $c$ with $F_{\psi_1}(u)\le c\,u^\lambda(\log(e/u))^q$ for $0<u\le1$. Hence, for $0<\varepsilon<e^{-1}$ and $\mathsf L=\log(1/\varepsilon)$,
\[
 0\le\frac{F_{\varphi_1\psi_k}(\varepsilon)}{\varepsilon^\lambda\mathsf L^q}\le c\,(2+2Nk)^q\,2^{(k-1)(D-2N\lambda)}\sup_{\|A\|\ge2^{k-1}}\varphi_1(A),\qquad k\ge1,
\]
and the right side is summable in $k$. Each term converges to $\mathcal R(\varphi_1\psi_k)/(\lambda\,q!)$ as $\varepsilon\downarrow0$, by the expansion applied to $\varphi_1\psi_k$, so dominated convergence for the series gives
\[
 \Pb(g\le\varepsilon)=\sum_{k\ge0}F_{\varphi_1\psi_k}(\varepsilon)\sim\kappa_\R\,\varepsilon^\lambda(\log1/\varepsilon)^q,\qquad \kappa_\R=\frac{\mathcal R(\varphi_1)}{\lambda\,q!}\in(0,\infty).
\]
Here $\mathcal R(\varphi_1)=\sum_k\mathcal R(\varphi_1\psi_k)$ by monotone convergence, which is finite by the majorant and positive because $\varphi_1\psi_0\ge(\min_{B_1}\varphi_1)\mathbf 1_{B_1}$. The Laplace asymptotics follows by the Abelian direction, and part (iii) follows from $\nop P^2\le g\le n\nop P^2$.

Finally, let $n=1$. By Lemma~\ref{lem:bartlett}, $g=\prod_{j}\chi^2_{d'_j}$ with independent factors. Hence
\[
 \E\,g^{-s}=2^{-Ns}\prod_j\Gamma(d'_j/2-s)/\Gamma(d'_j/2),
\]
which has a pole of order $\mu$ at $s=C/2$. Comparing its leading Laurent coefficient with $\tfrac C2\,\kappa_\R\,(\mu-1)!$, as in the proof of Corollary~\ref{cor:lct}, gives (iv). Part (v) is Proposition~\ref{prop:twofactors} below. \qed

\subsection{Two factors: the real constant in closed form}\label{sec:twofactors}

Let $N=2$. By Appendix~\ref{app:invariance} we may take $d=(n,m,p)$ with $n\le m$ and $n\le p$. Write
\[
 \Gamma_r(a)=\pi^{r(r-1)/4}\prod_{i=1}^r\Gamma\bigl(a-\tfrac{i-1}2\bigr),\qquad \Gamma_0=1,
\]
for the multivariate Gamma function. Put
\[
 E_k=\binom k2+k\,\frac{m-n+1}2+(n-k)\,\frac p2\quad(0\le k\le n),\qquad \tau=\frac{p-m+n-1}2 .
\]
The offsets are $(m-n,p-n)$, so $2E_k=E(k,n-k)$, and Lemma~\ref{lem:filling} gives $\lambda:=C/2=\min_kE_k$. Since $E_{k+1}-E_k=k-\tau$, the minimizer is unique unless $\tau\in\{0,\dots,n-1\}$, in which case $E_\tau=E_{\tau+1}$. Comparing with Definition~\ref{def:shell}, this tie happens exactly when the shell is open ($q=1$); otherwise $q=0$. Heuristically, $k$ is the number of eigenvalues of $A_1^{\top}A_1$ of order $1/t$ in the Laplace transform at large $t$, the others being of order one.

\begin{proposition}\label{prop:twofactors}
Let $N=2$, $\F=\R$ and $d=(n,m,p)$ as above.
\begin{enumerate}[label=(\alph*)]
\item If $q=0$, let $k$ be the minimizer of $E_k$. Then
\[
 \kappa_\R=\frac{2^{k(n+p-m-k)-np}}{\Gamma(\lambda+1)}\cdot
 \frac{\Gamma_k(\frac k2)\,\Gamma_k(\frac{p-m+n-k}2)\,\Gamma_{n-k}(\frac{m-p+k}2)}{\Gamma_k(\frac n2)\,\Gamma_k(\frac p2)\,\Gamma_{n-k}(\frac m2)} .
\]
\item If $q=1$, let $k=\tau$ and $r=n-k-1$. Then
\[
 \kappa_\R=\frac{2^{r(r+1)-mn}}{\sqrt\pi\,\Gamma(\lambda+1)}\cdot
 \frac{\Gamma_k(\frac k2)\,\Gamma_{k+1}(\frac{k+1}2)\,\Gamma_{r+1}(\frac{r+1}2)}{\Gamma_{k+1}(\frac n2)\,\Gamma_k(\frac p2)\,\Gamma_{r+1}(\frac m2)} .
\]
\end{enumerate}
Both expressions are symmetric under $m\leftrightarrow p$, which maps $k\mapsto n-k$ in (a) and $k\leftrightarrow r$ in (b).
\end{proposition}

\begin{proof}
We begin by reducing to a Wishart integral. Let $W=A_1^{\top}A_1$, a real Wishart matrix of size $n$ with $m$ degrees of freedom, and $g=\nF P^2$. Integrating out $A_2$,
\[
 \cZ(t)=\E\,e^{-\frac t2g}=\E\det(I_n+tW)^{-p/2}.
\]
The eigenvalues of $W$ have density $Z_W^{-1}\Delta_1(w)\prod_iw_i^{\alpha}e^{-w_i/2}$ on $(0,\infty)^n$, where
\[
 \Delta_1(w)=\prod_{i<j}|w_i-w_j|,\qquad \alpha=\tfrac{m-n-1}2,\qquad Z_W=\frac{n!\,2^{mn/2}\,\Gamma_n(\frac m2)\,\Gamma_n(\frac n2)}{\pi^{n^2/2}}
\]
\cite[Thm.~3.2.18]{Mui82}. Put $\gamma=p/2$, so that $E_k=\binom k2+k(\alpha+1)+(n-k)\gamma$ and $\tau=\gamma-\alpha-1$. Let
\[
 J_k=\int_{(0,\infty)^k}\Delta_1(y)\prod_iy_i^\alpha(1+y_i)^{-\gamma}\,dy,\qquad
 \Lambda_r(c)=\int_{(0,\infty)^r}\Delta_1(v)\prod_iv_i^{c-1}e^{-v_i/2}\,dv .
\]
We show that
\begin{equation}\label{eq:twofactor-limit}
 \lim_{t\to\infty}\frac{t^\lambda \cZ(t)}{(\log t)^q}=\mathsf A:=\begin{cases}\binom nk\,J_k\,\Lambda_{n-k}(\alpha+k-\gamma+1)/Z_W,&q=0,\\[3pt] \dfrac{n!}{k!\,r!}\,J_k\,\Lambda_{r}(1)/Z_W,&q=1.\end{cases}
\end{equation}
Then $\E\,e^{-sg}=\cZ(2s)$ and Karamata's theorem give $\Pb(g\le\varepsilon)\sim\mathsf A\,2^{-\lambda}\varepsilon^\lambda(\log1/\varepsilon)^q/\Gamma(\lambda+1)$, that is, $\kappa_\R=\mathsf A/(2^\lambda\Gamma(\lambda+1))$.

Consider first a closed shell. Write $\cZ(t)=(n!/Z_W)\int_{0<w_1<\dots<w_n}(\cdots)$, substitute $w_i=y_i/t$ for $i\le k$, and multiply by $t^{E_k}$. The integrand becomes
\[
 \Delta_1(y)\prod_{i\le k}y_i^\alpha e^{-y_i/2t}(1+y_i)^{-\gamma}\cdot\Delta_1(w)\prod_{l>k}w_l^\alpha e^{-w_l/2}\bigl(w_l+t^{-1}\bigr)^{-\gamma}\prod_{i\le k<l}\Bigl(w_l-\frac{y_i}t\Bigr)
\]
on the region $\{y_1<\dots<y_k<tw_{k+1},\ w_{k+1}<\dots<w_n\}$. It converges pointwise to its value at $t=\infty$. It is dominated by that limit, taken on the product of the two ordered cones, because each factor $w_l-y_i/t$ lies in $(0,w_l]$. The dominating function is integrable exactly when $\alpha+1>0$, $\gamma-\alpha-k>0$ and $\alpha+k-\gamma+1>0$. The last two conditions say that $E_k<E_{k-1}$ and $E_k<E_{k+1}$ (whenever these exist), i.e.\ that the minimizer is strict. Dominated convergence gives \eqref{eq:twofactor-limit}.

Consider next an open shell, so that $k=\tau$ and hence $k+\alpha-\gamma=-1$. Substitute $w_i=y_i/t$ for $i\le k$, keep $x=w_{k+1}$, and multiply by $t^{E_k}$. The factors depending on $x$ alone combine to
\[
 x^{k+\alpha}t^\gamma(1+tx)^{-\gamma}=x^{-1}\bigl(1+\tfrac1{tx}\bigr)^{-\gamma}.
\]
Hence
\[
 t^\lambda \cZ(t)=\frac{n!}{Z_W}\int_0^\infty x^{-1}\bigl(1+\tfrac1{tx}\bigr)^{-\gamma}e^{-x/2}H_t(x)\,dx,
\]
where $H_t(x)$ is the integral over $\{y_1<\dots<y_k<tx\}\times\{x<w_{k+2}<\dots<w_n\}$ of
\begin{multline*}
 \Delta_1(y)\prod_iy_i^\alpha e^{-y_i/2t}(1+y_i)^{-\gamma}\Bigl(1-\frac{y_i}{tx}\Bigr)\\
 \cdot\Delta_1(w)\prod_lw_l^\alpha e^{-w_l/2}\bigl(w_l+t^{-1}\bigr)^{-\gamma}(w_l-x)\prod_i\Bigl(w_l-\frac{y_i}t\Bigr).
\end{multline*}
As in the closed case, $0\le H_t\le H^*:=(J_k/k!)(\Lambda_r(1)/r!)$. Here $H^*$ is finite because $\gamma-\alpha-k=1$ and $\alpha+k+1-\gamma=0$. Moreover $H_t(t^{-\sigma})\to H^*$ for every $\sigma\in(0,1)$, since then $tx\to\infty$ and $x\to0$. Substituting $x=t^{-\sigma}$,
\[
 \frac{t^\lambda \cZ(t)}{\log t}=\frac{n!}{Z_W}\int_{-\infty}^\infty\bigl(1+t^{\sigma-1}\bigr)^{-\gamma}e^{-t^{-\sigma}/2}H_t(t^{-\sigma})\,d\sigma .
\]
For $t\ge2$ the integrand is at most
\[
 H^*\bigl[\mathbf 1_{[0,1]}(\sigma)+2^{-\gamma(\sigma-1)}\mathbf 1_{\sigma>1}+e^{-2^{|\sigma|}/2}\mathbf 1_{\sigma<0}\bigr],
\]
and pointwise it tends to $H^*\mathbf 1_{(0,1)}(\sigma)$. Dominated convergence gives \eqref{eq:twofactor-limit}. The logarithm is produced by a single intermediate eigenvalue in the window $t^{-1}\ll x\ll1$: this is the one particle--hole pair of the open shell.

It remains to evaluate the limits. For a symmetric function $f$ of the eigenvalues of a $k\times k$ matrix,
\[
 \int_{(0,\infty)^k}\Delta_1(y)f(y)\,dy=\frac{k!\,\Gamma_k(k/2)}{\pi^{k^2/2}}\int_{Y>0}f(Y)\,dY
\]
\cite[Thm.~3.2.17]{Mui82}. Combine this with the matrix beta and gamma integrals \cite[\S2.1]{Mui82},
\[
 \int_{Y>0}\frac{(\det Y)^{a-\frac{k+1}2}}{\det(I+Y)^{a+b}}\,dY=\frac{\Gamma_k(a)\Gamma_k(b)}{\Gamma_k(a+b)},\qquad
 \int_{V>0}(\det V)^{a-\frac{r+1}2}e^{-\operatorname{tr}V/2}\,dV=2^{ra}\Gamma_r(a).
\]
This gives
\[
 J_k=\frac{k!\,\Gamma_k(\frac k2)}{\pi^{k^2/2}}\cdot\frac{\Gamma_k(\frac{m-n+k}2)\,\Gamma_k(\frac{p-m+n-k}2)}{\Gamma_k(\frac p2)},\qquad
 \Lambda_r(c)=\frac{r!\,\Gamma_r(\frac r2)}{\pi^{r^2/2}}\cdot2^{r(c+\frac{r-1}2)}\,\Gamma_r\bigl(c+\tfrac{r-1}2\bigr).
\]
Insert these into $\mathsf A$. The splitting rule $\Gamma_n(a)=\pi^{j(n-j)/2}\Gamma_j(a)\Gamma_{n-j}(a-\tfrac j2)$ cancels $\Gamma_n(\frac m2)$ and $\Gamma_n(\frac n2)$. In case (b) also use its special cases $\Gamma_k(\frac{k+1}2)=\pi^{-(k+1)/2}\Gamma_{k+1}(\frac{k+1}2)$ and $\Gamma_r(\frac{r+1}2)=\pi^{-(r+1)/2}\Gamma_{r+1}(\frac{r+1}2)$. All powers of $\pi$ then cancel, except a factor $\pi^{-1/2}$ in case (b). Collecting the powers of $2$, using $2\lambda=nm-n^2+n+2kn-k^2-k$ in case (b), gives (a) and (b). The same splitting rule proves the symmetry under $m\leftrightarrow p$.
\end{proof}

\begin{remark}[Independent checks]\label{rem:twofactor-checks}
For $n=1$, (a) and (b) reduce to Theorem~\ref{thm:real}(iv). For $d=(2,2,2)$, (a) gives $\lambda=\tfrac32$ and $\kappa_\R=\pi/6$, which we can confirm independently by writing $g=w_1G_1+w_2G_2$, where $w_1<w_2$ are the eigenvalues of $W$ and $G_1,G_2\sim\chi^2_2$, and integrating over the region $w_1=O(\varepsilon)$. The leading term is
\[
 \tfrac1{16}\sqrt{2\pi}\cdot\tfrac43\sqrt{2\pi}\;\varepsilon^{3/2}=\tfrac\pi6\,\varepsilon^{3/2}.
\]
For $d=(2,2,4)$, (a) gives $\kappa_\R=1/16$. Here the minimizer is $k=2$, so that all of $A_1$ is small, and $g=\operatorname{vec}(A_1)^{\top}(I_2\otimes M)\operatorname{vec}(A_1)$ with $M=A_2^{\top}A_2$. The volume of the ellipsoid gives
\[
 \Pb(g\le\varepsilon)\sim(2\pi)^{-2}\tfrac{\pi^2}2\,\E[\det M^{-1}]\,\varepsilon^2=\tfrac1{16}\varepsilon^2,
\]
because $\det M\sim\chi^2_4\,\chi^2_3$ by Lemma~\ref{lem:bartlett}. For $d=(2,3,3)$, (a) and Theorem~\ref{thm:closed} both give $\kappa_\R=1/15$, and for the open shell $d=(2,2,3)$, (b) gives $1/16$ for both orders $(m,p)=(2,3)$ and $(3,2)$. Finally, the same proof with the complex Wishart density gives an independent derivation of $\cA$ for $N=2$, which for $d=(2,2,2)$, $(2,2,3)$ and $(2,2,4)$ returns $\cA=1$, $\tfrac12$ and $\tfrac1{12}$, in agreement with Theorem~\ref{thm:A}.
\end{remark}
\begin{corollary}\label{cor:realsquare}
Let $d=(n,n,n)$. If $n=2k$ is even, then $q=0$, $\lambda=\tfrac32k^2$ and
\[
 \kappa_\R=\frac{2^{-3k^2}}{\Gamma(\lambda+1)}\Bigl[\frac{\Gamma_k(k/2)}{\Gamma_k(k)}\Bigr]^3 .
\]
If $n=2k+1$ is odd, then $q=1$, $\lambda=(3n^2+1)/8$ and
\[
 \kappa_\R=\frac{2^{k(k+1)-n^2}}{\sqrt\pi\,\Gamma(\lambda+1)}\cdot\frac{\Gamma_k(\frac k2)\,\Gamma_{k+1}(\frac{k+1}2)^2}{\Gamma_{k+1}(\frac n2)^2\,\Gamma_k(\frac n2)} .
\]
For example, $\kappa_\R=\pi/6$ for $n=2$ and $\kappa_\R=1/(105\pi)$ for $n=3$.
\end{corollary}

\subsection{Comparison with Aoyagi's formula}

\begin{lemma}\label{lem:aoyagi}
Recall that $n=d_0\le d_1\le\dots\le d_N$, and put $\sigma_\ell=\sum_{i\le\ell}d_i$. For an open shell the index pair of \cite{Aoy24} at target rank $0$ is $(\ell,a)=(\mu,h)$, and for a closed shell it is $\ell=\mu+\#\{j:\nu_j=L_F+1\}$ and $a=\ell$. In both cases Aoyagi's threshold and multiplicity equal $C/2$ and $a(\ell-a)+1=q+1$.
\end{lemma}

\begin{proof}
Direct substitution gives
\[
 \sigma_\ell-\ell d_\ell=n-N_<(\nu_\ell),\qquad \sigma_\ell-\ell d_{\ell+1}=n-N_<(\nu_{\ell+1}).
\]
Hence Aoyagi's conditions $\ell d_\ell\le \sigma_\ell<\ell d_{\ell+1}$ and $d_\ell<d_{\ell+1}$ select $\ell=\mu$ for an open shell, with $\sigma_\mu=\mu(n+L_F)+h$, and the stated $\ell$ with $\sigma_\ell=\ell(n+L_F+1)$ for a closed shell. The integer $a=\sigma_\ell-\ell(\lceil \sigma_\ell/\ell\rceil-1)$ equals $h$, respectively $\ell$.

At target rank $0$, Aoyagi's threshold is
\[
 4\lambda_{\rm A}=\frac{a(\ell-a)}\ell+\frac{\sigma_\ell^2}\ell-\sum_{i\le\ell}d_i^2 .
\]
Put $y_i=d_i-(n+L_F)$, so that $\sum_{i\le\ell}y_i=h-n-L_F$. A short computation gives
\[
 4\lambda_{\rm A}=n^2+2nL_F+h-\sum_{\nu_i\le L_F}(L_F-\nu_i)^2 .
\]
On the other hand, using $\sum_{\nu_i\le L_F}(L_F-\nu_i)=n-h$,
\[
 2C=n(n+1)+\sum_{\nu_i\le L_F}(L_F-\nu_i)(L_F+\nu_i-1)+2hL_F
\]
is the same expression. The closed shell is identical, with $L_F\to L_F+1$ and $h\to0$.
\end{proof}

The expression for the multiplicity printed in \cite[Thm.~8.6]{PLR24} equals $q$, the power of the logarithm, whereas the order of the pole is $q+1$, as in Aoyagi's formula and in Theorem~\ref{thm:real}(i).

\section{\texorpdfstring{The real constant: proof of Theorem~\ref{thm:closed}}{The real constant: proof of Theorem D}}\label{sec:proofD}

Throughout this section we sort the sizes as in Section~\ref{sec:proofB}, so that $n=d_0\le d_1\le\dots\le d_N$ and $\nu_j=d_j-n$, and we put $\nu_0=0$. This is allowed for both fields by the width-permutation invariance of \cite[Sec.~III.B]{IK14}, which covers real, complex and quaternion entries (see also Appendix~\ref{app:invariance}), and $K_\beta(e)$ is invariant under relabelling the ladders. We fix a minimizer $e$ of Lemma~\ref{lem:filling}, with ranks $r_j=n-\sum_{i\le j}e_i$, and we write $\bar r_j=d_j-r_j$, so that $\bar r_0=0$, $\bar r_N=d_N$ and $\bar r_j-\bar r_{j-1}=\nu_j-\nu_{j-1}+e_j\ge0$. We call the $\mu$ ladders $j$ with $\nu_j\le L_F$, which carry a state at the Fermi level, \emph{tied}. We treat both fields at once, with $\beta=1$ for $\F=\R$ and $\beta=2$ for $\F=\C$. We write $\varphi_\beta$ for the standard Gaussian density on $\F^D$, $D=\sum_jd_jd_{j-1}$, and $\gamma_\nu$ for the squared norm of a standard Gaussian vector in $\F^\nu$, so that
\begin{equation}\label{eq:chimoment}
 \E\,\gamma_\nu^{-s}=\Bigl(\frac\beta2\Bigr)^{s}\frac{\Gamma(\beta\nu/2-s)}{\Gamma(\beta\nu/2)},\qquad s<\beta\nu/2 .
\end{equation}
In Section~\ref{sec:slice} we construct, along the stratum $S_e$ of Proposition~\ref{prop:geometry}, a transversal slice in which the quadratic part of $\nF P^2$ is block diagonal. In Section~\ref{sec:evaluation} we evaluate the resulting Morse--Bott integral in closed form, regroup it into the particle form \eqref{eq:Kbeta} and prove Proposition~\ref{prop:particle}. In Section~\ref{sec:divisors} we show by a principalization argument, which transfers the support of residue measures from $\C$ to $\R$, that for a closed shell the Morse--Bott integral is the true constant. In Section~\ref{sec:conjevidence} we turn to open shells, explain the residue rule and prove it for one logarithm, which completes the proof of Theorem~\ref{thm:closed}.

\subsection{A slice adapted to the layers}\label{sec:slice}

Let $A=(A_1,\dots,A_N)\in S_e$ and put $P_j=A_j\cdots A_1$. Let $V_j=\operatorname{Im}P_j\subset\F^{d_j}$ and $F_j=\ker P_j\subset\F^n$, of dimensions $r_j$ and $n-r_j$. The kernels form a flag $0=F_0\subset F_1\subset\dots\subset F_N=\F^n$, and we let $G_j=F_j\ominus F_{j-1}$ be the orthogonal complement of $F_{j-1}$ in $F_j$, so that $\F^n=G_1\oplus\dots\oplus G_N$ orthogonally with $\dim G_j=e_j$. The map $P_{j-1}$ is injective on $F_{j-1}^\perp\supset G_j$, and $\mathcal K_j=P_{j-1}(G_j)$ is the kernel of $A_j$ restricted to $V_{j-1}$. For $\delta_j\in\operatorname{Hom}(\mathcal K_j,V_j^\perp)$ we let $\iota_j(\delta_j)\in\F^{d_j\times d_{j-1}}$ be the map that equals $\delta_j$ on $\mathcal K_j$ and vanishes on the complement $P_{j-1}(F_j^\perp)\oplus V_{j-1}^\perp$ of $\mathcal K_j$. The dimensions add up correctly, since $\sum_je_j\bar r_j=\sum_je_j(d_j-r_j)=\codim S_e$ by the proof of Proposition~\ref{prop:geometry}.

\begin{lemma}\label{lem:slice}
Put $\mathsf N_j=P_{j-1}|_{G_j}:G_j\to \mathcal K_j$ and $\mathsf M_j=(A_N\cdots A_{j+1})|_{V_j^\perp}$.
\begin{enumerate}[label=(\alph*)]
\item The first-order variation of $P$ in the direction $(\iota_j(\delta_j))_j$ is $\sum_j\mathsf M_j\delta_j\mathsf N_j\Pi_{G_j}$, and $\nF{\sum_j\mathsf M_j\delta_j\mathsf N_j\Pi_{G_j}}^2=\sum_j\nF{\mathsf M_j\delta_j\mathsf N_j}^2$.
\item Under $(A,\delta)\mapsto(A_j+\iota_j(\delta_j))_j$, with $\delta_j$ written in orthonormal frames of $\mathcal K_j$ and $V_j^\perp$, Lebesgue measure on $\F^D$ pulls back to
\[
 (1+O(|\delta|))\,\operatorname{vol}_1(dA_1)\,\operatorname{vol}_2(dA_2\,|\,A_1)\cdots\operatorname{vol}_N(dA_N\,|\,A_{<N})\,d\delta .
\]
Here $\operatorname{vol}_j(\cdot\,|\,A_{<j})$ is the Riemannian measure on the layer stratum $\{A_j:\operatorname{rank}(A_j|_{V_{j-1}})=r_j\}$ for fixed $A_1,\dots,A_{j-1}$, and we write $\operatorname{vol}_e=\prod_j\operatorname{vol}_j$ for this iterated measure.
\item Write $A_i=A_i'\Pi_{V_{i-1}}+A_i''\Pi_{V_{i-1}^\perp}$ and $\mathsf B_i=\Pi_{V_i^\perp}A_i''|_{V_{i-1}^\perp}\in\F^{\bar r_i\times\bar r_{i-1}}$. Then
\begin{equation}\label{eq:Mchain}
 \mathsf M_j=\mathsf B_N\mathsf B_{N-1}\cdots\mathsf B_{j+1}.
\end{equation}
The maps $\mathsf N_j$ are invertible at every point of $S_e$, and at a generic point the $\mathsf M_j$ are injective, so that the quadratic form in (a) is nondegenerate.
\end{enumerate}
\end{lemma}

\begin{proof}
Since $P=0$ on $S_e$, the first-order variation of $P$ is $\sum_jA_N\cdots A_{j+1}\iota_j(\delta_j)P_{j-1}$. Consider $x\in G_i$. If $i<j$ then $P_{j-1}x=0$. If $i>j$ then $x\in F_j^\perp$, so that $P_{j-1}x\in P_{j-1}(F_j^\perp)$ is annihilated by $\iota_j(\delta_j)$. If $i=j$ then $\iota_j(\delta_j)P_{j-1}x=\delta_j\mathsf N_jx\in V_j^\perp$. Hence the $j$-th term is $\mathsf M_j\delta_j\mathsf N_j\Pi_{G_j}$, and the orthogonality of the $G_j$ gives the identity of norms.

For (b) we compare $\iota_j$ with the orthogonal normal slice $\delta_j\mapsto\delta_j\Pi_{\mathcal K_j}$ of the layer stratum. The difference is $\delta_j(\pi_j-\Pi_{\mathcal K_j})$, where $\pi_j$ is the projection onto $\mathcal K_j$ along $P_{j-1}(F_j^\perp)$. It vanishes on $\mathcal K_j$, and it is therefore tangent to the stratum, whose tangent space consists of the maps that send $\mathcal K_j$ into $V_j$. A shear along tangent directions has unit Jacobian. Moreover, the $j$-th component of $(A_1+\iota_1(\delta_1),\dots,A_N+\iota_N(\delta_N))$ depends on the base point through $A_1,\dots,A_j$ but on no $\delta_i$ with $i\ne j$. The Jacobian of the parametrization is therefore block lower triangular with respect to the layers, which gives (b).

For (c), $\mathsf N_j$ is the restriction to $G_j$ of the injective map $P_{j-1}|_{F_{j-1}^\perp}$. Since $A_i'$ maps $V_{i-1}$ onto $V_i$, we have $\Pi_{V_i^\perp}A_i=\mathsf B_i\Pi_{V_{i-1}^\perp}$, and induction on $i$ gives $\Pi_{V_i^\perp}A_i\cdots A_{j+1}|_{V_j^\perp}=\mathsf B_i\cdots\mathsf B_{j+1}$. For $i=N$ this is \eqref{eq:Mchain}, because $V_N=0$. The matrices $\mathsf B_i$ are free on $S_e$ and tall, since $\bar r_i\ge\bar r_{i-1}$, and so $\mathsf M_j$ is injective at generic points.
\end{proof}

By Lemma~\ref{lem:slice}, near a generic point $A\in S_e$ the function $\nF P^2$ is, to second order, a nondegenerate quadratic form in the $\beta C$ real coordinates of $\delta$. Its Gram determinant is
\[
 \prod_j\det(\mathsf M_j^*\mathsf M_j)^{\beta e_j}\,|\det\mathsf N_j|^{2\beta\bar r_j},
\]
where $\det$ is the determinant over $\F$. The set $\{\nF P^2\le\varepsilon\}$ therefore meets the slice, to leading order, in an ellipsoid of volume
\[
 \operatorname{vol}(B_{\beta C})\,\varepsilon^{\beta C/2}\prod_j\det(\mathsf M_j^*\mathsf M_j)^{-\beta e_j/2}\,|\det\mathsf N_j|^{-\beta\bar r_j},
\]
where $B_{\beta C}$ is the unit ball of $\R^{\beta C}$. This leads us to the \emph{Morse--Bott integral}
\begin{equation}\label{eq:MB}
 I_\beta(e;\chi)=\operatorname{vol}(B_{\beta C})\int_{S_e}\chi(A)\prod_{j=1}^N\det(\mathsf M_j^*\mathsf M_j)^{-\beta e_j/2}\,|\det\mathsf N_j|^{-\beta\bar r_j}\,d\operatorname{vol}_e(A),
\end{equation}
for a weight $\chi\ge0$ on $\F^D$, and we write $I_\beta(e)=I_\beta(e;\varphi_\beta)$. It is the candidate for the constant, and the next lemma makes this precise near generic points of $S_e$. Here and below we write $g=\nF P^2$ and $\lambda=\beta C/2$, and we call a point of $S_e$ generic if the conclusions of Lemma~\ref{lem:slice}(c) hold there.

\begin{lemma}\label{lem:localMB}
Let $z\in S_e$ be generic and let $\chi\ge0$ be smooth with support in a sufficiently small neighbourhood of $z$. Then $(\lambda-s)\int g^{-s}\chi\,dA\to\lambda\,I_\beta(e;\chi)$ as $s\uparrow\lambda$.
\end{lemma}

\begin{proof}
The differential $dP_z$ vanishes on $T_zS_e$, is injective on the slice by Lemma~\ref{lem:slice}(a), and the slice is transversal to $S_e$ by Lemma~\ref{lem:slice}(b), so $dP_z$ has real rank $\beta C$. Choose $\beta C$ real components of $P$ whose differentials at $z$ are independent. Their common zero set $Z'$ is a smooth submanifold of codimension $\beta C$ near $z$, and $S_e\subset\{P=0\}\subset Z'$. Since $S_e$ is smooth of the same codimension, it is open in $Z'$, and the three sets coincide near $z$. In the coordinates $(\sigma,\delta)$ of Lemma~\ref{lem:slice}(b), with $\sigma$ a coordinate on $S_e$, we have $g=\delta^*H(\sigma)\delta+O(|\delta|^3)$, where $H(\sigma)$ is the positive definite quadratic form of Lemma~\ref{lem:slice}(a), and Lebesgue measure equals $(1+O(|\delta|))\,d\operatorname{vol}_e(\sigma)\,d\delta$. The Morse lemma with parameters gives coordinates $x=H(\sigma)^{1/2}\delta+O(|\delta|^2)$ in which $g=|x|^2$ exactly, and then $d\delta=\det(H(\sigma))^{-1/2}(1+O(|x|))\,dx$ with $\det$ the real determinant. For a continuous compactly supported $f$ on $\R^{\beta C}$, polar coordinates give $(\lambda-s)\int|x|^{-2s}f(x)\,dx\to\frac12|S^{\beta C-1}|f(0)$. Since $\frac12|S^{\beta C-1}|=\lambda\operatorname{vol}(B_{\beta C})$ and $\det(H)^{1/2}$ is the product in \eqref{eq:MB}, the lemma follows.
\end{proof}

\subsection{Evaluation and the particle form}\label{sec:evaluation}

\begin{lemma}\label{lem:evaluation}
Put $\sigma_{i,a}=\sum_{j<i,\,\bar r_j\ge a}e_j$ and $\tau_{i,a}=\sum_{i\le j<N,\,r_j\ge a}(\bar r_{j+1}-\bar r_j)$. Then
\begin{multline}\label{eq:star}
 I_\beta(e)=\frac{(\beta/2)^{\beta C/2}}{\Gamma(\beta C/2+1)}\prod_{i=1}^{N-1}\Bigl[\Bigl(\frac\beta2\Bigr)^{\beta r_i\bar r_i/2}\prod_{t=1}^{r_i}\frac{\Gamma(\beta t/2)}{\Gamma(\beta(\bar r_i+t)/2)}\Bigr]\\
 \times\prod_{i=2}^{N}\prod_{a=1}^{\bar r_{i-1}}\E\,\gamma_{\bar r_i-a+1}^{-\beta\sigma_{i,a}/2}\;\prod_{i=1}^{N-1}\prod_{a=1}^{r_i}\E\,\gamma_{r_{i-1}-a+1}^{-\beta\tau_{i,a}/2},
\end{multline}
in the sense that both sides are finite or infinite together. All moments in \eqref{eq:star} are finite if and only if the shell is closed.
\end{lemma}

\begin{proof}
The Gaussian density factorizes over the layers and, within the layer $j$, over $A_j'$, $\Pi_{V_j}A_j''$ and $\mathsf B_j$. Consider first the output factors. By \eqref{eq:Mchain} they depend only on the $\mathsf B_i$, which are independent standard Gaussian matrices. For fixed $i$, the images of $\mathsf B_{i-1}\cdots\mathsf B_{j+1}$ in $\F^{\bar r_{i-1}}$, for $j<i$, form a flag of dimensions $\bar r_j$ that does not depend on $\mathsf B_i$. If $Z$ has full column rank and the columns of $O$ form an orthonormal basis of its image, then $\det(Z^*WZ)=\det(Z^*Z)\det(O^*WO)$. We apply this with $W=\mathsf B_i^*\mathsf B_i$ and $Z=\mathsf B_{i-1}\cdots\mathsf B_{j+1}$, and we write $\mathsf B_i$ in an orthonormal basis $Q$ of $\F^{\bar r_{i-1}}$ adapted to the flag. The Bartlett decomposition $\mathsf B_iQ=O'\mathsf T$ (Lemma~\ref{lem:bartlett} and its complex analogue) then gives $\det(O^*WO)=\prod_{a\le\bar r_j}t_{i,a}$ with $t_{i,a}=|\mathsf T_{aa}|^2\sim\gamma_{\bar r_i-a+1}$, and these variables are independent of $\mathsf B_1,\dots,\mathsf B_{i-1}$, because $\mathsf B_iQ$ is a standard Gaussian matrix for every $Q$ that depends only on them. Induction on $i$ gives $\det(\mathsf M_j^*\mathsf M_j)=\prod_{i>j}\prod_{a\le\bar r_j}t_{i,a}$ with independent $t_{i,a}$, which yields the first product in the second line of \eqref{eq:star}.

Consider next the constrained parts $A_j'$. On matrices of rank $r$ in $\F^{p\times q}$ the Riemannian measure is $d\operatorname{vol}_{\Gr(r,p)}(\operatorname{Im}A)\cdot\det(\mathsf C\mathsf C^*)^{\beta(p-r)/2}\,d\mathsf C$, where $\mathsf C\in\F^{r\times q}$ is $A$ written in an orthonormal basis of its image. Indeed, the fibre over a fixed image $V$ is open in $\operatorname{Hom}(\F^q,V)$, and moving $V$ by $\phi\in\operatorname{Hom}(V,V^\perp)$ moves $A$ by $\phi A$, orthogonally to the fibre and with Gram determinant $\det(\mathsf C\mathsf C^*)^{\beta(p-r)}$. Here $\Gr(r,p)$ carries the metric in which $T_V\Gr=\operatorname{Hom}(V,V^\perp)$ has the Frobenius norm. The integrand is invariant under rotations of the images, so the Grassmannians integrate out to $\operatorname{vol}\Gr(r_j,d_j)=\pi^{\beta r_j\bar r_j/2}\prod_{t\le r_j}\Gamma(\beta t/2)/\Gamma(\beta(\bar r_j+t)/2)$. After normalization by $\varphi_\beta$ this gives the bracket in the first line, and it leaves matrices $\mathsf C_j\in\F^{r_j\times r_{j-1}}$ with weight $c_j^{\beta\bar r_j}$ against the Gaussian measure, where $c_j=\det(\mathsf C_j\mathsf C_j^*)^{1/2}$. In these bases $P_j=\mathsf C_j\cdots\mathsf C_1$. Put $\mathsf D_j=\det(P_jP_j^*)^{1/2}$, the product of the nonzero singular values of $P_j$, with $\mathsf D_0=\mathsf D_N=c_N=1$. We claim that
\begin{equation}\label{eq:telescope}
 |\det\mathsf N_j|=\frac{\mathsf D_{j-1}\,c_j}{\mathsf D_j}.
\end{equation}
Indeed, the restriction $\Lambda=P_{j-1}|_{F_{j-1}^\perp}$ is an isomorphism onto $V_{j-1}$ with $|\det\Lambda|=\mathsf D_{j-1}$. With respect to the orthogonal decompositions $F_{j-1}^\perp=G_j\oplus F_j^\perp$ and $V_{j-1}=\mathcal K_j\oplus \mathcal K_j^\perp$ it is block upper triangular, because $\Lambda(G_j)=\mathcal K_j$, so $|\det\Lambda|=|\det\mathsf N_j|\,|\det\Lambda_{22}|$ with $\Lambda_{22}=\Pi_{\mathcal K_j^\perp}\Lambda|_{F_j^\perp}$. On the other hand $P_j|_{F_j^\perp}=A_j'\Lambda|_{F_j^\perp}=(A_j'|_{\mathcal K_j^\perp})\Lambda_{22}$, since $A_j'$ vanishes on $\mathcal K_j$, and $A_j'|_{\mathcal K_j^\perp}$ is an isomorphism onto $V_j$ with $|\det|=c_j$. Hence $\mathsf D_j=c_j|\det\Lambda_{22}|$, which gives \eqref{eq:telescope}.
Hence $\prod_j|\det\mathsf N_j|^{-\beta\bar r_j}=\prod_jc_j^{-\beta\bar r_j}\prod_{j=1}^{N-1}\mathsf D_j^{-\beta(\bar r_{j+1}-\bar r_j)}$, and the factors $c_j$ cancel against the weight of the measure. What remains is the Gaussian expectation of $\prod_{j=1}^{N-1}\mathsf D_j^{-\beta(\bar r_{j+1}-\bar r_j)}$ over independent standard matrices $\mathsf C_j$. Since $\mathsf D_j^2=\det(P_jP_j^*)$ is the Gram determinant of $P_j^*=\mathsf C_1^*\cdots\mathsf C_j^*$, and the images of $\mathsf C_2^*\cdots\mathsf C_j^*$ in $\F^{r_1}$ decrease with $j$, the argument used for the output factors applies to this transposed chain and gives $\mathsf D_j^2=\prod_{i\le j}\prod_{a\le r_j}t'_{i,a}$ with independent $t'_{i,a}\sim\gamma_{r_{i-1}-a+1}$. This yields the last product in \eqref{eq:star}.

It remains to collect the normalizations. With $\varphi_\beta=(2\pi/\beta)^{-\beta D/2}e^{-\beta\|A\|^2/2}$, a standard Gaussian block with $m$ entries in $\F$ contributes $(2\pi/\beta)^{\beta m/2}$, and the blocks $\Pi_{V_j}A_j''$, $\mathsf B_j$ and $\mathsf C_j$ together have $\sum_j(d_j\bar r_{j-1}+r_jr_{j-1})$ entries. Since $D-\sum_j(d_j\bar r_{j-1}+r_jr_{j-1})=\sum_j\bar r_jr_{j-1}=C+\sum_jr_j\bar r_j$, by $r_{j-1}=r_j+e_j$ and $\sum_je_j\bar r_j=C$, we are left with $\operatorname{vol}(B_{\beta C})(2\pi/\beta)^{-\beta C/2}=(\beta/2)^{\beta C/2}/\Gamma(\beta C/2+1)$ and, for each $j$, with $(2\pi/\beta)^{-\beta r_j\bar r_j/2}\operatorname{vol}\Gr(r_j,d_j)$. These are the prefactor and the bracket in \eqref{eq:star}. All integrands are nonnegative, so these manipulations are valid in $[0,\infty]$.

By \eqref{eq:chimoment}, the moment $\E\gamma_\nu^{-s}$ is finite if and only if $s<\beta\nu/2$. As $a$ runs through a block on which $\sigma_{i,a}$ or $\tau_{i,a}$ is constant, the index $\nu$ decreases, so the binding constraint is the last $a$ of each block. Put $f_j=\nu_j+e_j$. A short computation with $\bar r_{l-1}+r_{l-1}=n+\nu_{l-1}$ and $\bar r_i+r_{i-1}=n+f_i$ then turns the conditions into $f_l\le f_i$ for all $l<i$ with $\bar r_l>\bar r_{l-1}$, and $f_{i'}\le f_i$ for all $i<i'$ with $e_{i'}\ge1$. In a closed shell $f_j=L_F+1$ on every occupied ladder and $f_j=\nu_j\ge L_F+1$ on the others. Since the sizes are sorted, $f_i\ge\nu_i\ge\nu_l$ for $i>l$, so both conditions hold. In an open shell take tied ladders $j_0$ and $j_1$ with $\eta_{j_0}=0$ and $\eta_{j_1}=1$, so that $f_{j_0}=L_F<f_{j_1}$. The first condition fails if $j_1<j_0$, since $\bar r_{j_1}-\bar r_{j_1-1}\ge e_{j_1}\ge1$, and the second fails if $j_0<j_1$.
\end{proof}

We now regroup \eqref{eq:star} into the particle form \eqref{eq:Kbeta}.

\begin{lemma}\label{lem:particleform}
Let $e$ be any minimizer. In \eqref{eq:star}, replace every moment $\E\gamma_\nu^{-\beta\nu/2}$, which is infinite, by $(\beta/2)^{\beta\nu/2}/\Gamma(\beta\nu/2)=\lim_{s\uparrow\beta\nu/2}(\beta\nu/2-s)\,\E\gamma_\nu^{-s}$, the coefficient of the pole of \eqref{eq:chimoment} at $s=\beta\nu/2$. Then the right side of \eqref{eq:star} becomes $K_\beta(e)$, and the replaced moments correspond exactly to the pairs omitted in $\prod^*$. In particular $I_\beta(e)=K_\beta(e)$ for a closed shell.
\end{lemma}

\begin{proof}
Fix $i$. On the block $\bar r_{l-1}<a\le\bar r_l$ with $l<i$ we have $\sigma_{i,a}=r_{l-1}-r_{i-1}$. Writing $a=\bar r_{l-1}+b$ and using $\bar r_{l-1}+r_{l-1}=n+\nu_{l-1}$ and $\bar r_i+r_{i-1}=n+f_i$, the moment becomes a ratio of Gamma functions at $\beta(f_i-c)/2$ and $\beta(\bar r_i-a+1)/2$, with $c=\nu_{l-1}+b-1$. As $b$ runs from $1$ to $\bar r_l-\bar r_{l-1}=f_l-\nu_{l-1}$, the integer $c$ runs through $[\nu_{l-1},f_l-1]$. This interval is the union of $[\nu_{l-1},\nu_l-1]$ and the set of energies of the particles on ladder $l$. The first pieces add up to $[0,\nu_{i-1}-1]$, whose Gamma functions cancel against part of the denominators, and we obtain
\[
 \prod_a\E\gamma_{\bar r_i-a+1}^{-\beta\sigma_{i,a}/2}=\Bigl(\frac\beta2\Bigr)^{\frac\beta2\sum_a\sigma_{i,a}}\frac{\prod_{\ell(p)<i}\Gamma\bigl(\beta(f_i-\varepsilon_p)/2\bigr)}{\prod_{t=1}^{n-r_{i-1}}\Gamma\bigl(\beta(f_i+t)/2\bigr)} .
\]
In the same way, on the block $r_{l+1}<a\le r_l$ with $l\ge i$ we have $\tau_{i,a}=\bar r_{l+1}-\bar r_i$. The arguments become $\beta(f_i-\varepsilon_p)/2$ for the particles $p$ on ladder $l+1$, so that
\[
 \prod_a\E\gamma_{r_{i-1}-a+1}^{-\beta\tau_{i,a}/2}=\Bigl(\frac\beta2\Bigr)^{\frac\beta2\sum_a\tau_{i,a}}\frac{\prod_{\ell(p)>i}\Gamma\bigl(\beta(f_i-\varepsilon_p)/2\bigr)}{\prod_{t=e_i+1}^{r_{i-1}}\Gamma(\beta t/2)} .
\]
No moment is worse than borderline, because every Gamma argument is nonnegative: $f_i\ge L_F\ge\varepsilon_p$ for every pair with $i\ne\ell(p)$, and the cancelled arguments $f_i-c$ with $c\le\nu_{i-1}-1$ are at least $1$, since $f_i\ge\nu_i\ge\nu_{i-1}$. A factor $\Gamma(0)$, that is, a divergent moment, occurs exactly when $f_i=\varepsilon_p$, and the replacement turns it into $1$. Consider next the Gamma functions of $\beta(\nu_i+u)/2$, which come from the bracket in \eqref{eq:star} through $\bar r_i+t=n+\nu_i-r_i+t$ and from the first display through $f_i+t=\nu_i+e_i+t$. Together they give $\prod_{u=e_i+1}^n\Gamma(\beta(\nu_i+u)/2)$. The remaining Gamma functions of multiples of $\beta/2$ telescope to $\prod_jS_\beta(e_j)/S_\beta(n)$, because $r_{i-1}=r_i+e_i$ and $r_{N-1}=e_N$. Finally, the exponent of $\beta/2$ is $\beta/2$ times
\[
 C+\sum_{i<N}r_i\bar r_i+\sum_j(N-j)e_j\bar r_j+\sum_{j<N}j\,r_j(\bar r_{j+1}-\bar r_j)=C+(N-1)\sum_je_j\bar r_j=NC,
\]
by Abel summation and $\sum_je_j\bar r_j=C$. This is the prefactor of \eqref{eq:Kbeta}.
\end{proof}

\begin{proof}[Proof of Proposition~\ref{prop:particle}]
For $\beta=2$ every Gamma function in \eqref{eq:Kbeta} has an integer argument. Fix a particle $p$ and a ladder $i\ne\ell(p)$ with $f_i\ne\varepsilon_p$. If $\nu_i>\varepsilon_p$, then all particles on ladder $i$ lie above $\varepsilon_p$ and $\Gamma(f_i-\varepsilon_p)=\Gamma(\nu_i-\varepsilon_p)\prod_{p'\text{ on }i}(\varepsilon_{p'}-\varepsilon_p)$. If $\nu_i\le\varepsilon_p$, then ladder $i$ is occupied up to $f_i-1\ge\varepsilon_p$, and $\Gamma(f_i-\varepsilon_p)=\prod_{p'\text{ on }i,\ \varepsilon_{p'}>\varepsilon_p}(\varepsilon_{p'}-\varepsilon_p)$. In the omitted case $f_i=\varepsilon_p$ the last product is empty, so we may use it there as well. Now take the product over all pairs. The multiset of particle energies is the Fermi occupation $(n_L)$ for every minimizer, so the factors $\Gamma(\nu_i-\varepsilon_p)$ give $\prod_LR_L^{n_L}$. The differences give the Vandermonde factor $V$, except for the pairs on a common ladder, which contribute $\prod_jS_2(e_j)=\prod_j\prod_{l<e_j}l!$. Hence
\[
 \prod\nolimits^*\Gamma(f_i-\varepsilon_p)=\frac{V\prod_LR_L^{n_L}}{\prod_jS_2(e_j)},\qquad
 \frac{\prod_p\Gamma(\varepsilon_p+1)}{G}=\frac1{\prod_j\prod_{u=e_j+1}^n\Gamma(\nu_j+u)},
\]
where the second identity holds because ladder $j$ contributes $\Gamma(\nu_j+u)$ for $u\le e_j$ to the numerator and for $u\le n$ to $G$. Inserting both identities into \eqref{eq:Kbeta} gives $K_2(e)=\cA/(B\,C!)$, which is the claim, since $S_2(n)=S_n$ and $\kappa_F=\cA/C!$.
\end{proof}

\subsection{\texorpdfstring{Proof of Theorem~\ref{thm:closed} for closed shells}{Proof of Theorem D for closed shells}}\label{sec:divisors}

Let the shell be closed. Then $e$ is the unique minimizer, and Lemmas~\ref{lem:evaluation} and \ref{lem:particleform} and Proposition~\ref{prop:particle} give $I_\beta(e)=K_\beta(e)$ for both fields, with $I_2(e)=\kappa_F$. It remains to show that the real constant equals the Morse--Bott integral, and we do this by comparing the residues over $\R$ and over $\C$ through the divisors of a principalization.

We first encode the residues in a measure. Let $m=q+1$ be the order of the pole of $\E\,g^{-s}$ at $s=\lambda$, which is the same for both fields by Theorem~\ref{thm:real}(i) and Corollary~\ref{cor:lct}; here $m=1$, and we shall use $m=2$ in Section~\ref{sec:onelog}. The argument in our proof of Theorem~\ref{thm:real}(ii), which applies verbatim to complex entries after replacing the real dimension $D$ by $2D$ and Theorem~\ref{thm:real}(i) by Corollary~\ref{cor:lct}(ii), shows that
\[
 \mathcal R_\beta^{(m)}(\chi)=\lim_{s\uparrow\lambda}(\lambda-s)^m\int g^{-s}\chi\,dA,\qquad\chi\in C_c^\infty(\F^D),
\]
defines a positive Radon measure on $\F^D$ with $\mathcal R_\beta^{(m)}(\varphi_\beta)=\lambda\,(m-1)!\,\kappa$, where $\kappa=\kappa_\R$ for $\beta=1$ and $\kappa=\kappa_F$ for $\beta=2$. We call $\mathcal R_\beta^{(m)}$ the \emph{residue measure} of order $m$, and for $m=1$ we write $\mathcal R_\beta=\mathcal R_\beta^{(1)}$. We call a set $Y_0\subset\C^D$ \emph{real constructible} if it is a finite Boolean combination of zero sets of polynomials with real coefficients, and we put $Y_0(\R)=Y_0\cap\R^D$. The following lemma transfers the support of these measures from $\C$ to $\R$.

\begin{lemma}[Support transfer]\label{lem:transfer}
Let $Y_0\subset\C^D$ be real constructible. If $\mathcal R_2^{(m)}$ gives no mass to $\C^D\setminus Y_0$, then $\mathcal R_1^{(m)}$ gives no mass to $\R^D\setminus Y_0(\R)$.
\end{lemma}

\begin{proof}
Let $\pi\colon Y\to\R^D$ be a principalization of the ideal of entries of $P$ over $\R$, with divisors $E_i$, numerical data $(a_i,k_i)$ and complexification $\pi_\C$, as in the proof of Lemma~\ref{lem:comparison}. In the local form of that proof, $g^{-s}$ times the Jacobian of $\pi_\F$ is $\prod_i|u_i|^{\beta k_i-2sa_i}$ times a positive real-analytic function. Since $\E\,g^{-s}$ is finite for $s<\lambda$ and every compactly supported weight is bounded by a multiple of $\varphi_\beta$, positivity gives $(k_i+1)/a_i\ge C$ for every divisor with points over $\F$, and we call $E_i$ \emph{critical} if equality holds; this condition is the same for both fields. A noncritical coordinate then has $\beta k_i-2\lambda a_i>-\beta$, so that its factor is integrable uniformly for $s\le\lambda$, whereas a critical coordinate contributes $|u_i|^{-\beta+2(\lambda-s)a_i}$, and for continuous $f$ on $\F$
\[
 (\lambda-s)\int_{|u|\le1}|u|^{-\beta+2(\lambda-s)a}f(u)\,du\longrightarrow\frac{|S^{\beta-1}|}{2a}\,f(0)\qquad(s\uparrow\lambda),
\]
with the uniform bound $|S^{\beta-1}|\sup|f|/(2a)$. A partition of unity writes $\int g^{-s}\chi\,dA$ as a finite sum of chart integrals. A chart through which $p$ critical divisors pass contributes $O((\lambda-s)^{-p})$, and positivity shows that $p\le m$, since the pole of $\E\,g^{-s}$ has order $m$. Taking the displayed limit in the critical coordinates of the charts with $p=m$, and dominated convergence in the remaining ones, we obtain
\[
 \mathcal R_\beta^{(m)}(\chi)=\sum_I\int_{E_I(\F)}(\chi\circ\pi_\F)\,w_I\,d\sigma_I .
\]
Here $I$ runs over the sets of $m$ critical divisors whose intersection $E_I$ has points over $\F$, $\sigma_I$ is a smooth positive measure on $E_I(\F)$, and $w_I\ge0$, which collects the constants $|S^{\beta-1}|/(2a_i)$ for $i\in I$, the noncritical factors and the positive function of the local form, is locally integrable, independent of $\chi$, and positive on the dense open subset of $E_I(\F)$ that meets no other divisor.

Consider first complex entries, and let $Z$ be an irreducible component of $E_I$. By assumption the constructible set $Z\cap\pi_\C^{-1}(\C^D\setminus Y_0)$ is null for $w_I\sigma_I$. A constructible subset of an irreducible variety either contains a dense Zariski-open subset or lies in a proper Zariski-closed subset, and only the second alternative is compatible with being null, because a nonempty Zariski-open subset of $Z$ meets the dense open set on which $w_I>0$ in a nonempty open set. Hence $Z\cap\pi_\C^{-1}(\C^D\setminus Y_0)\subset W_Z$ for a proper Zariski-closed subset $W_Z\subsetneq Z$.

Consider now real entries, and a set $I$ of $m$ critical divisors with $E_I(\R)\ne\varnothing$. These divisors are defined over $\R$ (proof of Lemma~\ref{lem:comparison}), and their complexifications carry the same numerical data, so they are critical over $\C$. Let $Z_\R$ be a connected component of $E_I(\R)$, and $Z$ the irreducible component of $E_I(\C)$ that contains it. Since $Z$ is smooth, $Z_\R$ is a real-analytic manifold of dimension $\dim_\C Z$, and each of its nonempty open subsets is Zariski dense in $Z$. Hence $W_Z\cap Z_\R$ is a closed analytic subset of $Z_\R$ with empty interior, and it is null for $\sigma_I$. A point $z\in Z_\R$ with $\pi(z)\notin Y_0(\R)$ satisfies $\pi_\C(z)\notin Y_0$, so it lies in $W_Z$. Therefore the term of $\mathcal R_1^{(m)}$ indexed by $I$ gives no mass to $\R^D\setminus Y_0(\R)$, and summing over $I$ proves the lemma.
\end{proof}

We return to the closed shell, where $m=1$. Lemma~\ref{lem:localMB} and a partition of unity give an open set $U\subset\F^D$ containing the set $S_e^{\mathrm{gen}}$ of generic points of $S_e$, on which $\mathcal R_\beta=\lambda\,\mathcal M$, where $\mathcal M$ is the measure in \eqref{eq:MB}, so that $\mathcal M(\varphi_\beta)=I_\beta(e)$. The measure $\mathcal M$ is carried by $S_e^{\mathrm{gen}}$, because in the coordinates of Lemma~\ref{lem:slice} the non-generic points are those at which one of the products \eqref{eq:Mchain} of the free matrices $\mathsf B_i$ fails to be injective, which is a proper algebraic condition on the $\mathsf B_i$ and hence $\operatorname{vol}_e$-null. Therefore $\mathcal R_\beta\ge\lambda\mathcal M$, with equality exactly when $\mathcal R_\beta$ gives no mass to the complement of $S_e^{\mathrm{gen}}$. For complex entries $\lambda=C$ and $\mathcal R_2(\varphi_2)=C\kappa_F=C\,I_2(e)=\lambda\mathcal M(\varphi_2)$, so equality holds because $\varphi_2>0$. The set $S_e^{\mathrm{gen}}$ is real constructible: it is cut out by the rank conditions of Proposition~\ref{prop:geometry} together with $\operatorname{rank}(A_N\cdots A_{j+1})=d_j-r_j$ for all $j$, which express the injectivity of the maps $\mathsf M_j$ because $A_N\cdots A_{j+1}$ vanishes on $V_j$ when $P=0$. Lemma~\ref{lem:transfer} therefore gives equality for real entries as well, and
\[
 \kappa_\R=\frac{\mathcal R_1(\varphi_1)}{\lambda}=I_1(e)=K_1(e),
\]
which proves Theorem~\ref{thm:closed} for closed shells.

\subsection{Open shells}\label{sec:conjevidence}\label{sec:onelog}

For an open shell the Morse--Bott integral \eqref{eq:MB} diverges by Lemma~\ref{lem:evaluation}, but only mildly. The pairs omitted from $\prod^*$ are those in which $p$ is a particle at the Fermi level and $i$ is a tied ladder with no particle there, and there are exactly $h(\mu-h)=q$ of them. By Lemma~\ref{lem:particleform} each corresponds to a moment $\E\gamma_\nu^{-\beta\nu/2}$, which diverges logarithmically at the origin, while all other moments are finite. This matches the $q$ logarithms. The residue rule of Conjecture~\ref{conj:open} replaces each divergent moment by the coefficient of its pole and multiplies by the volume $B$ of the window of scales, as in Figure~\ref{fig:triangle}. A first consistency check is that the result does not depend on the minimizer, although the divergent moments do.

\begin{proposition}\label{prop:etaindep}
For every size vector $d$, $K_1(e)$ and $K_2(e)$ do not depend on the minimizer $e$.
\end{proposition}

\begin{proof}
Any two minimizers are joined by moves in which one particle at the Fermi level passes from a tied ladder $l$ to a tied ladder $i$ that has no particle there, so it suffices to compare the two sides of one move. Before the move $f_l=L_F+1$ and $f_i=L_F$, and after it the two values are exchanged. Let $e_l$ and $e_i$ be the occupations before the move. Pairs that involve neither ladder $l$ nor ladder $i$ are unchanged. A particle on a third ladder enters through its pairs with $l$ and with $i$, whose two factors are exchanged, so these are unchanged as well; this includes the case of a Fermi particle, where one of the two pairs is omitted. The moved particle has unchanged pairs with the other ladders, and its pairs with $l$ and $i$ are omitted before and after. The particles on ladder $i$, at gaps $x=1,\dots,e_i$ below $L_F$, contribute $\prod_{x=1}^{e_i}\Gamma(\beta x/2)/\Gamma(\beta(x+1)/2)$ to the ratio of pair products. The remaining particles on ladder $l$ contribute $\prod_{x=1}^{e_l-1}\Gamma(\beta(x+1)/2)/\Gamma(\beta x/2)$. Altogether the pair product changes by $\Gamma(\beta e_l/2)/\Gamma(\beta(e_i+1)/2)$. The superfactorials change by the inverse factor, the denominators of \eqref{eq:Kbeta} gain and lose the same factor $\Gamma(\beta(L_F+1)/2)$, and the prefactor depends only on $N$ and $C$.
\end{proof}

Conjecture~\ref{conj:open} holds when $n=1$ or $N=2$. For $n=1$ there is a single particle, at energy $L_F=C-1$ on a tied ladder $j_*$. Its pairs with the other $\mu-1$ tied ladders are omitted, and its pairs with the remaining ladders give $\Gamma((d_j'-C)/2)$. The superfactorials cancel, the denominators give $\prod_{j\ne j_*}\Gamma(d_j'/2)$, and with $B=1/(\mu-1)!$ this is Theorem~\ref{thm:real}(iv). For $N=2$ and $d=(n,m,p)$ with $n\le m\le p$, consider first an open shell, $k=\tau$, and the minimizer $e=(k+1,r)$ with $r=n-k-1$. Then $f=(L_F+1,L_F)$, and the tie reads $m-r=p-k$. The pairs give $S_1(r+1)/\Gamma(\frac12)$ and $S_1(k)$, the superfactorials give $S_1(k+1)S_1(r)/S_1(n)$, and the denominators are $\prod_{i<r}\Gamma(\frac{m-i}2)\prod_{i\le k}\Gamma(\frac{p-i}2)$. Expanding the multivariate Gamma functions in Proposition~\ref{prop:twofactors}(b) as $\Gamma_r(a)=\pi^{r(r-1)/4}\prod_{i<r}\Gamma(a-\frac i2)$ produces the same Gamma functions. The two denominators differ only by $\Gamma(\frac{p-k}2)/\Gamma(\frac{m-r}2)=1$, and the powers of $2$ agree because $C=mn-r(r+1)$ in this case. For a closed shell, Theorem~\ref{thm:closed} and Proposition~\ref{prop:twofactors}(a) are proved independently, and the same expansion shows that they agree. With $e=(k,n-k)$ the pairs give $\pi^{-k(k-1)/4}\Gamma_k(\frac{p-m+n-k}2)\,\pi^{-(n-k)(n-k-1)/4}\Gamma_{n-k}(\frac{m-p+k}2)$, the denominators give the same powers of $\pi$ times $\Gamma_{n-k}(\frac m2)\Gamma_k(\frac p2)$, and the superfactorials give $S_1(k)S_1(n-k)/S_1(n)=\Gamma_k(\frac k2)/\Gamma_k(\frac n2)$. The powers of $2$ agree because $-C=k(n+p-m-k)-np$.

For $q=1$ the conjecture is part of Theorem~\ref{thm:closed}, which we prove below by an exact local normal form along the intersection of the two top components. For $q\ge2$, $N\ge3$ and $n\ge2$ it is supported by three facts. Its complex counterpart is a theorem (Proposition~\ref{prop:particle} with Theorem~\ref{thm:A}), the rule is independent of the minimizer, and the computation for three $2\times2$ factors in Section~\ref{sec:threefactors} reproduces it from the window of scales.

We now prove Theorem~\ref{thm:closed} when $q=1$, that is, $\mu=2$ and $h=1$. With sorted sizes the two tied ladders are the first two, and the minimizers are $e=(a,b,0,\dots,0)$ and $e'=(a-1,b+1,0,\dots,0)$ with $a+b=n$, according to whether the Fermi particle sits on ladder $1$ or on ladder $2$. Thus $r_1=b$ and $r_2=0$, so that $P_2=A_2A_1$ vanishes on both components, and the layers $j\ge3$ are passive. The tie between $e$ and $e'$ reads $a-b=d_2-d_1+1$. We write $P=MX$ with $M=A_N\cdots A_3$ and $X=A_2A_1$, and we let $Y^\circ$ be the set of $A$ with $\operatorname{rank}A_1=b$, $\ker A_2=\operatorname{Im}A_1\oplus\ell$ for a line $\ell$, and $M$ injective. It is contained in $S_e\cap\overline{S_{e'}}$, and it is the open dense part of the locus of $S_e$ where $\mathsf B_2=A_2|_{V_1^\perp}$ has corank one. By Lemma~\ref{lem:evaluation}, applied with the minimizer $e$, the Morse--Bott integral $I_\beta(e)$ is a product of finite moments and a single borderline one, $\E\,t^{-\beta a/2}$ with $t=t_{2,\bar r_1}\sim\gamma_a$, which comes from $\det(\mathsf M_1^*\mathsf M_1)^{-\beta e_1/2}$; indeed $e_1=a=\bar r_2-\bar r_1+1$ by the tie. For a parameter $\sigma<\beta a/2$ let $I_\beta(e;\chi;\sigma)$ denote \eqref{eq:MB} with this exponent $-\beta e_1/2$ replaced by $-\sigma$.

\begin{lemma}\label{lem:normalform}
Let $y\in Y^\circ$. There are analytic coordinates $(z,u,v,w)$ on a neighbourhood $U$ of $y$, with $v,w\in\F^a$ and $z$ a coordinate along $Y^\circ$, in which the ideal of entries of $P$ is $(u)+(v\otimes w)$ and
\[
\begin{aligned}
 S_e\cap U&=\{u=w=0\}\cap U,\\
 S_{e'}\cap U&=\{u=v=0,\ w\ne0\}\cap U,\\
 \overline{S_{e'}}\cap U&=\{u=v=0\}\cap U,\qquad
 Y^\circ\cap U=\{u=v=w=0\}\cap U.
 \end{aligned}
\]
Moreover $g=\Phi^*\mathcal Q(z)\Phi$ with $\Phi=(u,vw)$, where $\mathcal Q(z)$ is positive definite and depends only on the base coordinates $z$, real-analytically in $z$ and $\bar z$, and on $S_e$ near $y$ we have $\det(\mathsf M_1^*\mathsf M_1)=\varpi(z)\,|\Lambda(z)v|^2$ with $\varpi>0$ continuous and $\Lambda(z)$ injective.
\end{lemma}

\begin{proof}
Since $M$ is injective near $y$, some $d_2\times d_2$ submatrix $M_R$ formed by a set $R$ of rows is invertible near $y$, and $M^\dagger=M_R^{-1}\circ(\text{restriction to the rows }R)$ is a left inverse that is rational in the entries, hence holomorphic over $\C$. So the ideal of entries of $P$ is that of $X$. Choose orthonormal bases adapted to $y$, with $V=\operatorname{Im}A_1(y)$ and $K=\ker A_1(y)$, and write $A_1$ in blocks with respect to $\F^n=K^\perp\oplus K$ and $\F^{d_1}=V\oplus V^\perp$. Its block $a_{11}$ is invertible near $y$, and $A_1=L_1\operatorname{diag}(a_{11},\mathsf S_1)U_1$ with the Schur complement $\mathsf S_1\in\F^{(d_1-b)\times a}$ and unipotent $L_1,U_1$ depending on the other three blocks. Right multiplication by $U_1^{-1}$ does not change the ideal, and $A_2\mapsto\tilde A_2=A_2L_1$ is a change of coordinates. Hence the ideal is generated by the entries of $\tilde A_2|_V$ and of $\tilde A_2|_{V^\perp}\mathsf S_1$. Next write $\tilde A_2|_{V^\perp}$ in blocks with respect to $V^\perp=\ell^\perp\oplus\ell$ and $\F^{d_2}=W\oplus W^\perp$, where $W=\operatorname{Im}A_2(y)$, and factor it as $L_2\operatorname{diag}(b_{11},v)U_2$. Here $b_{11}$ is invertible, $L_2$ and $U_2$ are unipotent, and $v\in\F^{d_2-(d_1-b-1)}=\F^a$ is the Schur complement of the $\ell$-column, whose length equals $a$ by the tie. Writing $U_2\mathsf S_1=\binom{\mathsf S'}{w}$ with $w\in\F^{1\times a}$ the $\ell$-row, we obtain
\[
 X\,U_1^{-1}=\bigl(\tilde A_2|_V\,a_{11},\;L_2\tbinom{b_{11}\mathsf S'}{vw}\bigr).
\]
The coordinates are $u=(\tilde A_2|_V,\mathsf S')$, $v$, $w$, and $z$, consisting of the remaining blocks of $A_1$ and $\tilde A_2$ together with the passive layers. The ideal is $(u)+(v\otimes w)$. On $\{u=0\}$ the rank identity $\operatorname{rank}A_1=b+\operatorname{rank}\mathsf S_1=b+\operatorname{rank}w$ holds. The stratum $S_e$ requires $\operatorname{rank}A_1=b$, hence $w=0$, whereas $S_{e'}$ requires $\operatorname{rank}A_1=b+1$, hence $w\ne0$, as well as $v=0$; both require $\tilde A_2|_V=0$. Therefore $\{u=v=0\}$ is the closure of $S_{e'}$ near $y$, and $Y^\circ=\{u=v=w=0\}$ is disjoint from $S_{e'}$. The displayed formula shows that $X=\Psi(z)\Phi$ for a linear map $\Psi(z)$ that is injective and depends only on $z$, so $g=\nF{M\Psi(z)\Phi}^2$ has the stated form. Finally, on $S_e$ we have $\mathsf M_1=M\mathsf B_2$, and the Gram determinant of the columns of $M\mathsf B_2$ is, up to a positive continuous factor coming from the choice of complement of $V_1$, the Gram determinant of the columns of $M\tilde A_2|_{V^\perp}U_2^{-1}=ML_2\operatorname{diag}(b_{11},v)$. This equals the Gram determinant of its first block times the squared distance of the last column $ML_2\iota_{W^\perp}v$ from the image of that block, which is $|\Lambda(z)v|^2$ with $\Lambda(z)=\Pi_{(ML_2W)^\perp}ML_2\iota_{W^\perp}$.
\end{proof}

\begin{lemma}\label{lem:localdouble}
Let $\chi\ge0$ be smooth with support in a sufficiently small neighbourhood of a point of $Y^\circ$. Then $(\beta a/2-\sigma)\,I_\beta(e;\chi;\sigma)$ converges, as $\sigma\uparrow\beta a/2$, to a limit $\mathcal R^\circ(\chi)$, and $(\lambda-s)^2\int g^{-s}\chi\,dA\to\lambda\,\mathcal R^\circ(\chi)$ as $s\uparrow\lambda$.
\end{lemma}

\begin{proof}
In the coordinates of Lemma~\ref{lem:normalform} write Lebesgue measure as $J\,dz\,du\,dv\,dw$ with $J>0$ analytic, and let $c_v=\beta a$ and $c_u=\beta C-c_v$ be the real dimensions of $v$ and of $u$, so that $2\lambda=c_u+c_v$. Let $Q_z$ be the $u$-block of $\mathcal Q(z)$ and $\tilde{\mathcal Q}_z$ its Schur complement on the $vw$-block, and for a unit vector $\theta$ let $\mathcal Q_{z,\theta}$ be the positive definite quadratic form $w\mapsto\tilde{\mathcal Q}_z(\theta w)$.

Consider first the Morse--Bott side, and write $v=r\theta$. At a point $(z,0,v,0)$ of $S_e$ the quadratic part of $g$ in the normal variables $(u,w)$ has real determinant $\det(Q_z)\,r^{2c_v}\det(\mathcal Q_{z,\theta})$. The Morse--Bott density with respect to $dz\,dv$ is therefore $\operatorname{vol}(B_{\beta C})\,\chi J\det(Q_z)^{-1/2}r^{-c_v}\det(\mathcal Q_{z,\theta})^{-1/2}$. By Lemma~\ref{lem:normalform}, replacing the exponent $-\beta a/2$ by $-\sigma$ multiplies it by $(\varpi(z)|\Lambda(z)v|^2)^{c_v/2-\sigma}$, which is $r^{c_v-2\sigma}$ times a continuous positive function. Since $dv=r^{c_v-1}dr\,d\theta$, we obtain
\[
 \mathcal R^\circ(\chi)=\tfrac12\operatorname{vol}(B_{\beta C})\int(\chi J)(z,0,0,0)\det(Q_z)^{-1/2}\int_{S^{c_v-1}}\det(\mathcal Q_{z,\theta})^{-1/2}\,d\theta\,dz .
\]

Consider next the zeta function. The substitution $v=\rho_1\theta_1$, $w=\rho_2\theta_2$, $u=\rho_1\rho_2u'$ gives $g=\rho_1^2\rho_2^2\,\Phi'^*\mathcal Q(z)\Phi'$ with $\Phi'=(u',\theta_1\theta_2)$, and
\[
 \int g^{-s}\chi\,J\,dz\,du\,dv\,dw=\int_0^R\!\!\int_0^R(\rho_1\rho_2)^{2(\lambda-s)-1}F(s;\rho_1,\rho_2)\,d\rho_1\,d\rho_2,
\]
where $F(s;\rho)$ is the integral of $(\Phi'^*\mathcal Q(z)\Phi')^{-s}(\chi J)(z,\rho_1\rho_2u',\rho_1\theta_1,\rho_2\theta_2)$ over $z$, $\theta_1$, $\theta_2$ and $|u'|<R/(\rho_1\rho_2)$. Since $\mathcal Q$ is bounded below by a positive constant and $|\theta_1\theta_2|=1$, the integrand is at most a constant times $(1+|u'|^2)^{-s}$, which is integrable over $\R^{c_u}$ for $s>c_u/2$. By dominated convergence, $F$ is continuous on $[0,R)^2\times[\lambda-\eta,\lambda]$. Splitting the square at a small $\rho_0$ then shows that $\int\!\!\int(\rho_1\rho_2)^{\alpha-1}(F(s;\rho)-F(s;0))\,d\rho=o(\alpha^{-2})$ as $\alpha=2(\lambda-s)\downarrow0$, so that $(\lambda-s)^2\int g^{-s}\chi\,J\,dz\,du\,dv\,dw\to F(\lambda;0)/4$. At $\rho=0$ the variable $u'$ runs over all of $\R^{c_u}$. Completing the square in $u'$ and using
\[
 \int_{\R^{c_u}}(x^*Qx+\xi)^{-\lambda}dx=\frac{\pi^{c_u/2}\,\Gamma(c_v/2)}{\Gamma(\lambda)}\det(Q)^{-1/2}\xi^{-c_v/2}
\]
and $\int_{S^{c_v-1}}(\theta^*Q'\theta)^{-c_v/2}d\theta=|S^{c_v-1}|\det(Q')^{-1/2}$, we find that $F(\lambda;0)/4$ is the double integral in the formula for $\mathcal R^\circ(\chi)$ multiplied by
\[
 \frac{\pi^{c_u/2}\,\Gamma(c_v/2)\,|S^{c_v-1}|}{4\,\Gamma(\lambda)}=\frac{2\pi^\lambda}{4\,\Gamma(\lambda)}=\frac\lambda2\operatorname{vol}(B_{\beta C}).
\]
Hence $F(\lambda;0)/4=\lambda\mathcal R^\circ(\chi)$.
\end{proof}

\begin{lemma}\label{lem:residuemass}
For every continuous $\chi$ with $|\chi|\le c\,\varphi_\beta$ the limit
\[
 \mathcal R^\circ(\chi)=\lim_{\sigma\uparrow\beta a/2}(\beta a/2-\sigma)\,I_\beta(e;\chi;\sigma)
\]
exists. It defines a positive Radon measure $\mathcal R^\circ$ on $\F^D$, which is carried by $Y^\circ$, coincides near points of $Y^\circ$ with the limit of Lemma~\ref{lem:localdouble}, and satisfies $\mathcal R^\circ(\varphi_\beta)=K_\beta(e)$.
\end{lemma}

\begin{proof}
We use the parametrization of the proof of Lemma~\ref{lem:evaluation}. Since $r_2=0$ we have $V_2=0$, so $A_2$ vanishes on $V_1$ and is given on $V_1^\perp$ by $\mathsf B_2\in\F^{d_2\times\bar r_1}$, while $M=\mathsf B_N\cdots\mathsf B_3$ and $\mathsf M_1=M\mathsf B_2$. Under the Gaussian weight, $\mathsf B_2$ is a standard Gaussian matrix, independent of the remaining variables of the parametrization, which we collect in $\omega$. These determine $V_1$, the frames in which $\mathsf B_2$ is written, the passive layers and all factors of \eqref{eq:MB} other than $\det(\mathsf M_1^*\mathsf M_1)$. We write $\mathsf B_2=O\mathsf T$ with $O$ in the Stiefel manifold and $\mathsf T$ upper triangular, as in Lemma~\ref{lem:bartlett}, so that $O$, the off-diagonal entries of $\mathsf T$ and the squared diagonal entries $t_{2,a'}$ are independent, and $t=t_{2,\bar r_1}\sim\gamma_a$ by the tie. Then
\[
 \det(\mathsf M_1^*\mathsf M_1)=\mathcal G\prod_{a'\le\bar r_1}t_{2,a'},\qquad \mathcal G=\det(O^*M^*MO),
\]
where $\mathcal G$ depends only on $O$ and $\omega$. With the product $\mathcal N>0$ of the prefactor and the brackets in \eqref{eq:star} we therefore have
\[
 I_\beta(e;\chi;\sigma)=\mathcal N\,\E\Bigl[\frac{\chi}{\varphi_\beta}(A)\,t^{-\sigma}\mathcal W_\sigma\Bigr],\qquad \mathcal W_\sigma=\mathcal W\,\mathcal G^{-\sigma}\prod_{a'<\bar r_1}t_{2,a'}^{-\sigma},
\]
where $\mathcal W\ge0$ collects the factors that do not depend on $\sigma$, and $A=A(\mathsf T,O,\omega)\in S_e$ is the point with these coordinates. It depends linearly on $\mathsf T$, whose last diagonal entry is $t^{1/2}$, hence continuously on $t\in[0,\infty)$, and we put $A_0=A(\mathsf T_0,O,\omega)$, where $\mathsf T_0$ is $\mathsf T$ with its last diagonal entry replaced by $0$.

The variable $t$ has density $c_at^{\beta a/2-1}e^{-\beta t/2}$ with $c_a=(\beta/2)^{\beta a/2}/\Gamma(\beta a/2)$, and for every bounded $f$ that is continuous at $0$,
\[
 (\beta a/2-\sigma)\,c_a\!\int_0^\infty f(t)\,t^{\beta a/2-\sigma-1}e^{-\beta t/2}\,dt\longrightarrow c_af(0)\qquad(\sigma\uparrow\beta a/2),
\]
with a bound $c'\sup|f|$ that is uniform for $\sigma$ near $\beta a/2$. We apply this to $f(t)=(\chi/\varphi_\beta)(A)$ for fixed values of the other variables. By Lemmas~\ref{lem:evaluation} and \ref{lem:particleform}, the parameter $\sigma$ changes only the exponents of the variables $t_{i,a'}$ with $a'\le\bar r_1$, and among all moments only that of $t$ diverges at $\sigma=\beta a/2$, because the shell has exactly one borderline moment. Since $x^{-\sigma}\le1+x^{-\beta a/2}$ for $x>0$ and $0\le\sigma\le\beta a/2$, the function $\mathcal W_\sigma$ is dominated by $\mathcal W(1+\mathcal G^{-\beta a/2})\prod_{a'<\bar r_1}(1+t_{2,a'}^{-\beta a/2})$. Expanding this product gives finitely many moments, each finite: those of $\mathcal W$ and of $\mathcal W\mathcal G^{-\beta a/2}$ are Morse--Bott moments other than that of $t$, the variables $t_{2,a'}$ are independent of $\mathcal W$ and $\mathcal G$, and $\E\,t_{2,a'}^{-\beta a/2}<\infty$ because the index of $t_{2,a'}$ is at least $a+1$. As $\mathcal W_\sigma\to\mathcal W_{\beta a/2}$ pointwise, Fubini's theorem and dominated convergence give
\[
 \mathcal R^\circ(\chi)=\mathcal N c_a\,\E\Bigl[\frac{\chi}{\varphi_\beta}(A_0)\,\mathcal W_{\beta a/2}\Bigr],
\]
so that $\varphi_\beta\mathcal R^\circ$ is the push-forward of the finite measure $\mathcal Nc_a\mathcal W_{\beta a/2}\,d\Pb$ under $A_0$. For $\chi=\varphi_\beta$ this is \eqref{eq:star} with the moment of $t$ replaced by $c_a$, which equals $K_\beta(e)$ by Lemma~\ref{lem:particleform}. For $\chi$ supported near a point of $Y^\circ$ the limit is the one of Lemma~\ref{lem:localdouble} by definition.

It remains to locate $A_0$. Almost surely the diagonal entries of $\mathsf T$ other than the last are positive, so $\mathsf T_0$ has rank $\bar r_1-1$ and $O\mathsf T_0$ has a one-dimensional kernel $\ell\subset V_1^\perp$. At $A_0$ we therefore have $\ker A_2=V_1\oplus\ell$, while $A_1$, and with it $\operatorname{rank}A_1=b$, is unchanged. Finally $M=A_N\cdots A_3$ is a product of independent standard Gaussian matrices with $d_N\ge\dots\ge d_2$, so it is injective almost surely. Hence $A_0\in Y^\circ$ almost surely, and $\mathcal R^\circ$ is carried by $Y^\circ$.
\end{proof}

We can now conclude as in Section~\ref{sec:divisors}, with pole order $m=2$. By Lemma~\ref{lem:localdouble} and a partition of unity there is an open set $U\supset Y^\circ$ on which $\mathcal R_\beta^{(2)}=\lambda\mathcal R^\circ$, and by Lemma~\ref{lem:residuemass} the measure $\mathcal R^\circ$ is carried by $Y^\circ$. Hence $\mathcal R_\beta^{(2)}\ge\lambda\mathcal R^\circ$, with equality exactly when $\mathcal R_\beta^{(2)}$ gives no mass to the complement of $Y^\circ$. For complex entries $\mathcal R_2^{(2)}(\varphi_2)=C\kappa_F=C\,K_2(e)=\lambda\mathcal R^\circ(\varphi_2)$, by Theorem~\ref{thm:A}, Proposition~\ref{prop:particle} and $B=1$, so equality holds. The set $Y^\circ=\{\operatorname{rank}A_1=b,\ A_2A_1=0,\ \operatorname{rank}A_2=d_1-b-1,\ \operatorname{rank}(A_N\cdots A_3)=d_2\}$ is real constructible, and Lemma~\ref{lem:transfer} gives equality for real entries as well. Therefore $\lambda\kappa_\R=\mathcal R_1^{(2)}(\varphi_1)=\lambda\mathcal R^\circ(\varphi_1)=\lambda K_1(e)$, and $K_1(e')=K_1(e)$ by Proposition~\ref{prop:etaindep}, which completes the proof of Theorem~\ref{thm:closed}.

The smallest examples with one logarithm and more than two factors, for $n=2$ and $n=3$, are $d=(2,2,3,4)$, with $\kappa_\R=\frac1{32}$, and $d=(3,3,3,5)$, with $\kappa_\R=\sqrt2/(630\pi^{3/2})$. In the first case we can also integrate out $A_2$ and the passive factor directly, as in the proof of Proposition~\ref{prop:twofactors}. With $\Theta=A_3^{\top}A_3$, whose eigenvalues we denote by $\sigma_i$, and with $w_k$ the eigenvalues of $A_1^{\top}A_1$, we find $\E\,e^{-(t/2)g}=\E\prod_{i\le3,\,k\le2}(1+t\sigma_iw_k)^{-1/2}$, where $w_1=y/t$ is small and $w_2$ runs through a logarithmic window. The coefficient of $t^{-2}\log t$ is then $\frac14\,\E\bigl[\det\Theta^{-1/2}\int_0^\infty y^{-1/2}\det(I+y\Theta)^{-1/2}dy\bigr]=\frac14$, and the factor $\frac1{2^2\Gamma(3)}$ again gives $\frac1{32}$.

\section{Applications and discussion}\label{sec:applications}

\subsection{Outage of Rayleigh-product MIMO channels}\label{sec:wireless}
Let $H=H_N\cdots H_1$ have independent standard complex Gaussian factors of sizes $d_j\times d_{j-1}$, each with entry density $\pi^{-1}e^{-|z|^2}$ (so that $\E|H_{j,ab}|^2=1$). Here $d_0$ is the number of transmit antennas, $d_N$ the number of receive antennas, and $d_1,\dots,d_{N-1}$ are scatterer or keyhole dimensions \cite{GBGP02,Mul02,YB07}. Let $\rho$ be the signal-to-noise ratio (SNR) on a linear scale. With optimal beamforming (BF) the effective SNR is $\rho\nop H^2$, so the outage probability at fixed rate $R$ is $P_{\rm out}^{\rm BF}=F_\op((2^R-1)/\rho)$. For an orthogonal space-time block code (OSTBC) of rate $r_c$ and normalization $c_0$ the effective SNR is proportional to $\rho\nF H^2$, which gives the outage probability $P_{\rm out}^{\rm OSTBC}=F_F(c_0(2^{R/r_c}-1)/\rho)$. Other normalizations of the channel reduce to these by rescaling $\rho$.

\begin{corollary}\label{cor:outage}
As $\rho\to\infty$,
\[
 P_{\rm out}^{\rm BF}\sim\kappa_\op\,(2^R-1)^C\,\rho^{-C}(\log\rho)^q,\qquad
 P_{\rm out}^{\rm OSTBC}\sim\kappa_F\,\bigl(c_0(2^{R/r_c}-1)\bigr)^C\rho^{-C}(\log\rho)^q .
\]
\end{corollary}

The diversity exponent $C$ and its dependence on the widths are due to Yang and Belfiore \cite{YB11,YB07}. Corollary~\ref{cor:outage} adds the logarithmic factor $(\log\rho)^q$ and the coding-gain constant for all sizes. Channels with a single antenna at one node, among them keyhole channels, were treated in \cite{JMWG08,GL07,Zho10,ZRJW12}, and for multi-layer channels the diversity--multiplexing tradeoff and closed-form outage approximations were obtained in \cite{Tar20,WZCT15}.

For illustration, four independent $2\times2$ complex Gaussian hops have $C=3$, $q=4$, $\kappa_\op=1/48$ and $\kappa_F=1/72$. At $30$~dB, that is $\rho=10^3$, the logarithmic factor of the leading term is already $(\log1000)^4\simeq2.28\times10^3$. The full logarithmic structure at order $\rho^{-C}$ is given by the polynomial $\mathcal P_\op$ of Theorem~\ref{thm:grass}: the outage probability equals $\rho^{-C}(2^R-1)^C\mathcal P_\op\bigl(\log\rho-\log(2^R-1)\bigr)+O(\rho^{-C-\eta})$, so the lower Pl\"ucker coefficients determine how much of the logarithmic factor survives at a given finite SNR.

\subsection{Deep linear networks}\label{sec:dln}
For zero-target regression with whitened inputs the population loss is the energy $H(A)=\tfrac12\nF P^2$ of Section~\ref{sec:picture}. With a standard Gaussian prior on all weights the evidence is its partition function $\cZ(t)=\E\,e^{-tH}=Z(t/2)$, and Theorem~\ref{thm:real} gives
\[
 -\log\cZ(t)=\tfrac C2\log t-q\log\log t-\log\bigl(2^{C/2}\Gamma(\tfrac C2+1)\kappa_\R\bigr)+o(1).
\]
For complex weights the same expansion holds with $C$ in place of $C/2$ and the explicit constant $-\log(2^C\cA)$. For real weights it is explicit whenever $q\le1$, by Theorem~\ref{thm:closed}, and in particular for every network with two layers, where Proposition~\ref{prop:twofactors} writes it in multivariate Gamma functions. Two layers give zero-target reduced-rank regression, whose threshold and multiplicity were found by Aoyagi and Watanabe \cite{AW05}. For one-dimensional output, $n=1$, it is given by Theorem~\ref{thm:real}(iv); in this case Hanin and Zlokapa computed the evidence exactly in terms of Meijer-$G$ functions \cite{HZ23}. For example, three square layers of width $3$ have $\kappa_\R=\pi^2/192$, so that $2^{3}\Gamma(4)\kappa_\R=\pi^2/4$ and $-\log\cZ(t)=3\log t-\log(\pi^2/4)+o(1)$.

\begin{remark}[Gaussian readout]\label{rem:readout}
Integrating out the last layer $A_N$ for fixed $P_{N-1}=A_{N-1}\cdots A_1$ gives
\[
 \cZ_\R(t)=\E\det(I_{d_0}+tP_{N-1}^{\top}P_{N-1})^{-d_N/2},\qquad
 \cZ_\C(t)=\E\det\bigl(I_{d_0}+\tfrac t2P_{N-1}^*P_{N-1}\bigr)^{-d_N},
\]
for the population loss $H$ and unit-variance entries in both cases. The evidence of an $N$-layer network is thus an average over the law of the $(N-1)$-fold product, and the width of the readout layer enters only as an exponent.
\end{remark}

\subsection{Discussion}\label{sec:open}
We have described the small-ball behaviour of Gaussian matrix products through a single filling rule. The unifying object is the Laplace transform $Z(t)$, which up to the rescaling $t\mapsto t/2$ is a low-temperature partition function on parameter space and, for deep linear networks, a Bayesian evidence, and our results describe its singular behaviour through an open Fermi shell whose determinant is graded by the Schubert cells of a Grassmannian. For complex entries the exponent, the logarithmic power and the constants are determined by the shell data, and the shell determinant has an explicit Schubert-graded Pl\"ucker expansion. For real entries the Frobenius small-ball probability has a positive leading term with exponent $C/2$ and logarithmic power $q$, the operator-norm probability has the same order, and the constant $B\,K_1(e)$, the particle form at $\beta=1$ multiplied by the shell factor, is proved for every number of factors when $q\le1$, which includes all products of two factors, and for vector chains. Its extension to $q\ge2$ is Conjecture~\ref{conj:open}, the main open problem. For real entries with $N,n\ge2$ the leading constant of the operator-norm probability is also open, and so, when $q\ge1$, are the lower coefficients of the pole of $\E\,\nF P^{-2s}$ at $s=C/2$.

In the proof of Theorem~\ref{thm:grass} the Grassmannian appears as $\Gr(h,\C[u]_{<\mu})$, and the shell polynomials are the Pl\"ucker coordinates of a translate of the span of the polynomials $\xi_j$ built from the principal parts of the last shell (Section~\ref{sec:proofC}). It would be interesting to see $\Gr(h,\mu)$ in the geometry of $\Sigma_d$ itself, for instance by identifying the window of scales, the triangle of Figure~\ref{fig:triangle} for three real $2\times2$ factors, with the dual complex of the log canonical places of the pair $(\C^D,C\cdot I_d)$.

There are a number of other interesting directions. Small balls around a matrix of rank $r>0$ shift the sizes by $r$ at the level of exponents \cite{YB07}, but the logarithms and constants require non-central product ensembles. The Newton-basis argument of Section~\ref{sec:proofA} uses only that the Mellin weights are products of Gamma functions. Products with truncated unitary factors \cite{KKS16}, whose weights are Gamma ratios, should therefore have a similar shell structure with finite ladders. It would also be interesting to know whether the particle form, in which every factor is attached to a pair formed by an occupied and an empty state, has a meaning for other values of $\beta$, and whether the Pl\"ucker coordinates $p_{T'}(0)$, which carry the subleading Laurent coefficients (Section~\ref{sec:proofC}), have a geometric meaning.

\backmatter
\pdfbookmark[1]{Acknowledgements}{acknowledgements}\bmhead{Acknowledgements}
JT is supported by National Natural Science Foundation of China under Grant No. 12405085 and by the Natural Science Foundation of Shanghai (Grant No. 24ZR1419300). GPT-6 Astra and Claude Opus 5.5 are used to prepare the numerics. JT thanks the organizers of East Asia Joint Workshop on Fields and Strings 2026 at KIAS, where this work is finalized. JT would like to thank Ying Zhang for her love and support.

\bmhead{Competing interests} There are no competing interests.

\bmhead{Data availability} The code and the numerical results mentioned in Section~\ref{sec:threefactors} are available at \numericsurl. All mathematical definitions and derivations needed for the main results are contained in the article.

\appendix
\section{Width-permutation invariance}\label{app:invariance}

We sketch an elementary proof of the following fact, which is contained in \cite[Sec.~III.B]{IK14} for all three Dyson indices: \emph{for $\F\in\{\R,\C\}$, the law of the multiset of nonzero singular values of $P$ depends only on the multiset $\{d_0,\dots,d_N\}$.} Write $\beta'=\tfrac12$ for $\F=\R$ and $\beta'=1$ for $\F=\C$, so that $\E\,e^{-\beta'\operatorname{tr}(BMB^*)}=\det(I+M)^{-\beta'p}$ for a standard Gaussian $B$ with $p$ rows and $M\ge0$.

\begin{lemma}
For independent standard Gaussian $A\in\F^{q\times d}$ and $B\in\F^{p\times q}$, the law of $(BA)^*(BA)$ is symmetric in $(p,q)$.
\end{lemma}

\begin{proof}
For $\Lambda\ge0$, let $a=\Lambda^{1/2}$, let $W_q=A^*A$, and let $W_p=H^*H$ with $H\in\F^{p\times d}$ an independent standard Gaussian matrix. Then
\[
 \E\,e^{-\beta'\operatorname{tr}(\Lambda(BA)^*BA)}=\E\det(I+aW_qa)^{-\beta'p}=\E\,e^{-\beta'\operatorname{tr}(aW_qaW_p)}.
\]
The last expression is symmetric under $W_p\leftrightarrow W_q$ by cyclicity of the trace.
\end{proof}

\begin{proof}[Sketch of the invariance]
We start from a necklace identity. Let $U$ and $V$ be fixed matrices of compatible sizes, and let $G$ be a standard Gaussian matrix of the shape of $UPV$. Up to the normalization of $G$,
\[
 e^{-\beta'\nF{UPV}^2}=\E_G\,e^{i\,\mathrm{Re}\operatorname{tr}(G^*UPV)} .
\]
Integrating out any single factor $A_k$ instead turns $\E\,e^{-\beta'\nF{UPV}^2}$ into the same functional of the open chain $A_{k-1}\cdots A_1VG^*UA_N\cdots A_{k+1}$.

Consider first the swaps of interior sizes. Take $U=I$, $V=\Lambda^{1/2}$ and integrate out $A_{j+2}$. This writes the Laplace transform of $P^*P$ as $\E\,e^{-\beta'\operatorname{tr}(X^*X\,RR^*)}$, where $X=A_{j+1}A_j$ and $R$ is independent of $X$ with a law not involving $d_j,d_{j+1}$. The lemma then gives invariance of the law of $P$ under $d_j\leftrightarrow d_{j+1}$ for $1\le j$ and $j+1\le N-1$.

To reach all permutations we apply this to the chains $PH$ and $G^*P$ representing
\[
 \Phi(t)=\E\prod_m\det(I+t_mP^*P)^{-\beta'}.
\]
This shows invariance of $\Phi$ under permutations of $(d_0,\dots,d_{N-1})$ and of $(d_1,\dots,d_N)$. For $N\ge2$ these subgroups generate the symmetric group $\mathfrak S_{N+1}$; for $N=1$ use $P\mapsto P^*$. Finally, $\Phi$ determines the law of the nonzero eigenvalues of $P^*P$ by the Stone--Weierstrass theorem.
\end{proof}

\bibliography{refs}
\end{document}